\documentclass{aa}  

\usepackage{graphicx}
\usepackage{longtable}
\usepackage{verbatim}
\usepackage{float}
\usepackage{placeins}
\usepackage{array,multirow,makecell}
\usepackage{xcolor, soul}
\usepackage{threeparttablex}
\usepackage[export]{adjustbox}
\usepackage{lscape}
\usepackage{dirtytalk}
\usepackage[autostyle]{csquotes}
\usepackage[version=4]{mhchem}
\usepackage{silence}
\newcommand{\pcms}{cm$^{-2}$}
\newcommand{\pcmc}{cm$^{-3}$}
\newcommand{\kms}{km s$^{-1}$}
\newcommand{\mjybeamkms}{mJy.beam$^{-1}$.km s$^{-1}$}

\usepackage{txfonts}
\usepackage{natbib}
\bibpunct{(}{)}{;}{a}{}{,} 
\usepackage{hyperref}
\hypersetup{
    colorlinks=True,
    citecolor=blue,
    urlcolor=magenta,      
    linkcolor=blue,
    }
\begin{document}

   \title{Complex Organic Molecules towards the central molecular zone of  NGC\,253}
\titlerunning{Complex species towards the CMZ of  NGC\,253 }
   \subtitle{}

   \author{M. Bouvier
          \inst{\ref{leiden}},
          S. Viti \inst{\ref{leiden},\ref{univ. Bonn},\ref{UCL}},
          J. G. Mangum
          \inst{\ref{NRAO}},
          C. Eibensteiner\inst{\ref{NRAO}}\thanks{Jansky Fellow of the National Radio Astronomy Observatory},
          E. Behrens
          \inst{\ref{Univ. Virginia},\ref{NRAO}},
          V. M. Rivilla \inst{\ref{CAB}},
           \'A. L\'opez-Gallifa \inst{\ref{CAB}},
          S. Mart\'in \inst{\ref{ESO-chile},\ref{alma-obs}},
          N. Harada \inst{\ref{NAOJ},\ref{dept. astr.japan}},
          S. Muller \inst{\ref{chalmers}},
          L. Colzi \inst{\ref{CAB}},
          K. Sakamoto \inst{\ref{IAA-Taiwan}}
          }
\authorrunning{Bouvier et al.}
   \institute{\label{leiden} Leiden Observatory, Leiden University, P.O. Box 9513, 2300 RA Leiden, The Netherlands\\
              \email{bouvier@strw.leidenuniv.nl}
              \and \label{univ. Bonn}
              Transdisciplinary Research Area (TRA) ‘Matter’/Argelander-Institut f\"ur Astronomie, University of Bonn
              \and \label{UCL}
              Physics and Astronomy, University College London, UK
                \and \label{NRAO}
              National Radio Astronomy Observatory, 520 Edgemont Road, Charlottesville, VA 22903-2475, USA
              \and\label{Univ. Virginia}
              Department of Astronomy, University of Virginia, P.~O.~Box 400325, 530 McCormick Road, Charlottesville, VA 22904-4325, USA
               \and \label{CAB}
            Centro de Astrobiolog\'ia (CAB), CSIC-INTA, Carretera de Ajalvir km 4, Torrej\'{o}n de Ardoz, 28850 Madrid, Spain
            \and  \label{ESO-chile}   
             European Southern Observatory, Alonso de C\'ordova, 3107, Vitacura, Santiago 763-0355, Chile
            \and \label{alma-obs}
             Joint ALMA Observatory, Alonso de C\'ordova, 3107, Vitacura, Santiago 763-0355, Chile
               \and \label{NAOJ}    
             National Astronomical Observatory of Japan, 2-21-1 Osawa, Mitaka, Tokyo 181-8588, Japan
             \and \label{dept. astr.japan}
            Department of Astronomy, School of Science, The Graduate University for Advanced Studies (SOKENDAI), 2-21-1 Osawa, Mitaka, Tokyo, 181-1855 Japan
             \and \label{chalmers}
             Department of Space, Earth and Environment, Chalmers University of Technology, Onsala Space Observatory, SE-43992 Onsala, Sweden 
             \and \label{IAA-Taiwan}
             Institute of Astronomy and Astrophysics, Academia Sinica, 
             11F of AS/NTU Astronomy-Mathematics Building, 
             No.1, Sec. 4, Roosevelt Rd, Taipei 106319, Taiwan % for K. Sakamoto
             }

   \date{}

% \abstract{}{}{}{}{} 
% 5 {} token are mandatory
 
  \abstract
  % context heading (optional)
  % {} leave it empty if necessary  
   {Interstellar complex organic molecules (iCOMs) may have a link to prebiotic species, key building blocks for life. In Galactic star-forming (SF) regions, spatial variations in the iCOMs emission could reflect the source physical structure or different chemical formation pathways. Investigating iCOMs in extragalactic SF regions may thus provide crucial information about these regions.
}
  % aims heading (mandatory)
   {As an active extragalactic SF region, the central molecular zone (CMZ) of the nearby galaxy NGC\,253 provides an ideal template for studying iCOMs under more extreme conditions. We aim to investigate the emission of a few selected iCOMs and understand if a difference between the iCOMs could reflect on the source's chemical or physical structure. }
  % methods heading (mandatory)
   {Using the high angular resolution ($\sim 27$ pc) observations from the ALCHEMI ALMA large program, we imaged the emission of selected iCOMs and precursors; \ce{CH3CHO}, \ce{C2H5OH}, \ce{NH2CHO}, \ce{CH2NH}, and \ce{CH3NH2}. We estimated the gas temperatures and column densities of the iCOMs using a rotational diagram analysis, and by performing a non-LTE analysis for \ce{CH2NH}.}
  % results heading (mandatory)
   {The iCOM emission concentrates mostly towards the inner part of the CMZ of NGC\,253 and can be reproduced with two gas components. Different emission processes can explain iCOM emission towards the CMZ of NGC\,253: at Giant Molecular Cloud (GMC) scales ($\sim 27$ pc), the iCOMs could trace large-scale shocks whilst at smaller scales (a few parsecs), both shock and heating processes linked with ongoing star formation may be involved. Using column density correlation trends and known formation pathways, we find that more than one formation path could be involved to explain the iCOM emission. Finally, we found chemical differences between the GMCs, such as a decrease of abundance for the N-bearing species towards one of the GMCs or different excitation conditions for \ce{NH2CHO} and \ce{CH3CHO} towards two of the GMCs. }
  % conclusions heading (optional), leave it empty if necessary 
   {}

   \keywords{astrochemistry - galaxies: ISM - Galaxies: starburst - ISM: molecules - methods: observational}

   \maketitle
%
%-------------------------------------------------------------------

\section{Introduction}

One of the major topics in astrochemistry is to understand how complex species can form and survive under the harsh conditions of the interstellar medium (ISM). In particular, interstellar Complex Organic Molecules (iCOMs) are carbon-bearing molecules of at least 6 atoms \citep{herbst2009, ceccarelli2017}. Other species which do not fit formally in this category are also relevant for prebiotic chemistry. This is the case of methanimine (\ce{CH2NH}), the simplest imine and an important precursor of glycine \citep[e.g.][]{theule2011, danger2011}. iCOMs have been of particular interest due to their potential in forming pre-biotic molecules, themselves important building blocks for life. Within our Galaxy, iCOMs are particularly abundant in various star-forming (SF) regions (e.g., \citealt{blake_1987, herbst2009, belloche2019, jorgensen2020, ceccarelli2023, chen2023}) and our Galactic Centre \citep[see e.g.,][references therein]{jimenez-serra2025}. However, despite numerous observations, understanding iCOM formation pathways is challenging. Two main chemical paradigms have been invoked in the literature: only grain surface synthesis and the combination of grain surface and gas phase synthesis (see \citealt{ceccarelli2023} and references therein for a review). 

Multiple observations of iCOMs in Galactic SF Regions, at both low and high angular resolution, revealed a dichotomy among iCOMs, in particular between O- and N-bearing species \citep[e.g.,][]{blake_1987, caselli1993, widicus_weaver_deep_2017, codella2017,suzuki2018, csengeri2019, bogelund_NvsO_2019, vanderwalt_2021,mininni2023, peng2022, busch2024, bouscasse2024}. The cause for such segregation between N-bearing and O-bearing species is not clear but several hypotheses have been proposed. These include a difference in thermal history \citep[e.g.,][]{caselli1993, viti_evaporation_2004, suzuki2018, garrod2022} or in the formation and destruction pathways of the species \citep[e.g.,][]{busch2024}, the intrinsic characteristics of the chemical species (e.g., binding energies; \citealt{bianchi2022}), or the influence of external forces (e.g., shocks; \citealt{tercero2018, csengeri2019, busch2024}). Studying iCOMs in external galaxies, under the assumption that they trace similar physical processes as in Galactic SF regions, may thus provide important information about the physics and the chemistry of the dense gas of extra-galactic SF regions. 

iCOMs have been studied in several external galaxies, such as \ce{CH3OH} in IC 342 \citep[][]{henkel1987}, NGC\, 6946 \citep{eibensteiner2022},  the Small Magellanic Cloud \citep{shimonishi_detection_2023} or PKS 1830-211 \citep{muller2021}; \ce{CH3CN} in M82 \citep[e.g.][]{mauersberger1991} and Arp 220 \citep{martin_2011}; \ce{CH3NH2} in PKS 1830-211 \citep{muller2011}; \ce{HC5N} in NGC\,4418 \citep{costagliola2015}; \ce{CH3OCH3} in the Large Magellanic cloud \citep{sewilo_detection_2018, golshan2024} or towards NGC\,1068 \citep{qiu2018}, and \ce{C2H5OH} in NGC\,253 \citep{martin_alchemi_2021}, to cite a few examples. Among these galaxies, the origin of iCOMs was constrained in only a few of them: \ce{CH3OH}, \ce{CH3OCH3}, and \ce{CH3OCHO} trace hot cores in the Small and Large Magellanic Clouds (SMC and LMC, respectively; \citealt{sewilo_detection_2018, shimonishi_detection_2023}), \ce{CH3OH} traces shocks in NGC\,253, NGC\,1068, and M\,82 (e.g. \citealt{martin_2006, humire_methanol_2022, huang2024}, Huang et al. subm.), and \ce{CH2NH} has been found to be enhanced in the outflow of IC 860 \citep{gorski2023}. Studies of iCOMs in extragalactic regions are thus still very limited to a few extragalactic sources, and a handful of iCOMs. 

While Galactic studies offer high-resolution data on molecular cloud structure and chemistry, they are limited to relatively moderate star formation conditions. NGC\,253 is a nearby (3.5 Mpc; \citealt{rekola_distance_2005}) relatively edge-on (inclination of 76$^\circ$; \citealt{mccormick_2013}) galaxy hosting a starburst nucleus in its inner kpc \citep{bendo_2015,leroy_alma_2015}.  NGC\,253 provides a unique laboratory to test what we know from Galactic chemistry studies under more extreme conditions — intense radiation fields ($G_0 \sim 10^4-10^5$; \citealt{harada_alchemi_2022}) and cosmic-ray ionisation rate ($\zeta > 10^{-14}$s$^{-1}$;e.g. \citealt{holdship_energizing_2022, behrens2024}), intense star formation (2$M_{\odot}$yr$^{-1}$; \citealt{leroy_alma_2015}) and strong associated feedback processes that drive large-scale outflows \citep{mccarthy1987}. By studying iCOMs in NGC 253, we can assess whether the same assumptions used in Galactic star-forming regions hold in more extreme environments, or whether additional chemical and dynamical processes dominate. This is particularly relevant for understanding the role of GMCs in starbursts and high-redshift galaxies, where similar conditions prevail but are much harder to observe in detail. So rather than merely confirming what we already know from the Milky Way, NGC 253 allows us to probe how universal our chemical diagnostics are and whether they remain valid in an active nuclear starburst environment.

The central molecular zone (CMZ) of NGC\,253 has been surveyed by the ALMA Comprehensive High-resolution Extragalactic Molecular Inventory (ALCHEMI; \citealt{martin_alchemi_2021}) large programme at a resolution of $1.6 \arcsec$ (27 pc). Previous ALCHEMI studies have revealed the chemical richness of the CMZ and how it can be used to better understand its structure and physics of the CMZ of NGC\,253 (e.g., \citealt{harada_starburst_2021, holdship_energizing_2022, humire_2020,  behrens_tracing_2022, behrens2024, huang_reconstructing_2023, tanaka_2023, harada_pca_2024, bouvier2024, kishikawa2024}, López-Gallifa et al. in prep.). Therefore, ALCHEMI provides an ideal template to study iCOMs toward a nuclear extragalactic starburst, at a high enough angular resolution to investigate the origin of their emission for the first time. In particular, we aim to study a selection of N- and O-bearing iCOMs detected towards the CMZ of NGC\,253 \citep{martin_alchemi_2021} and investigate whether spatial segregation between iCOMs at GMC scales could reflect different gas physical conditions (temperature, density), environmental conditions (e.g., heating, shocks) or evolutionary stage within the CMZ, and whether we can constrain their region of emission, and their formation pathways. In this study, we focussed on a sample of O- and N-bearing iCOMs, namely acetaldehyde (\ce{CH3CHO}), ethanol (\ce{C2H5OH}), formamide (\ce{NH2CHO}), methylamine (\ce{CH3NH2}), and methanimine (\ce{CH2NH}), all previously detected in \cite{martin_alchemi_2021}. In addition, some of these species could be chemically linked (\ce{CH3CHO} with \ce{C2H5OH}; e.g., \citealt{skouteris2018} and \ce{CH3NH2} with \ce{CH2NH}; e.g., \citealt{woon2002, dejesus2021, molpeceres_ch3nh2_2024}) which will allow us to address the question of their formation pathways. 

%structure paper
In Section \ref{sec:obs} we describe the observations. We present the emission distribution and the results from the  Gaussian fit of the lines in Sec.~\ref{sec:emission_maps_fit} and we derive the physical parameters in Sec.~\ref{sec:phys_params}. We discuss the possible chemical pathways involved in the formation of the surveyed iCOMs, their region of emission and the possible differences between the regions and galaxies in Sec.~\ref{sec:discussion}. We summarise our results in Sec.~\ref{sec:conclusions}.

%%%%%%%%%%%%%%%%%%%%%%%%%%%%%%%%%%%%%%%%%%

\section{Observations}\label{sec:obs}

We used the observations from the ALCHEMI Large Program (co-PIs.: S. Mart\'in, N. Harada, and J. Mangum). We only give here the relevant information for the present work. Further details about the the data acquisition, calibration and imaging are provided in the ALCHEMI summary paper \citep{martin_alchemi_2021}. The observations were performed during Cycles 5 and 6, under the project codes 2017.1.00161.L and 2018.1.00162.S. Bands 3 through 7 were covered, with frequencies ranging between 84.2 and 373.2 GHz. The observations were centred at the position $\alpha$(ICRS)$=00^h47^m33.26^s$ and $\delta$(ICRS)$=-25^\circ 17'17.7''$.  The common region imaged corresponds to a rectangular area of size $50'' \times 20''$ (850 $\times$ 340 pc) with a position angle of 65$^\circ$, covering the CMZ of NGC\,253. The data cubes have a common beam size of $1.6\arcsec\times 1.6\arcsec$ ($\sim 27$ pc), a maximum recovered scale $\gtrsim 15''$ ($\gtrsim 255$ pc), and a spectral resolution of $\sim 10$ km s$^{-1}$. Finally, the adopted flux calibration uncertainty is 15\%, following \cite{martin_alchemi_2021}.

We used the continuum-subtracted FITS image cubes provided by ALCHEMI containing the targeted species in this work. We produced velocity-integrated maps (see Sec.~\ref{sec:emission_maps_fit}) from the processed FITS image cubes using the packages MAPPING and CLASS from GILDAS\footnote{\url{http://www.iram.fr/IRAMFR/GILDAS}}. We used these same packages to extract the spectra and perform Gaussian line fitting. The spectra were extracted from beam-sized regions before being converted to brightness temperature following \cite{bouvier2024}. For each spectrum, we fitted a zeroth order polynomial to the line-free channels from the continuum-subtracted data cubes to retrieve the information about the spectral root-mean-square (rms). 

%%%%%%%%%%%%%%%%%%%%%%%%%%%%%%%%%%%%%%%%%%

\section{Emission distribution and Gaussian line fitting results}\label{sec:emission_maps_fit}

The CMZ of NGC\,253 is a region of 300 pc $\times$ 100 pc where 10 Giant Molecular Clouds (GMCs; named from GMC\,1 to GMC\,10) of a size of about 30 pc in diameter have been identified in both molecular and continuum emission \citep{sakamoto_2011, leroy_alma_2015}. At a higher angular resolution, it was found that the GMCs located in the inner 160 pc in length of the CMZ (corresponding to GMCs 3, 4, 5 and 6) can be resolved into multiple star-forming clumps called proto-super star clusters (hereafter pSSCs; e.g., \citealt{ando_2017, leroy_forming_2018}), which are compact ($\leq 1$ pc), massive ($M\geq 10^5$M$_\odot$), and young ($\leq 3$ Myr) clusters with ongoing star formation.

In the following, we focus our analysis in the inner part of the CMZ, which includes the regions named GMC\,3 to GMC\,7 \citep{leroy_alma_2015}. This is where all the five iCOMs investigated in this work are the most intense. However, whilst the peak of the iCOMs emission coincides well with the location of GMC\,6 and GMC\,7, it is slightly shifted ($\lesssim 1.6^{\prime\prime}$) with respect to GMC\,3 and GMC\,4 (see Figure~\ref{fig:sup-icoms}a) and seems to be more consistent with the location of pSSCs. Since later on (see Section~\ref{sec:lines}) we centred the region where we extracted the spectra on the coordinates of the pSSC\,2 and pSSC\,5 sources, we use the name of these pSSCs for the regions that would correspond to GMC\,3 and GMC\,4, respectively. Note that the angular resolution of ALCHEMI does not allow distinguishing each pSSC, which means that our position named pSSC\,2 in fact includes pSSCs 1 to 3, and our position named pSSC\,5 includes pSSCs 4 to 7 (see \citealt{leroy_forming_2018} for the pSSCs distribution). The region GMC\,6 hosts pSSC 14, but since the coordinates used to extract the spectra are centred on the coordinates of GMC\,6 and not on those of pSSC\,14 (which slightly differs), we kept the name GMC\,6. No pSSCs have been detected towards GMC\,7. 

Finally, the emission of the iCOMs studied in this work is relatively faint towards GMC\,5, which lies close to the kinematic centre. Moreover, this region shows particular absorption features due to the strong radio continuum emission \citep{meier_alma_2015, humire_methanol_2022} and would require a specific radiative transfer modelling, which is beyond the scope of this paper. We thus excluded this GMC from our analysis. As a result, in the rest of the paper, we focus on positions that we named as GMC\,7, GMC\,6, pSSC\,2, and pSSC\,5, whose coordinates are listed in Table~\ref{tab:region_selected}. Below, we describe and compare the emission distribution of the selected species towards these four regions in Sec.~\ref{sec:maps}, and we describe the extraction and analysis of the spectral lines in Sec.~\ref{sec:lines}.

\subsection{Emission distribution}\label{sec:maps}

To perform moment maps, we looked for transitions isolated enough (i.e. with either no contribution from other adjacent species or with a contamination from adjacent features that are below 3$\sigma$ level) within the range of velocity on which we integrate the emission. We then integrated the emission over the full line profile, taking into account the velocity shift across the regions, hence within the range [70 ; 380] \kms. The transitions used to perform these moment maps are labelled in the spectra of the brightest region, GMC\,6, region in Figures~\ref{fig:spec_ch3cho_GMC6} to ~\ref{fig:spec_c2h5oh_GMC6}. The corresponding spectral parameters of the transitions are shown in Table~\ref{tab:lines}.

\begin{figure*}
    \centering
    \includegraphics[width=1\linewidth]{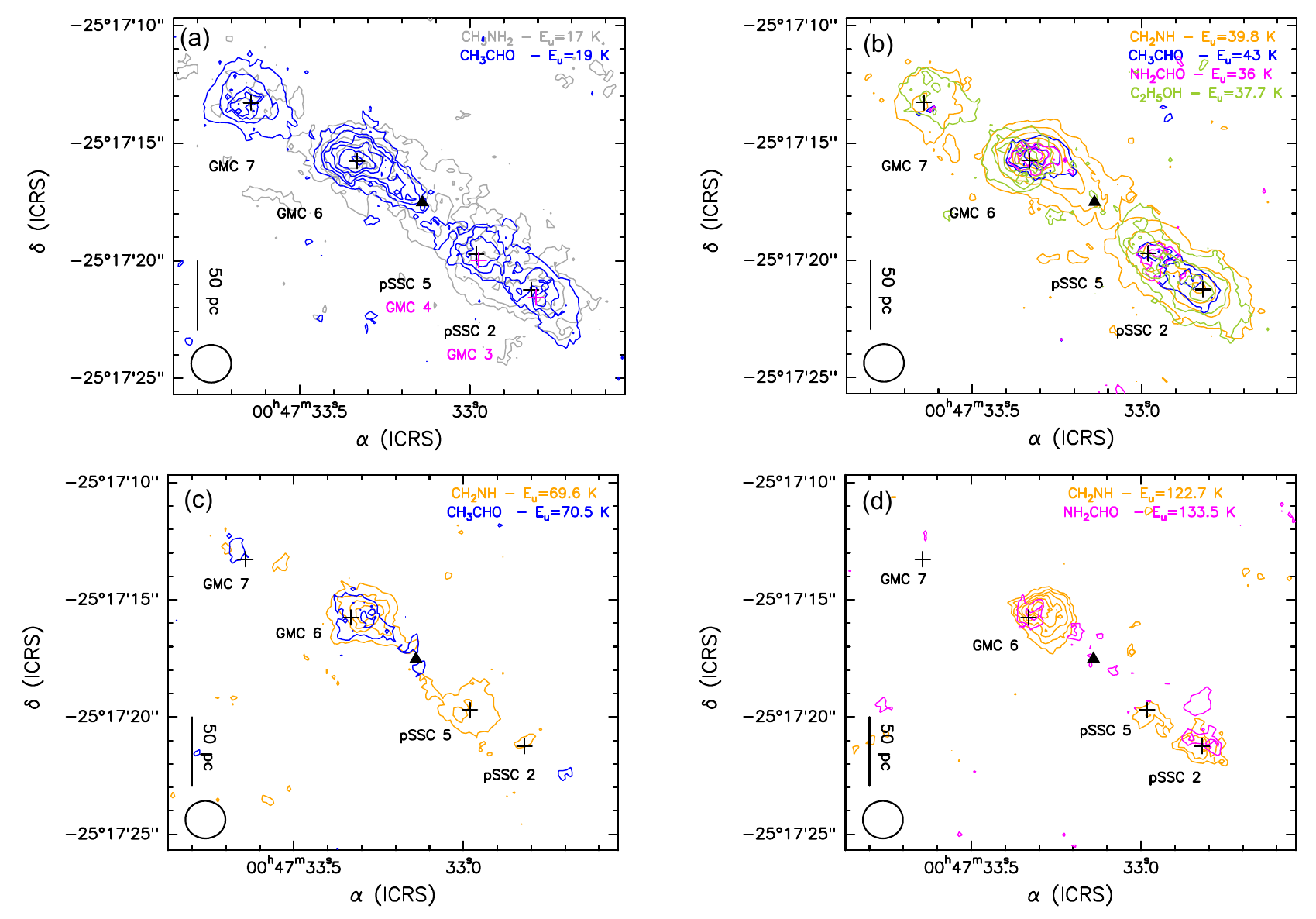}
    \caption{Overlap of contours corresponding to the velocity integrated species emission, ordered by their upper energy level, $E_{\mathrm{u}}$. A spatial scale of 50 pc corresponds to $\sim 3\arcsec$. The kinematic centre ($\alpha_{\text{ICRS}}=00^h47^m33.14^s$ and $\delta_{\text{ICRS}}=-25^\circ 17'17.52''$; \citealt{muller-sanchez2010}) is labelled by a filled black triangle. The black crosses mark the positions analysed in this work and correspond to GMC\,7, GMC\,6, pSSC\,5 and pSSC\,2.  For reference, the position of GMC\,3 and GMC\,4 \citep{leroy_alma_2015} are indicated by magenta crosses in the panel (a). The beam size if depicted in the lower left corner of each  box. Contours of \ce{CH3NH2}, \ce{CH3CHO}, \ce{CH2NH}, \ce{NH2CHO}, and \ce{C2H5OH} are in grey, blue, orange, magenta, and green, respectively. \textit{(a)} Contours overlap of \ce{CH3NH2} ($3_{-1,1}-2_{0,0}; E_{\text{u}}=17$ K) and \ce{CH3CHO} ($1_{1,5}-1_{0,5}$ (E+A); $E_{\text{u}}=19$ K). Contours start at $3\sigma$ with steps of 3$\sigma$ with $1\sigma=42$ and 25 \mjybeamkms, respectively. \textit{(b):} Contours overlap of \ce{CH2NH} ($4_{1,3}-3_{1,2}; E_{\text{u}}=39.8$ K) with $1\sigma=78$ \mjybeamkms, \ce{CH3CHO} ($9_{2,9}-8_{1,8}$ (E+A); $E_{\text{u}}=43$ K) with $1\sigma=55$ \mjybeamkms, \ce{NH2CHO}  ($8_{0,8}-7_{0,7}$; $E_{\text{u}}=36$ K) with $1\sigma=58$ \mjybeamkms, and \ce{t-C2H5OH} ($6_{4}-5_{3}$; $E_{\text{u}}=37.7$ K) with $1\sigma=64$ \mjybeamkms. Contours start at $3\sigma$ with steps of 3$\sigma$ for \ce{CH3CHO} and \ce{t-C2H5OH}, $8\sigma$ for \ce{CH2NH}, and  $2\sigma$ for \ce{NH2CHO}. \textit{(c):} Contours overlap of \ce{CH2NH} ($6_{1,6}-5_{1,5}; E_{\text{u}}=69.6$ K) with $1\sigma=42$ \mjybeamkms and \ce{CH3CHO} ($11_{2,9}-10_{2,8}$ (E+A); $E_{\text{u}}=70.5$ K) $1\sigma=25$ \mjybeamkms. Contours start at $3\sigma$ with steps of 5$\sigma$ and $3\sigma$ for \ce{CH2NH} and \ce{CH3CHO}, respectively. \textit{(d)} Contours overlap of \ce{CH2NH} ($8_{1,7}-8_{0,8}; E_{\text{u}}=122.7$ K) with $1\sigma=61$ \mjybeamkms and \ce{NH2CHO} ($15_{2,14}-14_{2,13}$; $E_{\text{u}}=133.5$ K) $1\sigma=78$ \mjybeamkms. Contours start at $3\sigma$ with steps of 2$\sigma$.}
    \label{fig:sup-icoms}
\end{figure*}

Figure~\ref{fig:sup-icoms} shows the integrated line emission distribution of the five iCOMs, grouped in four $E_{\mathrm{u}}$ ranges, to minimize biases due to different excitation conditions. For each $E_{\mathrm{u}}$ range, we superposed the contour maps of the various species to visually show the overlap between their emission distribution. Since we have a limited sample of moment 0 maps for the five iCOMs, we could not always compare simultaneously their emission distribution at each $E_{\mathrm{u}}$ range. Panel (a) of Figure~\ref{fig:sup-icoms} shows the contour maps of the \ce{CH2NH} and \ce{CH3NH2} transitions close to $E_{\mathrm{u}}=20$ K. Panel (b) shows the contour overlap of all species except \ce{CH3NH2} at $E_{\mathrm{u}}=30-40$ K.  Panel (c) shows the contour maps of the \ce{CH2NH} and \ce{CH3CHO} transitions at $E_{\mathrm{u}}\sim 70$ K. Finally, panel (d) shows the contour maps of \ce{CH2NH} and \ce{NH2CHO} around $E_{\mathrm{u}} > 100$ K.

From Figure~\ref{fig:sup-icoms}, we can see a difference in the spatial emission of the iCOMs across the four regions. First, \ce{CH3NH2} shows more extended emission than \ce{CH3CHO} towards the part of the CMZ encompassing the regions GMC\,6 to pSSC\,2, as seen in  Figure~\ref{fig:sup-icoms}, panel (a). On the other hand, the emission of \ce{CH3CHO} dominates towards GMC\,7 compared to that of \ce{CH3NH2}. For the transitions with upper energy levels around $\sim 30-40$ K (see Figure~\ref{fig:sup-icoms}b), we can see that each GMC seems to show a different chemical composition: the \ce{CH2NH} and \ce{C2H5OH} emission dominate GMC\,7 and show the most extended emission throughout the regions. However, there is a lack of \ce{C2H5OH} emission towards the kinematic centre (indicated by a black triangle in Figure~\ref{fig:sup-icoms}) compared to \ce{CH2NH}. On the other hand, the emission of \ce{CH3CHO} and \ce{NH2CHO} is more compact around the GMCs. Both species show emission towards GMC\,6 and have faint emission toward GMC\,7 but whilst \ce{CH3CHO} emission covers both pSSC\,5 and pSSC\,2, \ce{NH2CHO} lacks emission towards pSSC\,2.  At $E_{\mathrm{u}} \sim 70$ K (see Figure~\ref{fig:sup-icoms}c), both \ce{CH2NH} and \ce{CH3CHO} show emission towards GMC\,6 with \ce{CH2NH} showing a slightly more spatially extended emission compared to \ce{CH3CHO}. On the other hand, only \ce{CH2NH} emits towards pSSC\,5 and the two species are fainter towards pSSC\,2 and GMC\,7. For $E_{\mathrm{u}} > 100$ K (see Figure~\ref{fig:sup-icoms}d), we see that both \ce{CH2NH} and \ce{NH2CHO} emit towards GMC\,6 and pSSC\,2 but \ce{CH2NH} shows a more extended emission. \ce{CH2NH} only emits towards pSSC\,5 and both species are absent at GMC\,7. Interestingly, whilst \ce{NH2CHO} showed fainter emission for $E_{\mathrm{u}} \sim 30-40$ K towards pSSC\,2, it shows stronger emission for $E_{\mathrm{u}} > 100$ K. Whilst the emission of the iCOMs seems to show a spatial disparity within the inner GMCs of  NGC\,253, one should keep in mind that fainter emission could be due to either a true lower abundance or a sensitivity bias if the transition is not very intense (Band 5 observations were less sensitive compared to the rest of the data; see \citealt{martin_alchemi_2021}). Hence, whether there is a disparity among the species across the region will be investigated after deriving the physical parameters (Sec.~\ref{sec:phys_params}).

Finally, we can see an evolution in the emission morphology compared to $E_{\mathrm{u}}$: Looking at \ce{CH3CHO} and \ce{CH2NH} for which we imaged transitions from 19 K to 70.5 K and from $\sim 40$ K to $\sim 123$ K, respectively, we see the emission being more compact with increasing $E_{\mathrm{u}}$. This shows that transitions with different $E_{\mathrm{u}}$ may be excited under different conditions and it is important to compare spatial extent of species which have similar upper level energies. All the velocity-integrated maps obtained for each species are shown in Figures~\ref{fig:ch3nh2-maps} to \ref{fig:c2h5oh-maps}.

\begin{table}
    \centering
        \caption{Selected regions and coordinates.}
    \label{tab:region_selected}
    \begin{tabular}{lcc}
    \hline \hline
    \multirow{2}{*}{Region} & R.A. (ICRS) & Dec. (ICRS) \\
    & ($00^h47^m$)&($-25^\circ:17'$) \\
    \hline
       GMC\,7  & $33^s.6432$ & $13\arcsec .272$ \\
        GMC\,6 & $33^s.3312$ & $15\arcsec .756$ \\
        pSSC\,5 &$32^s.9811$ & $19\arcsec .710$ \\
        pSSC\,2 & $32^s.8199$ & $21\arcsec .240$\\
        \hline
    \end{tabular}
    \tablefoot{\say{GMC} stands for giant molecular Cloud and \say{pSSC} stands for proto-super star cluster.}
\end{table}

\subsection{Spectra and line parameters}\label{sec:lines}
We extracted the spectra towards GMC\,7, GMC\,6, pSSC\,5, and pSSC\,2 from regions of a beam-size (1.6$\arcsec \times 1.6\arcsec$) and centred on the coordinates given in Table~\ref{tab:region_selected}. We then centred each spectrum at the local standard of rest velocity (LSRK) corresponding to the rest frequency of each transition used in this work before performing a Gaussian line fitting to the detected transitions. A threshold of 3$\sigma$ on the intensity peak was set to consider a line to be detected. Transitions which were contaminated by more than 15\% by a transition from another species (i.e., contributing to 15\% or more of the total integrated intensity of our desired transition), or that were too blended to perform a multiple Gaussian fit, were not taken into account. To assess whether a line is contaminated, we used the existing line identification of ALCHEMI 7m and 12m data (\citealt{martin_alchemi_2021} and priv. comm.) and performed an additional inspection using WEEDS \citep{maret_2011}, an extension of CLASS, which uses both the Jet Propulsion Laboratory (JPL; \citealt{pickett_1998}) and the Cologne Database for Molecular Spectroscopy (CDMS; e.g. \citealt{muller_2005, endres2016}) databases. However, in the case of \ce{CH3CHO}, the E- and A- states are often completely blended together, the spectral resolution of 10 \kms \ being insufficient to disentangle the two states. Similarly, for \ce{NH2CHO} and \ce{CH3NH2}, where two transitions with similar spectroscopic parameters are completely blended together. In these cases, we still kept these transitions and performed a single line Gaussian fit. We explain how we used these transitions for the analysis in Sec.~\ref{subsec:methodology}. We present the list of all of the transitions per species and per region used in this work, as well as their spectroscopic parameters, in Table~\ref{tab:lines}. We detect only the anti- form of \ce{C2H5OH} among the different rotamers, but this is not surprising as the anti form is the most stable rotamer \citep{pearson2008, bianchi2019}, so the name \ce{C2H5OH} refers to this form only. 

From the Gaussian line fitting procedure, we extracted the integrated intensities ($\int T_B.d\text{V}$) in K~\kms, the line widths (FWHM) and the peak velocity ($V_{\mathrm{peak}}$), both in \kms. The extracted spectra and the best Gaussian fits are shown in Figure~\ref{fig:spec_ch3cho_GMC6} to ~\ref{fig:spec_c2H5OH_SSC5}. The Gaussian line-fitting results and the rms for each transition and region are shown in Table~\ref{tab:lines}.

%%%%%%%%%%%%%%%%%%%%%%%%%%%%%%%%%%%%%%%%%%

\section{Derivation of physical parameters}\label{sec:phys_params}

\subsection{Methodology}\label{subsec:methodology}
For each species and each of the four selected regions, we detected enough transitions ($>7$ for most species and regions) with a large enough range of $E_{\mathrm{u}}$ covered to derive the column densities and rotational temperatures using the rotational diagram (RD) method \citep{goldsmith_1999}. This method assumes Local Thermodynamic Equilibrium (LTE) and optically thin emission. These assumptions will be revisited later on in the case of \ce{CH2NH} (see Section~\ref{sec: nonlte}). We also assume the Rayleigh-Jeans approximation holds. We included a calibration error of 15\% (see Sec.~\ref{sec:obs}) and the spectral rms in the integrated flux uncertainties. The best fit was calculated by minimizing the reduced chi-square, $\chi_r^2$. 

As a first approximation, we set the beam-filling factor\footnote{We used the formula $\eta_{ff}=\left({\theta_S^2}/(\theta_S^2+\theta_B^2)\right)$
with $\theta_S$ as the source size and $\theta_B$ as the beam size of the ALCHEMI observations.} to 1/2, assuming that our emission comes from the GMC scales (hence from a region of the size of the beam or 1.6$\arcsec$). However, in some regions and for some species, we saw a clear deviation in the rotation diagram for the higher $E_{\mathrm{u}}$ transitions. This behaviour can be either due to the presence of multiple gas components with different physical conditions or to non-LTE or opacity effects \citep{goldsmith_1999}. We favoured the first option, i.e. the presence of multiple gas components. Our choice is supported by the fact that higher upper-energy level transitions showed often a different FWHM or $V_\mathrm{peak}$ compared to the lower upper-energy transitions (see discussion in Sec.~\ref{subsec:rds} below) and the difference in emission distribution of the species depending on the excitation conditions (see Sec.~\ref{sec:maps}). For these cases, we assumed that the first component, traced by the lower upper-level energy transitions, arises from the GMCs whilst the second component, traced by the higher upper-level energy transitions, arises from the pSSCs. For the latter component, we calculated the beam-filling factor using a size for the pSSC clusters of 0.12$\arcsec$ (corresponding to $\sim 2$ pc), estimated by \cite{leroy_forming_2018}. This assumption is supported by the emission being more compact with increasing $E_{\mathrm{u}}$, sometimes unresolved by the ALCHEMI beam (See Figure~\ref{fig:sup-icoms}).

Finally, as mentioned in Sec.~\ref{sec:lines}, most of the E- and A- states of \ce{CH3CHO} and some transitions for \ce{NH2CHO}, \ce{CH3NH2} and \ce{C2H5OH} are too blended to perform a multiple Gaussian fit. As the blended transitions have the same spectroscopic parameters ($E_{\mathrm{u}}, A_{ij}, g_{\mathrm{u}}$), we thus assumed that each transition accounts for 50\% of the total line integrated intensity. Furthermore, for \ce{CH3CHO},  we assumed a \ce{CH3CHO-E}/\ce{CH3CHO-A} ratio of 1 since the kinetic temperature is higher than the energy difference of $\sim 0.1$ K between the levels of the different symmetry states \citep{matthews1985}. Therefore, we included each pair of blended transitions in the RDs by attributing 50\% of the total integrated intensity given in Table~\ref{tab:lines} to each individual transition.

\subsection{Rotational Diagrams}\label{subsec:rds}

\begin{figure*}
    \centering
    \includegraphics[width=1\linewidth]{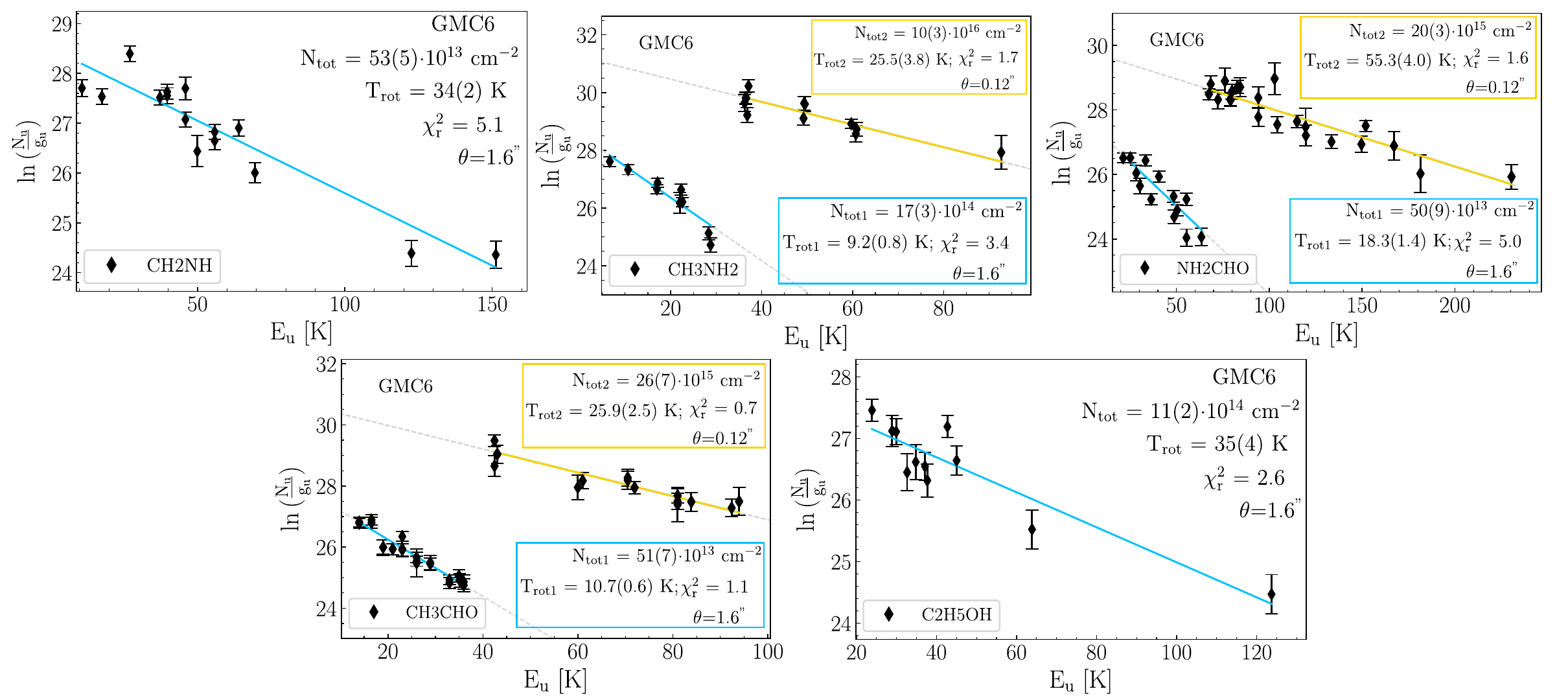}
    \caption{Rotation diagrams of each species towards  GMC\,6. The parameters $E_{\mathrm{u}}$, $N_{\mathrm{u}}$, and $g_{\mathrm{u}}$ are the level energy (with respect to the ground state), column density and degeneracy of the upper level, respectively. The error bars on $\text{ln}(N_{\mathrm{u}}/g_{\mathrm{u}}$) include a calibration error of 15\% (see Sec.~\ref{sec:obs}). The blue (and orange if a second component is fitted) solid lines represent the best fits and the dashed grey lines are the extrapolations of the fit for the full range of $E_{\mathrm{u}}$ covered. The source size assumed is indicated for each component (see Sec.~\ref{subsec:methodology}).}
    \label{fig:rds_GMC6}
\end{figure*}

In Figure~\ref{fig:rds_GMC6}, we show as an example the rotation diagrams obtained for each species towards GMC\,6, which is the region showing the most intense emission (see Sec.~\ref{sec:maps}). The RDs at other positions are presented in Appx.~\ref{appx:RDs}.   
In the case of GMC\,6, two temperature components were needed for \ce{CH3NH2}, \ce{NH2CHO}, and \ce{CH3CHO}. Two temperature components are also needed for pSSC\,2 (\ce{CH3NH2} and \ce{NH2CHO}) and for pSSC\,5 (\ce{CH3NH2} and \ce{CH3CHO}). For \ce{CH3NH2} and \ce{CH3CHO}, the deviation occurs near $E_{\mathrm{u}} \sim 40$ K and it is closer to 60 K for \ce{NH2CHO}. When two temperature components are present in the RDs, we made the assumption that they are not arising from the same spatial component. The low $E_{\mathrm{u}}$ transitions being more spatially extended compared to the higher $E_{\mathrm{u}}$ transitions (see Section~\ref{sec:maps}), we assumed that whilst the former mainly arises from GMC scales, the latter arises from more compact scales, such as the pSSC scales, the next identified structures within the GMCs. As a caveat, this implies that the contribution from the warm compact component to the cold extended one is negligible.

For the species showing two temperature components, we looked at the Gaussian fit results to check for any shift in the mean FWHM or $V_\mathrm{peak}$ as a function of the upper-level energy. In the following, we discuss first about the Gaussian fit results in the context of the two temperature components (Sec.~\ref{subsec:two-T-comp}) before discussing the results from the RDs (Sec.~\ref{subsec:rds_results}).

\subsubsection{Gaussian fits and temperature components}\label{subsec:two-T-comp}

\begin{figure*}
    \centering
    \includegraphics[width=1\linewidth]{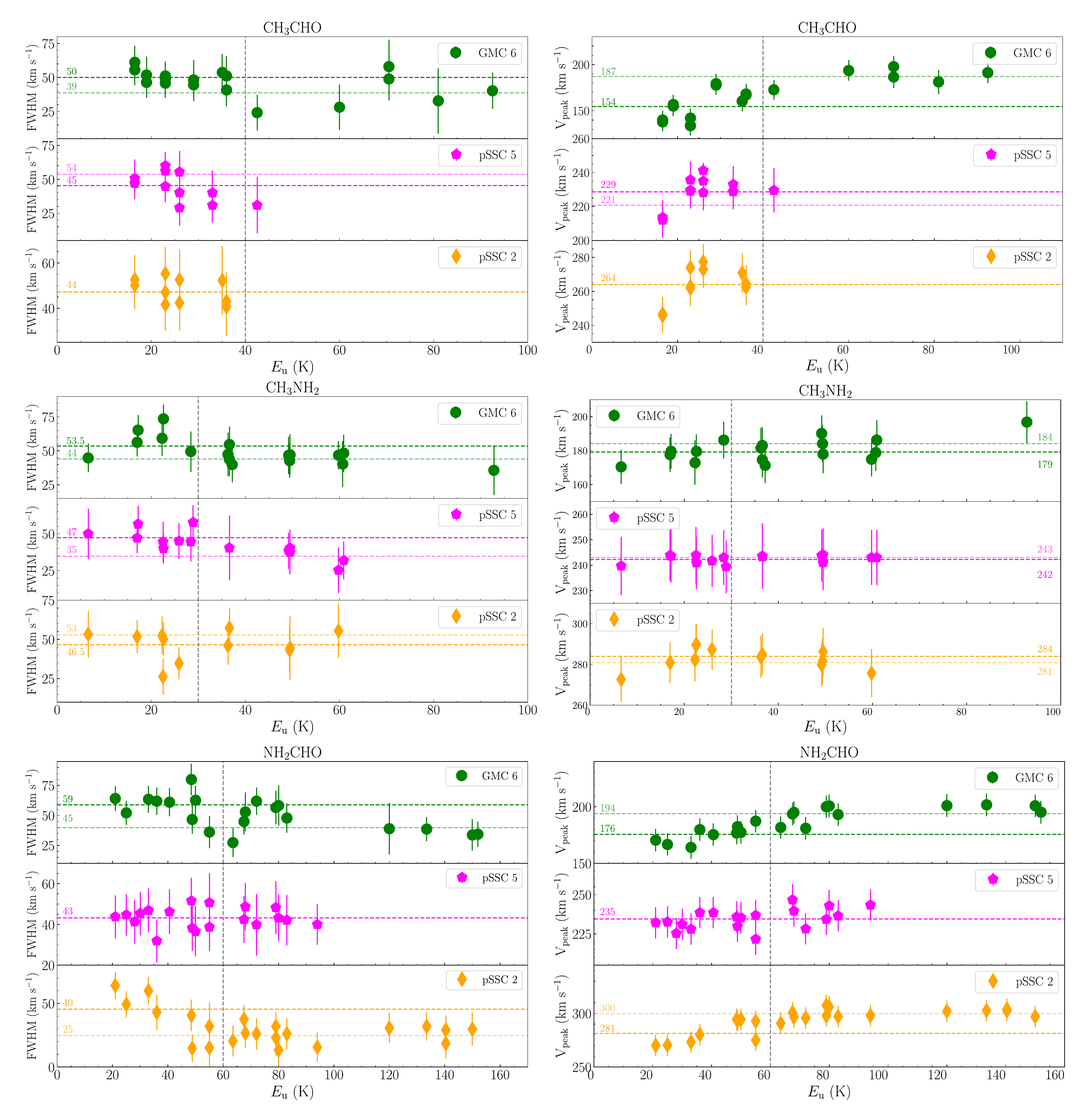}
    \caption{Linewidths (FWHM; left-hand side) and peak velocities ($V_{\mathrm{peak}}$; right-hand side) as a function of upper level energies ($E_{\mathrm{u}}$) for \ce{CH3OH} (top panels), \ce{CH3NH2} (middle panels), and \ce{NH2CHO} (bottom panels) and for GMC\,6 (filled green circles), pSSC\,5 (filled magenta pentagons) and pSSC\,2 (filled orange diamonds). The vertical dashed grey lines indicate where the deviation in the RDs occurs, i.e. $\sim$ 40 K, 30 K, and 60 K for \ce{CH3CHO}, \ce{CH3NH2}, and \ce{NH2CHO}, respectively. Only the most clear Gaussian fits have been used (see text). If a second component has been identified in the RDs, it is shown in lighter color. Mean FWHM and $V_{\mathrm{peak}}$ are indicated on the plots by horizontal dashed lines accompanied with their corresponding values.}
    \label{fig:av-GF}
\end{figure*}

To check if the presence of the two temperature components seen in the RDs for \ce{CH3CHO}, \ce{CH3NH2}, and \ce{NH2CHO} can be identified through a shift in the mean FWHM or $V_\mathrm{peak}$ as a function of the upper-level energy, we used only relatively isolated transitions, where a single Gaussian fit was performed to get the most reliable measurement for FWHM and $V_\mathrm{peak}$. In the case of auto-blended transitions, we kept only those for which the two lines are separated enough to perform two separate Gaussian fits. Figure~\ref{fig:av-GF} shows the FWHM and $V_{\mathrm{peak}}$ as a function of $E_{\mathrm{u}}$ towards GMC\,6, pSSC\,5, and pSSC\,2, where a second gas component has been fitted in the RDs. The uncertainties were derived by adding in quadrature the errors from the Gaussian fits with the spectral resolution of 10 \kms. The mean FWHM and $V_\mathrm{peak}$ for each component are summarised in Table~\ref{tab:av_GF}. We discuss individual species below:

\paragraph{\texorpdfstring{\ce{CH3CHO}}{CH3CHO} (Figure~\ref{fig:av-GF}, top panels):}
We see a clear difference in the mean $V_\mathrm{peak}$ for GMC\,6, where we see an increase in the $V_\mathrm{peak}$ with increasing $E_{\mathrm{u}}$. However, we cannot see this shift toward pSSC\,5 where a second temperature component has also been fit in the RD (see Figure~\ref{fig:rds_SSC5}), due to the lack of reliable Gaussian fit measurements.

\paragraph{\texorpdfstring{\ce{CH3NH2}}{CH3NH2} (Figure~\ref{fig:av-GF}, middle panels):}
A slight decrease in the linewidths with increasing $E_{\mathrm{u}}$ is seen towards GMC\,6 and pSSC\,5, although it is quite marginal. No variation in the FWHM is seen towards pSSC\,2 and no difference in the $V_\mathrm{peak}$ is seen towards any of the regions in \ce{CH3NH2}.  There is thus no strong indication of two different components for \ce{CH3NH2}, and the deviation in the RD could be due to non-LTE effects \citep{goldsmith_1999}. 

\paragraph{\texorpdfstring{\ce{NH2CHO}}{NH2CHO} (Figure~\ref{fig:av-GF}, bottom panels):} An increase in $V_\mathrm{peak}$ is relatively clear with increasing $E_{\mathrm{u}}$ towards the two regions where two components were fitted, i.e. GMC\,6 and pSSC\,2. For pSSC\,2, we also observe simultaneously a decrease in the FWHM with an increasing $E_{\mathrm{u}}$, and this can be seen more moderately towards GMC\,6. Interestingly, the shift in $V_\mathrm{peak}$ and in FWHM seems to occur closer to 40, compared to what is seen in the RDs, which could indicate that the majority of the \ce{NH2CHO} emission could be associated with the second higher temperature component.

In conclusion, with the current spectral resolution as a main contributor to the uncertainties, the presence of two components is only clear for \ce{NH2CHO} whilst it is not clear for \ce{CH3CHO} and \ce{CH3NH2}. In addition, one should note that optically thick lines can induce larger line widths. Higher spectral and angular resolution observations are needed to identify possible different gas components more clearly.

\subsubsection{Results from the rotational diagrams}\label{subsec:rds_results}

\begin{figure*}
    \centering
    \includegraphics[width=1\linewidth]{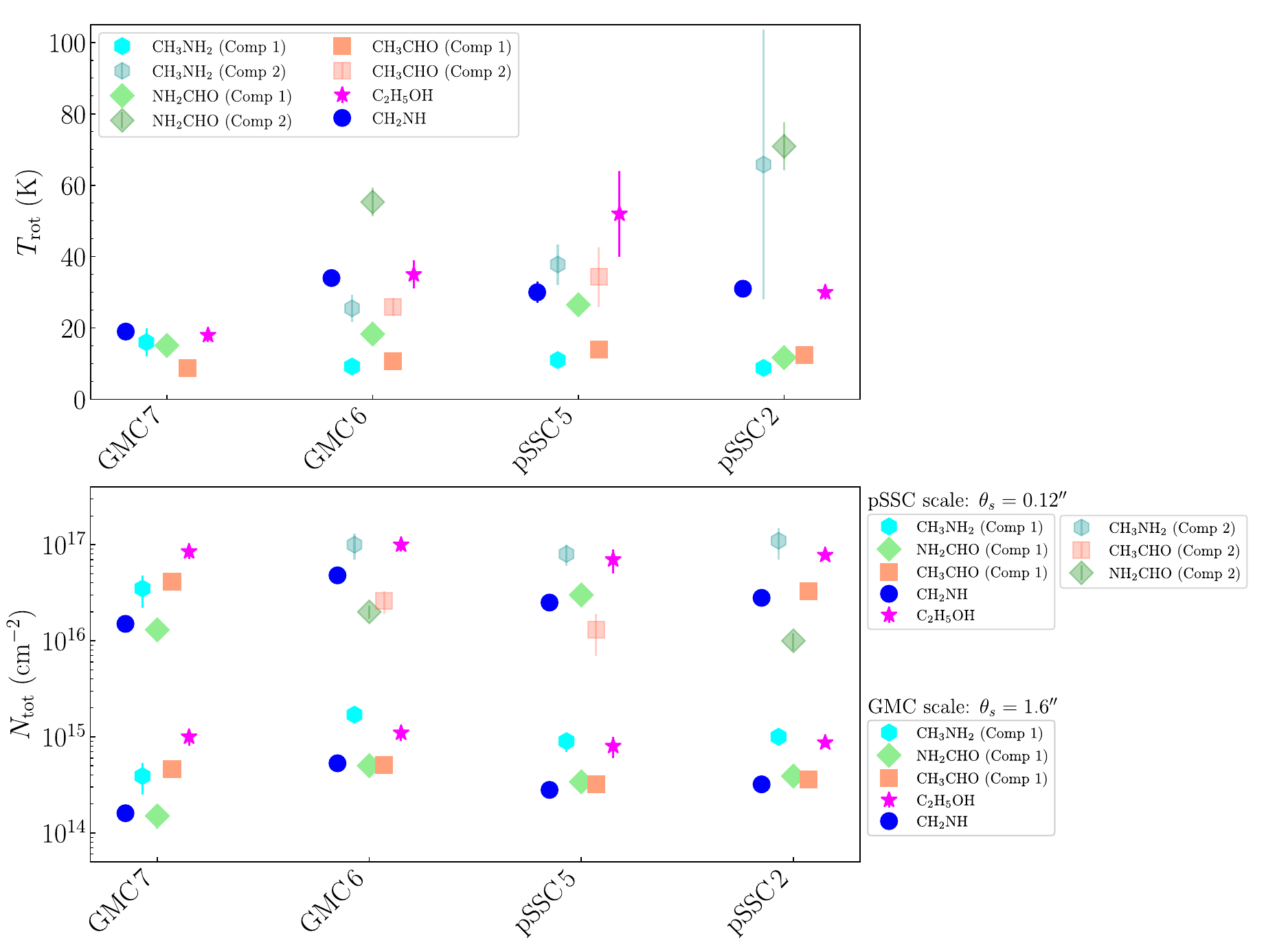}
    \caption{Derived rotational temperature ($T_{\text{rot}}$; Top panel) and total column density ($N_{\text{tot}}$; bottom panel) for each species and region. The order of the regions follows the layout of the GMCs/SSCs of the CMZ from north-east to south-west as shown in Figure~\ref{fig:sup-icoms}. \textit{Top:} If two gas components were fitted in the RDs, the $T_{\text{rot}}$ of the two components are indicated. \textit{Bottom:} The column densities are derived for GMC scales ($\theta_s=1.6\arcsec$; bottom part) and pSSC scales ($\theta_s=0.12\arcsec$; top part). The species (\ce{CH2NH} and \ce{C2H5OH}) and region (GMC\,7) where only one gas component was fitted, we calculated the column densities for the two scales. For the other regions and species, we attributed the first component to an emission coming from GMC scales and the second component to an emission coming from pSSC scales. In the cases of \ce{NH2CHO} and \ce{CH3CHO}, only one component was fitted for pSSC\,2 and pSSC\,5, respectively.  }
    \label{fig:summary_LTE}
\end{figure*}

Figure~\ref{fig:summary_LTE} shows an overview of the derived $T_{\text{rot}}$ and $N_{\text{tot}}$ for each species and region listed in Table~\ref{tab:results_RDs}. Due to the fact that we observe deviations from the fits in the RDs, it is clear that the derived $T_{\mathrm{rot}}$ are not representative of the true gas temperature. For the rotational temperatures (top panel of Figure~\ref{fig:summary_LTE}), we notice some differences between the regions. GMC\,7 shows only a cold ($T_{\text{rot}}\leq 20$ K) component compared to the other regions. On the other hand, the other regions have two components, a cold ($T_{\text{rot}}\leq 20$ K) and a warmer component ($T_{\text{rot}}\sim 30-80$ K).  Finally, none of the regions shows a hot ($T_{\text{rot}}> 100$ K) component, which would have been expected if the iCOMs are thought to be emitted from a hot-core-like ($T > 100$ K; see e.g. \citealt{kurtz2000}) region. On the other hand, this result is similar to the Milky Way Galactic Centre (GC), where widespread emission (a few tens to hundreds of parsecs) from iCOMs has been measured \citep[e.g.][]{requena-torres_organic_2006, li_2017_widespread_coms, li_2020_widespread_coms}. There, low rotational temperatures ($T_{\mathrm{rot}}\leq 20$ K; e.g., \citealt{requena-torres_organic_2006, zeng2018, rivilla2022}) were also derived, indicating that the GC iCOM emission is sub-thermally excited. The extended iCOM emission toward the GC is not associated with hot cores and can be explained by widespread recurrent low-velocity shocks \citep{requena-torres_organic_2006, zeng2020}.

For the column densities (bottom panel of Figure~\ref{fig:summary_LTE}), we describe below the behaviour separating the emission from the cold component (derived with $\theta_s=1.6\arcsec$; i.e. GMC scale) and the warm component (derived with $\theta_s=0.12\arcsec$, i.e. pSSC scale) transitions. Only one temperature component was fitted for all species towards GMC\,7, \ce{CH2NH} and \ce{C2H5OH} towards all regions, \ce{NH2CHO} towards pSSC5, and \ce{CH3CHO} towards pSSC2. In these cases, since we do not know the region of emission of these species, we calculated column densities for the two emission sizes, i.e. for the GMC scale ($\theta_s=1.6\arcsec $) and the pSSC scale ($\theta_s=0.12\arcsec$). 
 
\paragraph{GMC-scales:} \ce{C2H5OH} and \ce{CH3CHO} show relatively constant column densities throughout the regions. On the other hand, \ce{NH2CHO}, \ce{CH3NH2}, and \ce{CH2NH} show a decrease in $N_{\text{tot}}$ towards GMC\,7 and the column density of \ce{CH3NH2} increases towards GMC\,6. Overall, \ce{C2H5OH} and \ce{CH3NH2} show the highest column densities, except towards GMC\,7 where the column density of \ce{CH3NH2} decreases.

\paragraph{pSSC-scales:} For \ce{C2H5OH} and \ce{CH2NH}, the trend in column density is the same as at GMC-scales since the two species have only one component. In the case of \ce{CH3NH2}, its column density is constant across  the regions except towards GMC\,7, where it decreases. In the cases of \ce{NH2CHO} and \ce{CH3CHO}, if we consider only the second components (high $E_{\mathrm{u}}$) towards GMC\,6 and pSSC\,5 for \ce{CH3CHO} and pSSC\,2 for \ce{NH2CHO}, then the column density of both species is the highest towards GMC\,6. If we consider that in the other regions where only one component is detected (low $E_{\mathrm{u}}$ transitions), then the column density of \ce{CH3CHO} is the lowest towards pSSC\,5 and that of \ce{NH2CHO} is the lowest towards GMC\,7 and pSSC\,2.

We note that since the four regions do not have the same \ce{H2} column densities, the abundance trends could be different than the column density trends. An abundance study is beyond the scope of this paper.  We simply stress here that there are variations of the relative column densities between the species across the regions, indicating possible chemical differences, such as towards GMC\,7, where the relative amount between O-bearing and N-bearing iCOMs changes compared to the other regions. We discuss  chemical variations in more details in Sec.~\ref{subsec:chemical_variations}.

\subsection{Non-LTE analysis: \texorpdfstring{\ce{CH2NH}}{CH2NH}}\label{sec: nonlte}

\ce{CH2NH} is the only species from this work for which collisional excitation rates are available. Therefore, we can perform a non-LTE analysis to constrain more accurately the gas physical parameters traced by \ce{CH2NH} toward the CMZ of NGC\,253. 

We performed this analysis using the large velocity gradient (LVG) code \textit{grelvg} \citep{ceccarelli_2003}. We used the \ce{CH2NH}-\ce{H2} collisional rates from \cite{xue2024} available in the Excitation of Molecules and Atoms for Astrophysics (EMAA) database\footnote{\url{https://emaa.osug.fr/}}. The rates were calculated for a temperature in the range 10 to 150 K and for the transitions up to $E_{\mathrm{u}}=143$ K. Therefore, the maximum temperature tested is limited by the available rate coefficients, and the \ce{CH2NH} transition at 312.336 GHz with $E_{\mathrm{u}}=151.3$ K, that we detect in this work, was excluded from the analysis. For each region we ran a grid of models ($\sim 10,000$) varying the column density, the gas density and temperature within $10^{13}-10^{18}$ \pcms, $10^4-10^8$ \pcmc, and $10-150$ K, respectively. The column densities and gas densities were both sampled in logarithmic scale and the gas temperature was sampled in linear scale.  We left the source size as a free parameter.
Then, for the geometry, we chose a uniform sphere \citep{osterbrock1974} as the most physically realistic choice.
The ortho-to-para ratio of \ce{H2} was set to the statistical value of 3. Finally,  the average linewidths of \ce{CH2NH} chosen is 50 \kms, which is the average linewidth across the regions (see Table~\ref{tab:av_GF}).  We also included a 15\% calibration uncertainty in the integrated intensity (see Sec~\ref{sec:obs}).  

 From the RD analysis, there is no clear indication of a multiple temperature components for \ce{CH2NH} (See e.g., Figure~\ref{fig:rds_GMC6}). We thus first included all the lines in the LVG code. However, this resulted in a bad fit ($\chi^2_r>5$) for GMC\,6, pSSC\,5, and pSSC\,2. In these regions, the two transitions of \ce{CH2NH} at 225.554 and 172.267 GHz are not well fit with the other transitions. These lines have the lowest $E_{\mathrm{u}}$ ($< 20$ K) and could thus be coming from a cooler or more extended component compared to what the other transitions ($E_{\mathrm{u}}= 27-151$ K) trace. Since we cannot perform the LVG analysis with only two lines, we cannot constrain the physical parameter of this potentially cooler component. We thus ran the LVG with the rest of the lines ($E_\mathrm{u}\geq 27$ K).

\begin{figure}
    \centering
    \includegraphics[width=1\linewidth]{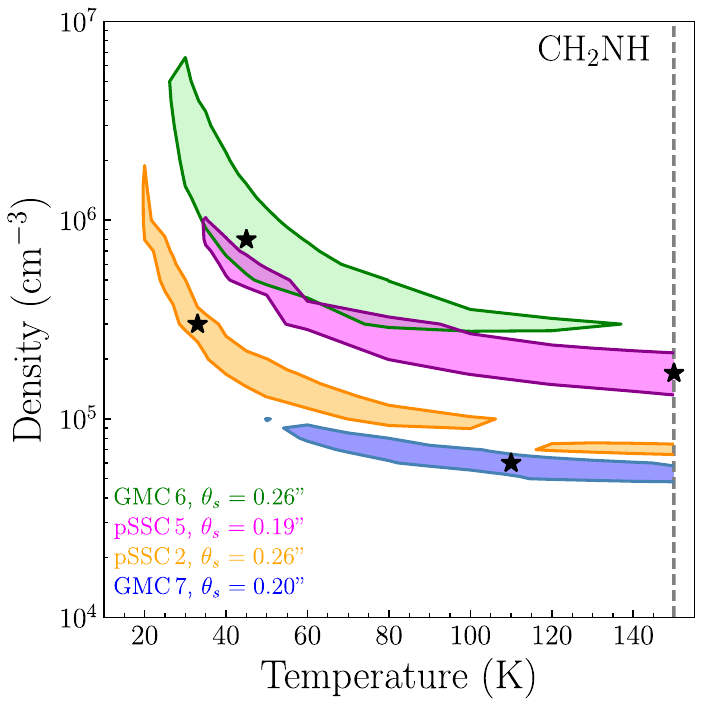}
    \caption{Volume density and kinetic temperature contour plots showing the results from the LVG analysis for \ce{CH2NH} towards GMC\,7 (green), GMC\,6 (blue), pSSC\,5 (magenta) and pSSC\,2 (orange). The contours show the $1\sigma$ solutions obtained for the minimum $\chi^2_r$ value in the column density parameter. The best fits are marked by black stars, while the best fit for the emission size ($\theta_s$) are indicated in the bottom-left corner of the figure. The vertical grey line indicates the maximum temperature at which the rates are calculated.}
    \label{fig:ch2nh_lvg}
\end{figure}

\begin{table*}
    \centering
        \caption{Best fit results and $1\sigma$ confidence level range from the non-LTE LVG analysis of \ce{CH2NH}.}
    \label{tab:lvg_params}
    \begin{tabular}{lcccccccc}
    \hline \hline
    Parameters   & \multicolumn{2}{c}{GMC\,7} & \multicolumn{2}{c}{GMC\,6} & \multicolumn{2}{c}{pSSC\,5} & \multicolumn{2}{c}{pSSC\,2}\\
    & Best Fit& Range &Best Fit& Range & Best Fit& Range & Best Fit& Range \\
    \hline
    $N_{\ce{CH2NH}}$ ($\times10^{16}$ \pcms) & $1.5$ & $0.7-4.5$ &$1.5$ & $0.7-4$ &$2$ & $0.2-5$ &$2$ & $0.2-5$ \\
    $n(\text{H}_{2})$  ($\times10^5$ \pcmc) & $0.6$ & $0.5-1$ & $8$ & $3-60$ & 1.7 & $1-10$ & $3$ & $0.7-20$\\
    $T_{\text{kin}}$ (K) &110 & $\geq 45$ & 45 & $25-130$ & 150 & $\geq 30$ & 33 & $20-100$ ; $\geq 120$  \\
    Source size ($''$) & 0.20 & $0.15-0.24$ & $0.26$ & $0.16-0.34$ & 0.19 & $0.17 - 0.60$ & $0.26$ & $0.22-0.60$\\
    \hline
    $\tau$ Range & \multicolumn{2}{c}{$0.6-3.4$} & \multicolumn{2}{c}{$0.04-1.3$} & \multicolumn{2}{c}{$0.03-2.4$} & \multicolumn{2}{c}{$0.01-2.3$} \\
    $\chi^2_r$ & \multicolumn{2}{c}{0.3} &\multicolumn{2}{c}{1.3} & \multicolumn{2}{c}{1.4}&\multicolumn{2}{c}{1.7}  \\
         \hline
    \end{tabular}
\end{table*}

 Figure ~\ref{fig:ch2nh_lvg} shows the results of the LVG analysis for each of the four regions. The best fit for the \ce{CH2NH} column densities is similar towards the four regions with values in the range $(1.5-2)\times10^{16}$ \pcms. The kinetic temperatures are not always well constrained and are limited by the maximum temperature available from the collisional rates (150 K). We thus have an overall lower limit of $T_{\text{kin}}\geq (20-50)$ K across the different regions. For the derived gas densities, they lie mostly between $n(\text{H}_2)=(10^5-10^6)$ \pcmc\ except for GMC\,7 where $n(\text{H}_2)\leq 10^5$ \pcmc. In all the regions, several lines were found to be optically thick with $\tau > 1$ (see Table~\ref{tab:lvg_params}). Finally, the best fit for the emission size is in the range $0.19\arcsec -0.26\arcsec$ corresponding to a linear scale of $\sim 3-4$ pc in diameter. The best-fit solutions and ranges obtained for each source are reported in Table \ref{tab:lvg_params}. The $1\sigma$ confidence levels constrained correspond to a 68\% confidence range.

We corrected the rotation diagram for the optical depth and source size for all the regions using the population diagram method \citep{goldsmith_1999}, taking the output values of the LVG analysis for these parameters to understand whether the optical depth effects are dominant. For consistency, we removed the transitions that were left out from the LVG analysis (see above). The corrected RDs are shown in \ref{appx:pop_diagrams}. We still find some deviation from a straight line in the corrected RDs, with rotational temperatures below the minimum gas temperature predicted by the LVG analysis. Therefore, \ce{CH2NH} is also affected by non-LTE effects. This result is consistent with that of \cite{faure2018}, who found that non-LTE effects can be expected in the density range of $n_{\text{H}_2}=10^3-10^7$ \pcmc \ for this species. Please note however, that in the discussion below we will use the column densities derived with the RDs for \ce{CH2NH} as the goal of performing population diagrams was only to evaluate the part of non-LTE effects.

 %%%%%%%%%%%%%%%%%%%%%%%%%%%%%%%%%%%%%%%%%%

\section{Discussion}\label{sec:discussion}

As mentioned in the introduction, investigating iCOMs in more extreme star-formation conditions to those that can be found in our own Galaxy allows us to probe how universal our chemical diagnostics are and whether they remain valid in extragalactic environments. Below, we qualitatively discuss whether the known proposed formation pathways for iCOMs could also be valid in NGC\,253 and whether the iCOMs in this study can be chemically linked (Sec.~\ref{sec:chemical_links}). We then discuss their possible region of emission and which physical processes they could trace (Sec.~\ref{sec:origin}). Finally, we focus on the small chemical differences within the regions (Sec.~\ref{subsec:chemical_variations}). Figure~\ref{fig:summary} shows a schematic which summarizes our main hypothesis and scenarios concerning iCOMs towards the CMZ of NGC\,253, focussing in particular on the discussion from sections ~\ref{sec:chemical_links} and ~\ref{sec:origin}.

\subsection{Chemical link and formation pathways for iCOMs}\label{sec:chemical_links}

Understanding which chemical pathways to the formation of a specific species are dominant within an environment is not straightforward. From observations, a method that is regularly used is to look at the abundance correlations between two species suspected to be chemically linked, or to have a similar precursor \citep[e.g.,][]{yamamoto2017, coletta2020, li2024_correlations}. However, it is worth noting that a correlation between two species does not always imply a direct chemical link. It may also indicate that they went through similar physical processes of formation across different environments \citep[e.g.,][]{quenard2018, belloche2020}. 

Figure~\ref{fig:correlations} shows the relation between the column densities of the iCOMs, at both GMC scales (corresponding to a source size of $1.6\arcsec$) or pSSC scales (corresponding to a source size of $0.12\arcsec$). We list below several trends from Figure~\ref{fig:correlations} that will be discussed below. We refer to \ce{CH3CHO} and \ce{C2H5OH} as O-bearing iCOMs, \ce{CH2NH} and \ce{CH3NH2} as N-bearing iCOMs. \ce{NH2CHO} lies in between the two categories.

\begin{itemize}
    \item At GMC scales, O-bearing species are well correlated over the four regions and present a constant ratio [\ce{C2H5OH}/\ce{CH3CHO}] of 3 (panel a). The correlation disappears at pSSC scales (panel f).
    \item At both GMC- and pSSC scales, there is an enhancement in O-bearing species compared to N-bearing species (including \ce{NH2CHO}) towards GMC\,7 (panels b, c, h, k).
    \item At GMC scales, \ce{CH3NH2} and \ce{NH2CHO} are relatively well correlated over the four regions and present a constant ratio [\ce{CH3NH2}/\ce{NH2CHO}] of $\sim 3$ (panel d)). At pSSC scales, the ratio increases towards GMC\,7 and pSSC\,2 (panel i).
    \item The ratio [\ce{CH3CHO}]/[\ce{NH2CHO}] varies greatly from GMC scales (panel b) to pSSC scales (panel g).
    \item The abundance ratio between \ce{CH3NH2} and \ce{CH2NH} (panels e, j) is  close to 3 at both scales, but slightly lower for GMC\,6 at pSSC scales.
\end{itemize}

In the following subsections, we look for correlation between the derived column densities of a few pairs of iCOMs, which are thought to share chemical roots. We first remind the known formation pathways of the different species before discussing whether a formation pathway is most probable for the CMZ of NGC\,253, and whether a chemical link between the pairs of species discussed is possible. A summary of possible formation pathways for each iCOM is shown in Figure~\ref{fig:summary}.

\subsubsection{\texorpdfstring{\ce{CH3CHO}}{CH3CHO} and \texorpdfstring{\ce{C2H5OH}}{C2H5OH}}\label{subsec:link_o-bearing}

The formation pathways of \ce{CH3CHO} are debated. Both gas phase and grain surface reactions are proposed. Among the viable gas-phase formation pathways, \ce{CH3CHO} could be formed either from the radical \ce{C2H5} previously formed on the grain and ejected in the gas phase \citep[e.g.][]{charnley2004, VH2013, BG2020, codella_outflow_2020, desimone2020, vazart2020}, or from \ce{C2H5OH} (known as the \say{ethanol tree}; e.g., \citealt{skouteris2018, vazart2020}). However, whilst the first pathway has been validated in both cold and warm environments, the validity of the second one has not been checked in cold environments \citep[e.g.][]{vazart2020}. Alternatively, various grain surface reactions have been proposed for the formation of \ce{CH3CHO} \citep[e.g.][]{bennett2005, garrod2008, ruaud2015,lamberts2019, BG2020, martindomenech2020, ibrahim2022} although the efficiency of some of these reactions are debated \citep[e.g.][]{rimola2018, enriqueromero016, enriqueromero2021}.
For \ce{C2H5OH}, grain-surface reactions are mostly proposed but there is no consensus on the formation path \citep[e.g.,][]{JG2020, chuang2020, perrero2022, enriqueromero2022, garrod2022, molpeceres2024}. One of the proposed pathways involves \ce{CH3CHO} as a precursor \citep{bisschop2007, chuang2020, fedoseev2022}, where it was expected that the solid-state abundances of \ce{C2H5OH} are less than those of \ce{CH3CHO} \citep{sivaramakrishnan2009, fedoseev2022}. This formation pathway could be ruled out since recent observations made with the James Webb Space Telescope showed relatively similar ice abundances for the two species \citep{rocha2024, chen2024}, as found previously in comets \citep{rubin2019}. However, a direct comparison of the content seen in the ice and in the gas towards two low-mass protostars showed that this is not what is seen in the gas phase, where \ce{C2H5OH} seems to be more abundant than \ce{CH3CHO} \citep{chen2024}. On the other hand, gas phase reactions starting from \ce{H3O+}+\ce{C2H4} and leading to the formation of \ce{C2H5OH} through \ce{C2H5OH2+} as a precursor are reported in both KIDA\footnote{\url{https://kida.astrochem-tools.org/}} and UMIST\footnote{\url{https://umistdatabase.net/}} databases, and could be a viable pathway if \ce{H3O+} is abundant enough (which may be the case due to the high CRIR towards these regions; e.g. \citealt{holdship_energizing_2022, behrens2024}). In the following, we investigate whether ethanol (\ce{C2H5OH}) and acetaldehyde (\ce{CH3CHO}) are correlated towards NGC\,253.

Table~\ref{tab:av_GF} shows that the mean FWHM and peak velocities of \ce{C2H5OH} are generally more consistent with that of the colder component of \ce{CH3CHO}. Optical depth effects could alter these results but further similarities can be seen when looking at the column densities of the two species: From Figure~\ref{fig:summary_LTE}, we see that both the column densities derived for GMC scales of \ce{C2H5OH} and \ce{CH3CHO} are relatively constant throughout the four regions. In addition, their abundance ratio of $\sim 2.3$ is also constant across the regions and show the best correlation at GMC scales (see Figure~\ref{fig:correlations}a). Therefore, a correlation between the cold component of \ce{CH3CHO} (governed by the low $E_{\mathrm{u}}$ transitions) and \ce{C2H5OH} could indicate that the two species are chemically linked (see Fig~\ref{fig:summary}, hypothesis i). Since the abundance of \ce{C2H5OH} is higher than that of \ce{CH3CHO}, the two species could be linked via the gas-phase production of \ce{CH3CHO} through the ethanol tree. This hypothesis, if true, implies that the ethanol tree pathway proposed by \cite{skouteris2018} would also be valid at low temperatures, since the iCOMs are likely sub-thermally excited. On the other hand, we cannot rule out that the two species were formed on the grains before being ejected through shocked processes (See Sec.~\ref{sec:origin}). The higher abundance of \ce{C2H5OH} compared to that of \ce{CH3CHO} could then indicate more fast or important destruction or reprocessing of \ce{CH3CHO} in the gas phase compared to \ce{C2H5OH}. For \ce{C2H5OH}, one proposed formation path for its precursor \ce{C2H5OH2+} is via the gas phase reaction between \ce{H3O+} and \ce{C2H4}. Considering the high CRIR derived towards NGC\,253 \citep[e.g.][]{holdship_energizing_2022, behrens_tracing_2022}, the abundance of \ce{H3O+} could be high enough for the gas phase production of ethanol to be viable.

When looking at the correlations at the pSSC scales in Figure~\ref{fig:correlations}f, the warm component of \ce{CH3CHO} (traced by the high $E_{\mathrm{u}}$ transitions) towards GMC\,6 and pSSC\,5 is no longer correlated with ethanol. The difference in peak velocity and FWHM for this high $E_{\mathrm{u}}$ transitions compared to the low$E_{\mathrm{u}}$ transitions (see Table~\ref{tab:av_GF}) indicates that additional \ce{CH3CHO} is produced in these two regions, and likely through a different pathway. If the cold component of \ce{CH3OH} is formed in the gas phase, the warm \ce{CH3CHO} component could, on the other hand, form on the icy grain mantles via thermal desorption (if the gas temperature is high enough, see Sec.~\ref{subsec:SSC_scales}), or via non-thermal processes such as shocks \citep[e.g.][]{ruaud2015, lu2024}. We investigate the origin of the high excitation of \ce{CH3CHO} in GMC\,6 and pSSC\,5 in Sec.~\ref{subsec:chemical_variations}.

\subsubsection{\texorpdfstring{\ce{NH2CHO}}{NH2CHO}}\label{subsec:link_formamide}

Both grain surface \citep[e.g.,][]{raunier2004, garrod2008, lopezsepulcre2015, fedoseev2016, rimola2018, dulieu2019, martindomenech2020, chuang2022} and gas phase \citep[e.g.,][]{kahane2013, barone2015, vazart2016, skouteris2017} formation pathways have been suggested for formamide (\ce{NH2CHO}), but which one is the most efficient is highly debated, as in the case of \ce{CH3CHO}. We refer to \cite{lopezsepulcre2019} and \cite{lopezsepulcre024} for a comprehensive review of the different formation paths proposed for formamide. Here, we will only investigate the potential chemical link of \ce{NH2CHO} with HNCO and \ce{H2CO}, both species thought to be chemically linked with \ce{NH2CHO}, either as a parent species on the ice grain mantle or in the gas phase (\ce{H2CO}; e.g., \citealt{fedoseev2016, dulieu2019, kahane2013, vazart2016, aikawa2020, lopezsepulcre024}), or as a daughter species  (HNCO; \citealt{brucato2006, gorai2020, haupa2022, chuang2022}). Both \ce{H2CO} and HNCO were studied towards the CMZ of NGC\,253 (\citealt{ mangum_fire_2019, huang_reconstructing_2023}, Huang et al. subm.).

HNCO traces low-velocity shocks throughout the CMZ of NGC\, 253 and the gas temperatures towards the four regions investigated in the present study range between $\sim 20-100$ K \citep{huang_reconstructing_2023}. The rotational temperatures we derived range between $\sim 10-30$ K and $\sim 50-80$ K, for the cold and warm component, respectively, consistent with that of HNCO. A word of caution is needed, however, since our rotational temperature could be underestimated if optical depth effects or non-LTE effects are at play, as we found to be the case for \ce{CH2NH} (see Sec.~\ref{sec: nonlte}). Using transitions with similar energy levels, we compared the emission distribution between \ce{NH2CHO} and \ce{HNCO} around $\sim 40$ and $\sim 140$ K as shown in Figure \ref{fig:nh2cho_hnco_contour_maps}. \ce{NH2CHO} is much less intense compared to HNCO, as it is much less abundant, but the two species do not peak towards the same regions. This is also seen when comparing the derived column densities of \ce{NH2CHO} and HNCO: whilst HNCO has a relatively constant column density across the regions \citep{huang_reconstructing_2023}, the cold component of \ce{NH2CHO} decreases towards GMC\,7 whilst the warm component is lower towards pSSC\,2. Neither of the two component of \ce{NH2CHO} seem thus to be chemically linked with HNCO.

As for HNCO, we also compared the emission distribution of \ce{NH2CHO} and \ce{H2CO} at similar energy levels, around 45 and 100 K, as shown in Figure \ref{fig:nh2cho_h2co_contour-maps}. Even though \ce{NH2CHO} is much less intense than \ce{H2CO} due to its lower abundance, we can see a good correlation between the two species at low $E_{\mathrm{u}}$ with a peak of emission towards GMC\,6 and pSSC\,5. Therefore, the cold component of \ce{NH2CHO} and \ce{H2CO} could be chemically linked: \ce{H2CO} being likely a gas phase product (Huang et al. subm.), \ce{NH2CHO} would thus form in the gas phase, from the reaction \ce{H2CO} + \ce{NH2}
 \citep{kahane2013, barone2015, vazart2016, lopezsepulcre024}. This scenario is however challenged by the different temperatures traced by the two species since the rotational temperature for the cold component of \ce{NH2CHO} is less than 30 K,  while \ce{H2CO} was found to trace a warmer gas ($T_\text{kin}\geq 70$ K at scales 1.6--5$\arcsec$ scales and $T_\text{kin}\geq 300$ K at scales $\leq 1\arcsec$; \citealt{mangum_fire_2019}, Huang et al. subm.). the low temperature of \ce{NH2CHO} could result from sub-thermal excitation. Unless the $T_{\mathrm{rot}}$ are underestimated, we cannot conclude whether \ce{NH2CHO} and \ce{H2CO} are chemically linked and whether the cold component of \ce{NH2CHO} is a gas-phase or a grain surface product.

On the other hand, the difference of peak velocities and FWHM for the low and high $E_{\mathrm{u}}$ transitions (see Figure~\ref{fig:av-GF}) could indicate that \ce{NH2CHO} truly traces two different gas components. Indeed, this warmer ($T_{\mathrm{rot}}> 50$ K) component, clearly present in GMC\,6 and pSSC\,2, shows narrower mean FWHM compared to the cold gas component and warmer rotational temperatures (see Figure~\ref{fig:summary_LTE}). Hence, another formation pathway could be involved for the warm component of \ce{NH2CHO}, different from that of the cold component. As for \ce{CH3CHO}, we further investigate the origin of this emission in Sec.~\ref{subsec:chemical_variations}.

\subsubsection{\texorpdfstring{\ce{CH2NH}}{CH2NH} and\texorpdfstring{ \ce{CH3NH2}}{CH3NH2}}\label{subsec:link_N-bearing}

From the literature, \ce{CH3NH2} is proposed to form on the ice grain surfaces through radical-radical reactions \citep[e.g.][]{garrod2008, KK2011, forstel2017, enriqueromero2022} or from the successive hydrogenation of \ce{CH2NH} \citep[e.g.][]{woon2002, theule2011, sil2018, dejesus2021, molpeceres_ch3nh2_2024}.  
Whilst the first pathways over-predicts \ce{CH3NH2} abundances \citep{garrod2008}, the second possible pathway has been supported by chemical models reproducing observed abundances of \ce{CH3NH2} towards high-mass protostars \citep{suzuki2016, suzuki2023}. This second pathway would indicate that \ce{CH2NH} is a precursor of \ce{CH3NH2} and would also form on the ice grain mantles. However, studies have shown that the conversion of \ce{CH2NH} to \ce{CH3NH2} would be too fast to account for the gas phase abundance of \ce{CH2NH} \citep[e.g.][]{theule2011, suzuki2016}. In addition, very recently, \cite{molpeceres_ch3nh2_2024} proposed a grain surface formation route for \ce{CH3NH2} from the \ce{CNH3} chemisorbate, without passing through \ce{CH2NH} as an intermediate species. From the recent literature, it does seem that the two species are not necessarily chemically linked as it was thought before.
Alternatively from the grain surface production, the neutral-neutral gas phase reaction  \ce{CH3} + \ce{NH3} $\rightarrow$ \ce{CH3NH2} + H has also been proposed as a viable formation pathway in warm ($T>100$ K) regions \citep{bocherel1996, halfen2013}. Similarly for \ce{CH2NH}, various gas-phase formation pathways (ion-molecules or neutral-neutral reactions involving \ce{CH3} or \ce{NH3}; see e.g. Table 11 of \citealt{suzuki2016}) have been proposed as a viable alternative for the formation of \ce{CH2NH} \citep[e.g.][]{bocherel1996, turner1999, suzuki2016, sil2018}.

\begin{figure*}
    \centering
    \includegraphics[width=1\linewidth]{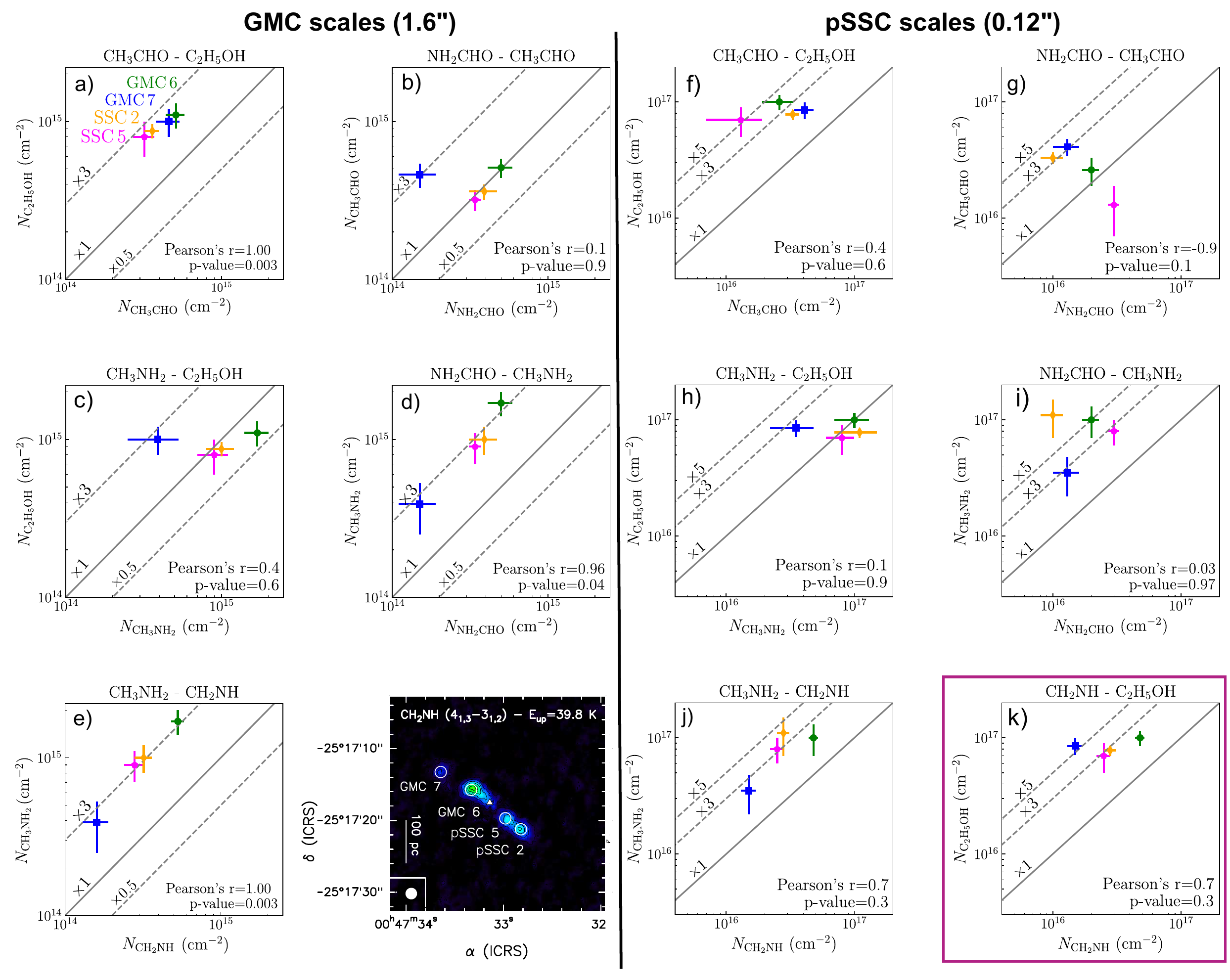}
    \caption{Relation between the column densities of the iCOMs derived from the rotation diagrams at GMC scales ($\theta=1.6\arcsec$; Left-hand side of the figure) and at pSSC scales ($\theta=0.12\arcsec$; Right-hand side of the figure), for each region (GMC\,7: blue; GMC\,6: green; pSSC\,5: magenta; pSSC\,2: orange). Pearson's coefficient and p-values are indicated in the bottom right part of each plot. Dashed and full grey lines indicate abundance ratio factors. The panel k) is highlighted as the correlation between \ce{CH2NH} and \ce{C2H5OH} is the same at both GMC-scale and pSSC-scale. For reference for the location of each region, we added a map of the \ce{CH2NH} ($4_{1,3}-3_{1,2}$) transition in the bottom-left part of the figure.  }
    \label{fig:correlations}
\end{figure*}

From Table~\ref{tab:av_GF}, both the peak velocities and FHWM of \ce{CH2NH} and \ce{CH3NH2} are consistent with each other (and considering the spectral resolution of 10 \kms) in all the four regions investigated. From the RDs, we found similar rotational temperatures between \ce{CH2NH} and the warm component of \ce{CH3NH2} in all the regions but GMC\,6 (see Figure~\ref{fig:summary_LTE}), although it was shown that temperatures were underestimated in the case of \ce{CH2NH}. For the column densities, since \ce{CH2NH} was found to emit at smaller scales than GMC scales (see Sec.~\ref{sec: nonlte}), we compare the column densities of \ce{CH2NH} and of the warm component of \ce{CH3NH2} at pSSC scales only. From Figure~\ref{fig:summary_LTE}, the column densities of the warm component of \ce{CH3NH2} and of \ce{CH2NH} are relatively constant across the regions, as seen also in their abundance ratio: Figure~\ref{fig:correlations} shows very good correlation  (panel e) between \ce{CH2NH} and the cold component of \ce{CH3NH2}, whilst the correlation becomes weaker (high p-value, panel j) correlation between \ce{CH2NH} and the warm component of \ce{CH3NH2}. This weak correlation might be due to the slight decrease in the abundance ratio towards GMC\,6, which seems to come from a higher column density in \ce{CH2NH}. However, the LVG results did not confirm this increase of \ce{CH2NH} towards GMC\,6 and the correlation between the two species could be strong for both components of \ce{CH3NH2}.

If the correlation between \ce{CH2NH} and \ce{CH3NH2} is true, the two species could  be chemically linked (see Fig~\ref{fig:summary}, hypothesis ii). Since the grain formation pathways linking \ce{CH3NH2} and \ce{CH2NH} are not viable to explain gas-phase abundance of \ce{CH2NH} (see above), the two species can be chemically linked if they share the same precursor. On the grain, the new proposed formation route of the two species from \ce{CNH3} indicates that both species can have HCN and HNC as precursors, without \ce{CH2NH} being a precursor of \ce{CH3NH2}. In the gas phase, formation pathways involving \ce{CH3} and \ce{NH3} are viable for both species but only for temperatures above 100 K. Hence, should the two species be chemically linked, both grain surface and gas phase reaction can account for the correlation and the chemical link. Our current results, however, do not allow us to distinguish between the two possibilities. Alternatively, the correlation could indicate that the two species trace similar physical processes of formation across the various regions \citep[e.g.,][]{quenard2018, belloche2020}.

\subsection{Possible origins of the iCOMs in the CMZ of NGC 253}\label{sec:origin}

In the previous section, we have discussed the iCOMs possible formation pathways and whether any pair of iCOMs has a chemical link. We discuss below the physical process that could be responsible for the iCOMs emission towards the CMZ of NGC\,253.

A first inspection of the emission distribution of the iCOMs in this study shows that the emission is (1) mostly concentrated towards the inner part of the CMZ, within $6\arcsec$ around the kinematic centre, (2) compact around each GMC, and (3) sometimes unresolved for the higher upper-level energies, depending the species. From the images, we thus hypothesize that the emission could be emitted on GMC scales (of the size of the order of the ALCHEMI beam size or 1.6$\arcsec$) and unresolved pSSC scales ($\sim 0.12\arcsec$ or a few parsecs), which are the next biggest sub-structures identified within the GMCs \citep[e.g.][]{ando_2017, leroy_forming_2018}. As we could not derive the emission size of the iCOMs, except for \ce{CH2NH}, we discuss how they would be produced at GMC and pSSC scales (Sections \ref{subsec:GMC_scales} and \ref{subsec:SSC_scales}, respectively) in the following.

\subsubsection{iCOMs production at GMC-scales}\label{subsec:GMC_scales}

Several studies derived the volume density and kinetic temperature of the molecular gas in the disk of NGC\,253, and found at least two components \citep[e.g.][]{rosenberg_radiative_2014,gorski_survey_2017, perez-beaupuits_thorough_2018,mangum_fire_2019, tanaka_2023}. The most recent study by \cite{tanaka_2023} (consistent with previous studies), derived a gas temperature and density for the low-density component (dominated by the low-J lines of CO and its isotopologues) of 85 K and $2\times10^3$ \pcmc, respectively, and for the high-density component (dominated by high-density tracers with $E_{\mathrm{u}}\leq 70$K) of 110 K and $\sim 3\times10^4$ \pcmc, respectively. The dust temperatures were estimated to be around 35--40 K \citep{leroy_alma_2015, martin_alchemi_2021}. \cite{leroy_alma_2015} derived line widths for the GMCs of NGC\,253 of $\sim 20-40$ \kms\ using observations of bulk gas tracers (e.g., low-J lines of CO, HCN, HCO$^+$, and isotopologues).

On average, the derived rotational temperatures ($T_{\text{rot}}\leq 60$K) for the iCOMs are thus below the gas temperatures found at GMC scales. At the densities measured (i.e. $n_{\mathrm{\ce{H2}}}\sim 10^3-10^4$ \pcmc), the dust and gas are likely not coupled. Most recent laboratory experiments and quantum chemical calculations of binding energies showed that \ce{CH3CHO} and \ce{CH3NH2} would desorb at a lower dust temperature than water ($\sim 100-120$ K; \citealt{chaabouni2018, ferrero2022, molpeceres2022}), whilst \ce{C2H5OH} would co-desorb with water at around 140 K \citep{perrero2024}. On the other hand, \ce{NH2CHO} is refractory to water desorption and would desorb at higher dust temperatures ($\geq 170 - 300$ K; e.g., \citealt{urso2017, chaabouni2018, ligterink2018, martindomenech2020, ferrero2020, chuang2022}). Therefore, the dust temperature derived at GMC scales ($35-40$ K) is too low to desorb the iCOMs, if they formed on the icy grain mantles. The iCOMs are very likely sub-thermally excited and are released into the gas phase via non-thermal processes if they formed on the grains. With  the CRIR being relatively high ($\zeta \sim 10^{-14}-10^{-12}$ s$^{-1}$; e.g., \citealt{holdship_energizing_2022, behrens_tracing_2022, behrens2024}) towards the CMZ of NGC\,253, CR-induced grain heating and sputtering desorption mechanisms \citep[e.g.][]{roberts2007, dartois2019, paulive2022, arslan2023} could be an important process to desorb these iCOMs. 

Another important non-thermal process in NGC\,253 releasing the species from the ice grain mantles into the gas phase could be shock sputtering. On average, we measured line widths mostly around $40-60$ \kms (see  Table~\ref{tab:av_GF}) for the cold gas component (component 1) of the iCOMs. These values are on average larger than the line widths derived by \cite{leroy_alma_2015} for the GMCs (between $20-40$ K) which could indicate that the species trace a more turbulent gas. However, as already mentioned, we cannot exclude possible optical depth effects as is the case for \ce{CH2NH}. If we consider moderate to high optical depth ($\tau\sim0.6-3$) as we found for \ce{CH2NH}, we could overestimate linewidths between 12\% and 30\%\footnote{using the formula $\sigma_{\mathrm{obs}}=\frac{\sigma_0}{\sqrt{\mathrm{ln}2}}\sqrt{\mathrm{ln}\left(\frac{\tau}{\mathrm{ln}\left(\frac{2}{1+e^{-\tau}}\right)}\right)}$ with $\sigma_{\mathrm{obs}}$, $\sigma_0$ and $\tau_\nu$ the observed and intrinsic velocity dispersions, and the line opacity, respectively \citep{burton1992}. The velocity dispersion ($\sigma_v$) is linked to the FWHM by $\sigma_v=\mathrm{FWHM}/2\sqrt{2 \ \mathrm{ln}(2)}$}. In the most optically-thick case derived ($\tau=3$), for a FWHM of 60 \kms the opacity-corrected linewidth would still be above that of GMCs but it would not be the case for a FWHM of 40 \kms. 

\paragraph{O-bearing iCOMs:} From Sec.~\ref{subsec:link_o-bearing}, we found that \ce{CH3CHO} and \ce{C2H5OH} could be chemically linked. If this is the case, then the cold component of \ce{CH3CHO} could be produced in the gas phase from \ce{C2H5OH}, which would have been sputtered from the grain through shocks. Alternatively, both species could be formed on the grain and be linked with \ce{CH3CHO} as a precursor of \ce{C2H5OH} \citep[e.g.,][]{chuang2020, fedoseev2022} but the gas phase abundance ratio would thus not be representative of the ice abundances (see Sec.~\ref{subsec:link_o-bearing}). In any case, \ce{CH3CHO} and \ce{C2H5OH} tracing shocks would be consistent with the fact that both species are detected in protostellar shocks in our Galaxy \citep[e.g.][]{arce2008, lefloch2017, holdship2019, csengeri2019, codella_outflow_2020, desimone2020, busch2024}.

\paragraph{N-bearing iCOMs, including \texorpdfstring{\ce{NH2CHO}}{NH2CHO}:}

Since \ce{NH2CHO} has not been detected in cold clouds \citep{lopezsepulcre2019} and the rotational temperatures for the cold component are low ($<30$ K) enough to point to sub-thermal excitation, the cold component of \ce{NH2CHO} is also very likely tracing shocked gas. This would be consistent with what is seen and modelled in Galactic shocked regions \citep[e.g.,][]{arce2008, burkhardt2019, codella2017, lopezsepulcre024}. The correlation of \ce{CH3NH2} with \ce{NH2CHO} at GMC scales (see Figure~\ref{fig:correlations}d) could indicate similar physical conditions across the regions for the formation environment of the two species \citep[e.g.,][]{quenard2018, belloche2020}, since no chemical reaction linking the two species has been reported in the literature. On the other hand, we found that \ce{CH3NH2} and \ce{CH2NH} could be chemically linked or could trace the same physical process (see Sec.~\ref{subsec:link_N-bearing}). Therefore, \ce{CH3NH2} and \ce{CH2NH} are also very likely tracing shocked gas.  Recently, \cite{zeng2018} found abundant N-bearing species towards G+0.693-0.027, a molecular cloud in the Galactic Centre. They mentioned that the rich chemistry of this source is due to the fact that it is dominated by low-velocity shocks and the high cosmic-ray ionisation rates \citep{rivilla2022_POp, sanz-novo2024}. They report a ratio [\ce{CH3NH2}]/[\ce{NH2CHO}] and [\ce{CH3NH2}]/[\ce{CH2NH}] of 5, which is of the same order of magnitude as our value of $\sim 3$, for both ratios.

Overall, at GMC scales, sputtering due to large-scale shocks is likely the main process releasing the iCOMs into the gas phase. Signatures of shocks have been previously found within the CMZ of NGC\,253 were already found in previous studies towards NGC\,253 \citep[e.g.,][]{garcia-burillo2000, meier_alma_2015, haasler_first_2022, harada_alchemi_2022, huang_reconstructing_2023}. Large-scale shocks traced by iCOMs would also be consistent with the origin of \ce{CH3OH}, the simplest iCOM, which is also tracing shocks within NGC\,253 (e.g. \citealt{humire_methanol_2022}, Huang et al. subm.).  We compared the emission distribution of \ce{CH3OH} with that of \ce{CH3CHO} and \ce{CH3NH2} around $E_{\mathrm{u}}=20$ K and found a relatively good correlation, except towards GMC\,7 where we retrieve the O- vs N- disparity. The contour maps of the above-mentioned transitions are shown in Figure~\ref{fig:ch3oh_sup_mols}. Following the possible shock scenarios by \cite{huang_reconstructing_2023} (see their Figure 12), the shocks traced by the iCOMs could be due to either star-formation within each GMC (via shocks induced by outflows or sporadic shocks due to scattered star-formation episodes; see scenario 2a in Figure~\ref{fig:summary}) or cloud-cloud collisions (see scenario 1 in Figure~\ref{fig:summary}).

The intersection of different orbits (such as Lindblad resonances; \citealt{iodice2014}) can cause large-scale cloud-cloud collisions \citep[see][and references therein]{harada_alchemi_2022,humire_methanol_2022}. However, the regions we observe are not located close to such intersections, which indicates that the large-scale shocks traced by the iCOMs are either due to cloud-cloud collisions occurring within the GMCs (scenario 1 in Figure~\ref{fig:summary}) or to star formation activity (scenario 2a in Figure~\ref{fig:summary}). From the LVG analysis, we found that \ce{CH2NH} is emitted on size scales between 0.15$\arcsec-0.6\arcsec$ which corresponds to 2.5--10 pc in linear scales, supporting the large-scale shocks scenario within GMCs. The situation could resemble that of our Galactic centre, where widespread emission of iCOMs are detected throughout the CMZ and are likely caused by low-velocity shocks due to cloud-cloud collisions \citep[e.g.,][]{requena-torres_organic_2006, jones2011, li_2017_widespread_coms, li_2020_widespread_coms, zeng2020}. Chemical modelling will help disentangle between the different shock scenarios.
Finally, both O- and N-bearing iCOMs are well correlated, except towards GMC\,7, where O-bearing species are more abundant than N-bearing species. We discuss this  further in Sec.~\ref{sec:GMC7}.

\subsubsection{iCOMs production at pSSC-scale}\label{subsec:SSC_scales}

\begin{figure*}
    \centering
    \includegraphics[width=1\linewidth]{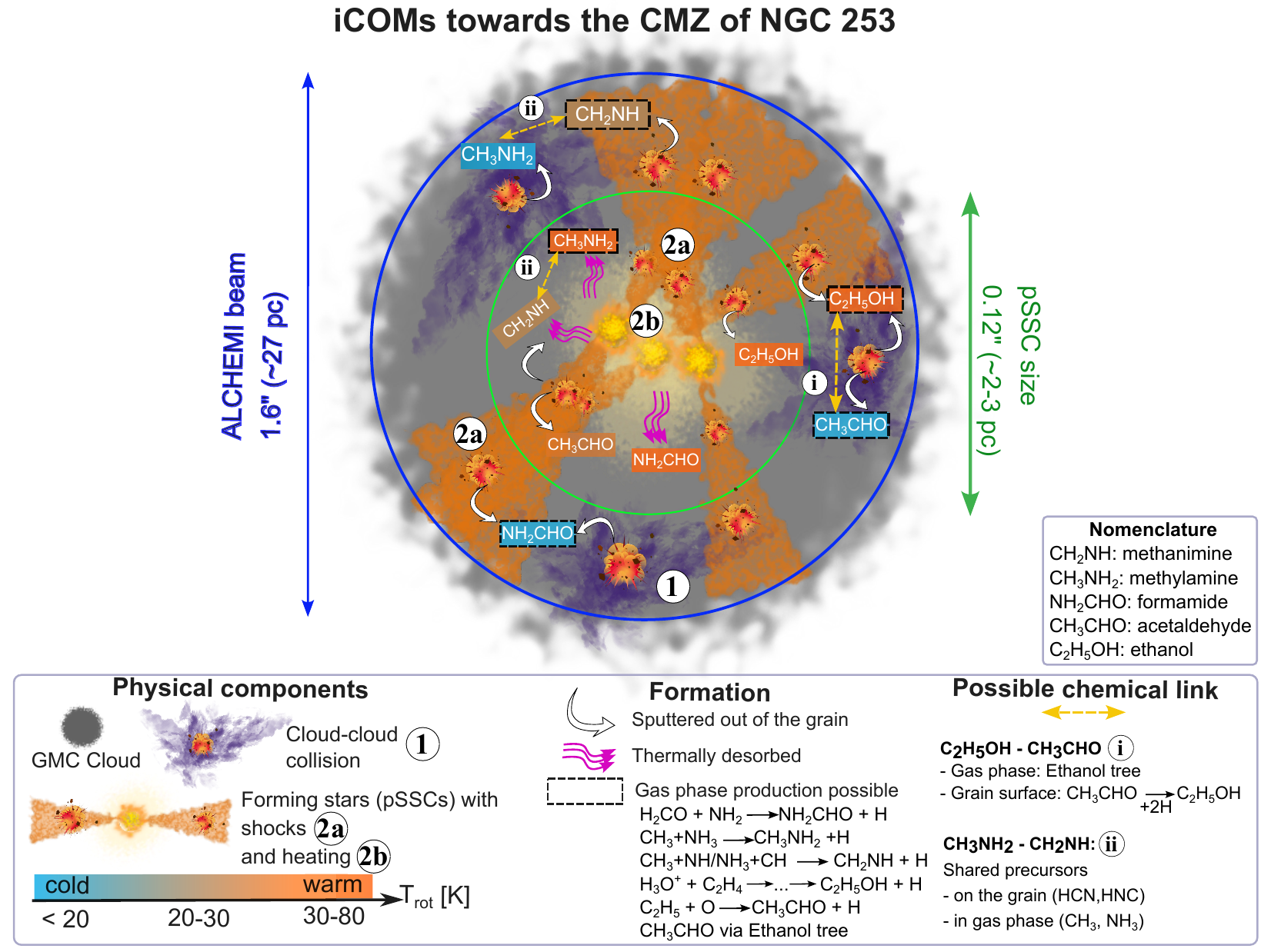}
    \caption{Schematic (not to scale) of summarising the possible formation pathways and chemical links between the iCOMs (Sec.~\ref{sec:chemical_links}) and their emission origins (Sec.~\ref{sec:origin}) within a GMC. The temperatures reported are averaged over the region but there can be variation from one region to another. We assume the size of a GMC is comparable or larger than the size of the ALCHEMI beam of 1.6$\arcsec$ (or $\sim 27$ pc). \textit{Formation and chemical links:} Sputtering from the ice grain mantle and gas phase formation pathways are both possible to explain the iCOM emission at GMC scales, whilst at SSC scales, the dust temperature can be high enough to allow thermal desorption. \ce{CH3CHO} and \ce{C2H5OH} can also be chemically linked (i), both on the grain (with \ce{CH3CHO} as the precursor of \ce{C2H5OH}) or via the Ethanol tree in the gas phase (with \ce{C2H5OH} as a precursor for \ce{CH3CHO}). \ce{CH3NH2} and \ce{CH2NH} could be chemically linked (ii) by sharing similar precursors, both on the grain or in the gas phase (if the gas temperature is higher than 100 K; e.g. \citealt{halfen2013}).  \textit{Origin:} At the scale of GMCs, the iCOMs are likely tracing large-scale shocks, due to either cloud-cloud collision (1) or shocks related to star formation activity (scenario 2a; see also \citealt{huang_reconstructing_2023}). At pSSC scales, iCOMs could trace either star formation-related heating process (scenario 2b; \ce{CH3NH2}, \ce{NH2CHO}) and shocks (scenario 2a; \ce{CH3CHO}, \ce{CH2NH}, \ce{C2H5OH}), associated or not (i.e. if cloud-cloud collision, scenario 1) with star formation. }
    \label{fig:summary}
\end{figure*}

The temperatures of the pSSCs previously derived are $\geq 100$ K \citep{rico-villas_super_2020, krieger_molecular_2020a}. The PSSCs show clear outflow features \citep{gorski_2019, levy_outflows_2021}, and some of them host Super Hot Cores (SHCs; \citealt{rico-villas_super_2020}) from the detection of vibrationally excited \ce{HC3N} emission and densities and temperatures of $\sim 10^6$ \pcmc\ and $\sim 200-300$ K, respectively. Such high temperatures are consistent with the kinetic temperatures derived at scales $\leq 1\arcsec$ ($\leq 17$ pc) using \ce{H2CO} observations \citep{mangum_fire_2019}. The densities of the pSSCs are high enough to consider that the dust and gas are coupled. Therefore, the gas temperature at pSSC scales would be enough to thermally desorb the iCOMs (Scenario 2b in Figure~\ref{fig:summary}), should they be formed on grains. Sputtering from shocks due to outflows driven by the forming-stars is also a possible mechanism releasing the iCOMs into the gas phase or to enable a rich gas-phase chemistry leading to the formation of iCOMs (Scenario 2a in Figure~\ref{fig:summary}). 

As for the GMC scales, the rotational temperatures derived for the iCOMs are much lower ($T_{\text{rot}}\leq 60$ K) than the gas temperature we expect at pSSCs scales, which could indicate that they are tracing a less dense gas. Therefore, the iCOMs are very likely subthermally excited on pSSC scales as well. Due to the limited temperature range for which collisional rates were calculated, we could not constrain the upper limit of the gas kinetic temperature, except for GMC\,6, where the range is $T_{\text{rot}}=25-130$ K (see Sec.~\ref{sec: nonlte}). \cite{suzuki2016} estimated an upper limit for the desorption temperature of \ce{CH2NH} of 160 K, consistent with the mean binding energy of 5534 K derived by \cite{ruaud2015}. If all the iCOMs trace the same gas component, then \ce{C2H5OH} and \ce{NH2CHO} in GMC\,6 cannot be thermally desorbed if they formed on the grain. However, higher temperatures are derived in the other regions, which could enable thermal desorption.

\paragraph{O-bearing iCOMs:} Both \ce{CH3CHO} and \ce{C2H5OH} are detected in shocked regions associated with outflows from protostars \citep[e.g.,][]{arce2008, lefloch2017}, as well as the warm envelope of protostars \citep[e.g.,][]{ belloche2013, fuente2014, jorgensen2018,  bonfand2019, lee2019, gorai2024, moller2024}. The abundance ratio [\ce{C2H5OH}]/[\ce{CH3CHO}] ranging between $\sim 2.3-5$ (see Figure~\ref{fig:correlations}f) corresponds well to the abundance ratio found in low- and high-mass star-forming regions, and shocks linked to protostellar outflows \citep[e.g.][]{lefloch2017, chen2023, busch2024}. 

\paragraph{N-bearing iCOMs, including \texorpdfstring{\ce{NH2CHO}}{NH2CHO}:} The measured abundance ratio of $[\ce{CH3NH2}]/[\ce{NH2CHO}]= 3-10$  towards GMC\,7 and pSSC\,5 (see Figure~\ref{fig:correlations}i) is similar to the ratios measured in high-mass protostars \citep[][]{bogelund2019, nazari2022}. The highest ratios derived towards GMC\,6 and pSSC\,2 (5 and 10, respectively) are in line with the range of values predicted by the chemical models of hot cores by \cite{garrod2022} (see their Table 19). The increase in the abundance ratio is due to the smaller column density of \ce{NH2CHO} in these regions (see Figure \ref{fig:summary_LTE}). We investigate possible physical causes in Sec. \ref{subsec:chemical_variations} below. Finally, the abundance ratio between the warm component of \ce{CH3NH2} and \ce{CH2NH} is close to 3 in all four regions (see Figure~\ref{fig:correlations}j), which falls on the lower side of ratios measured in massive protostars (\citealt{bogelund2019} and references therein, \citealt{suzuki2023}). When comparing to the Hot Core models by \cite{garrod2022}, our measured ratios are well below those predicted by the model (see their Table 17). The discrepancy could arise from the lack of destruction routes of \ce{CH3NH2} in the model \citep{garrod2022}.  To the best of our knowledge, between the two species, only \ce{CH2NH} has been detected in shocked or outflowing regions \citep[e.g.][]{widicus_weaver_deep_2017, gorski2023}. From the PCA analysis performed by \cite{harada_pca_2024}, \ce{CH2NH} was correlated with both shock tracers (SiO) and high-energy transitions of \ce{HC3N}, thought to be linked with young starbursts. The warm compact component of \ce{CH3NH2} could not probe the same gas component as \ce{CH2NH}, which can be partially supported by the weak correlation between the two species (see Figure~\ref{fig:correlations}).

Overall, if iCOMs are emitted on pSSC scales, the gas could be warm enough to desorb those forming on the icy grain mantles (scenario 2b in Fig.~\ref{fig:summary}). Alternatively iCOMs could be tracing shocks, associated (scenario 2a in Fig.~\ref{fig:summary}) or not (scenario 1 in Fig.~\ref{fig:summary}) with the ongoing star formation. With our current angular resolution, we cannot really distinguish between the different scenarios. Higher angular resolution observations and chemical modelling will be necessary to distinguish between the various processes.

\subsection{Chemical variations across regions}\label{subsec:chemical_variations}

In the previous sections, we found some chemical differences between the four regions. On the one hand, we found that GMC\,7 presents an enhancement in O-bearing species compared to N-bearing species. On the other hand, we found that whilst high $E_{\mathrm{u}}$ transitions of \ce{CH3CHO} are excited only towards GMC\,6 and pSSC\,5, those for \ce{NH2CHO} are excited in GMC\,6 and pSSC\,2, indicating an anti-correlation between the two species towards pSSC\,5 and pSSC\,2. The \ce{H2} density between the four regions does not seem to change significantly based on previous studies on various molecular tracers \citep[e.g.,][]{behrens_tracing_2022, huang_reconstructing_2023, bouvier2024}. Hence, the chemical differentiation seen could be real and we investigate in the following what could cause it.

\subsubsection{The O- versus N-bearing dichotomy towards GMC 7}\label{sec:GMC7}

The contour maps in Figure \ref{fig:sup-icoms} and the correlation plots in panels b), c), g), and j) of Figure \ref{fig:correlations} show a dichotomy between the O and N-bearing species toward GMC\,7. Looking at Table~\ref{tab:av_GF}, there is no particular shift in $V_{\mathrm{peak}}$ or change in FWHM between N- and O-bearing iCOMs towards GMC\,7 which would highlight a different origin of emission between the two types of iCOMs. If we consider the column densities derived at GMC scales (see bottom part of Figure~\ref{fig:summary_LTE}), GMC\,7 shows a decrease in N-bearing species compared to the other regions, whilst the amount of O-bearing species stay relatively constant. What causes the lack of N-bearing species towards GMC\,7, compared to the other regions?

A first factor could be related to a difference in physical conditions in this GMC. In Section~\ref{subsec:GMC_scales}, we hypothesised that most of the iCOMs could be linked to shocks occurring within the GMCs. From the analysis of shock species (e.g. HNCO, SiO, and S-bearing species; \citealt{huang_reconstructing_2023, bouvier2024}), it was shown that shocks are likely less intense towards GMC\,7 compared to the other regions. Stronger shocks would lead to the presence of warmer gas. However, from the gas temperatures derived for \ce{CH2NH} in Sec.~\ref{sec: nonlte}, GMC\,7 does not show a clear temperature difference compared to the other regions. On the other hand, the \ce{H2} column density in GMC\,7 is lower compared to the other three regions by approximately one order of magnitude \citep[e.g.,][]{mangum_fire_2019}. A correlation between the amount of formed iCOMs and the \ce{H2} column density was found in low-mass and high-mass protostellar cores, indicating that iCOMs formation is more advanced in denser cores \citep{yang_peaches_2021, baek2022}. However, in \cite{baek2022}, whilst the abundance of \ce{CH3CHO} was found to be relatively constant with $N_{\ce{H2}}$, \ce{C2H5OH}, \ce{CH3NH2}, and \ce{NH2CHO} were found to grow proportionally with $N_{\ce{H2}}$. Hence, the difference in column density does not seem to explain the O- vs N- dichotomy.

A second factor leading to a difference in O- versus N-bearing iCOMs could be evolutionary. \cite{huang_reconstructing_2023} estimated a timescale for the history of the shocks for each GMC and suggested that shocks towards GMC\,7 are older ($\sim 10^4$ yr) than towards the three other GMCs ($\sim 10^3$ yr). Hence, the lower abundance in N-bearing species, if they all trace the same shock event, could be linked to the fact that the shock is older. However, this would contradict what is observed in the Galactic prototypical L1157-B1 shock. There, an anti-correlation between \ce{CH3CHO} and \ce{NH2CHO} was observed \citep{codella2017}, but the authors showed that \ce{CH3CHO} was tracing the youngest shock, whilst \ce{NH2CHO} is tracing the older part of the shock, as the species needs time to be formed in the gas phase. We would thus have expected GMC\,7 to be more abundant in N-bearing species if it traced older shocks, which is not the case. We should note that the timescale estimation of the shocks in the GMCs of NGC\,253 relies on the assumption that both HNCO and SiO are tracing the same shock episode, which is not necessarily the case.

Studies of high-mass star-forming regions have shown that less evolved sources have a lower amount of N-bearing iCOMs compared to O-bearing ones \citep[e.g.][]{beuther2009, suzuki2018, bogelund_NvsO_2019, baek2022}. This is explained by the fact that ice or gas-phase nitrogen chemistry takes longer to initiate compared with the chemistry of O-bearing species \citep[e.g.][]{charnley1992, beuther2009, garrod2022}. Interestingly, \cite{dedes2011} observed three high-mass star-forming regions and found an overabundance of O-bearing species compared to N-bearing ones. However, whilst one of the sources is very young, the other two are relatively evolved with  HII regions. Hence, a low N-bearing iCOM abundance does not necessarily imply a younger region. In addition, it is not known whether pSSCs are present within GMC\,7, as it was not included in \cite{ando_2017} and \cite{leroy_forming_2018}. Understanding the core population could bring additional information on whether the GMC is younger or older compared to the other GMCs. Observations of Radio Recombination Lines (RRLs) and 3mm continuum emission (due to free-free emission), both arising from ionised gas that traces ongoing star formation showed that while there is no clear RRL emission towards GMC\,7 but faint 3mm continuum is detected  \citep[e.g.][]{rodriguez-rico2006,bendo_2015}. In addition, only one radio continuum source (unclassified) has been identified in the vicinity of GMC\,7, contrarily to the other three regions \citep{ulvestad_1997}. Young star-formation could thus be much less active towards GMC\,7, which is supported by the lack of line emission usually associated with young starburst towards this region \citep{harada_pca_2024, kishikawa2024}. Finally, performing chemical modelling will also be essential to help us understand the lack of N-bearing species towards GMC\,7.

\subsubsection{The \texorpdfstring{\ce{CH3CHO}}{CH3CHO} - \texorpdfstring{\ce{NH2CHO}}{NH2CHO} excitation condition anti-correlation in pSSC 5 and pSSC 2}\label{subsec:environment}

The regions GMC\,6, pSSC\,5, and pSSC\,2 show a relatively similar chemical content in terms of iCOMs. There is, however, a small difference between pSSC\,5 and pSSC\,2, when it comes to the warmer and more compact component (governed by high $E_{\mathrm{u}}$ transitions) that we identified in the rotational diagram analysis (see Sec.~\ref{subsec:rds}). Indeed, whilst \ce{NH2CHO} transitions with $E_{\mathrm{u}}$ up to 200 K are detected towards GMC\,6 and pSSC\,2, it is not the case towards pSSC\,5. On the other hand, the situation is reversed for \ce{CH3CHO} where high $E_{\mathrm{u}}$ transitions, up to 80--100 K, are lacking towards pSSC\,2. We investigate below whether known differences between these two regions could explain the lack of high excitation transitions for these two species.

Since the chemical difference concerns transitions for which the emission is mostly unresolved by the ALCHEMI beam, we consider that the difference might come from the small-scale constitution of the GMCs, i.e. at pSSC scales. Our ALCHEMI beam encompasses several of the pSSCs identified by \cite{leroy_forming_2018}. Our regions pSSC\,5 and pSSC\,2 encompass pSSCs 4 to 7 and pSSCs 1 to 3, respectively. In the case of GMC\,6, only one pSSC has been identified (pSSC\,14). The pSSC's chemical richness and physical properties have been investigated by a few studies at high angular resolution \citep[e.g.][]{krieger_molecular_2020a, levy_outflows_2021,mills_clustered_2021}. From these studies, no significant differences in the chemical composition have been found towards the pSSCs of interest for this work. Additionally, all host SHCs so the difference is likely not coming from a difference in source population between the pSSCs. On the other hand, only pSSC 14, pSSC\,4 and pSSC\,5 show clear outflow features \citep{levy_outflows_2021}. If on the scales of pSSCs, \ce{CH3CHO} and \ce{NH2CHO} do not probe the same gas component, then we could explain the lack of high excitation transitions of \ce{CH3CHO} towards our region pSSC\,2 if \ce{CH3CHO} traces shocked gas linked to the pSSC's outflow whilst \ce{NH2CHO} traces the forming pSSCs themselves. If \ce{NH2CHO} is linked to the pSSCs and their hosting SHCs, then a difference between the properties of the sources is needed to explain the lack of high excitation of the species towards pSSC\,5. Looking at the gas temperature derived for the SHCs \citep{rico-villas_super_2020}, the pSSCs composing our region pSSC\,5 seem to have a slightly lower temperature compared to those of regions GMC\,6 and pSSC\,2. As already mentioned, due to the binding energy range of \ce{NH2CHO} being the highest among the iCOMs studied here, the region pSSC\,5 could thus be less efficient in releasing \ce{NH2CHO} if it forms on the grain. Moreover, a cooler gas temperature could reflect a younger source, indicating that the pSSC\,5 is younger than the two others. However, there is no clear consensus on the evolutionary age pattern of the pSSCs  \citep[e.g.][]{rico-villas_super_2020, krieger_molecular_2020a, mills_clustered_2021, levy_outflows_2021}. This hypothesis does not, in addition, explain why narrower line widths for \ce{NH2CHO} are detected solely towards pSSC\,2.

%%%%%%%%%%%%%%%%%%%%%%%%%%%%%%%%%%%%%%%%%%

\section{Conclusions}\label{sec:conclusions}

We used ALCHEMI ALMA large program measurements to investigate the emission and formation pathways of selected iCOMs, \ce{CH3CHO} (acetaldehyde), \ce{C2H5OH} (ethanol), \ce{NH2CHO} (formamide), \ce{CH2NH} (methanimine), and \ce{CH3NH2} (methylamine), towards the CMZ of NGC\,253. We performed radiative transfer analyses to derive rotational temperatures and column densities for each iCOM under study. For \ce{CH2NH}, we performed a non-LTE LVG analysis as this was the only iCOM in our study with available collisional rates. We studied the relation between the iCOMs at both GMC and pSSC scales, to see whether chemical and physical differences between iCOMs and regions are present, and how it could relate to the source's physical or chemical structure. We summarise our main findings below.

\begin{itemize}
    \item The iCOM emission is mostly concentrated towards the inner part of the CMZ (i.e. within 6$\arcsec$ or $\sim 100 $ pc around the kinematic centre), between GMC\,7 and pSSC\,2 (close to GMC\,3). At similar upper-level energy, \ce{CH2NH} and \ce{CH3NH2} are usually more spatially extended compared to \ce{CH3CHO}, \ce{C2H5OH}, and \ce{NH2CHO}. Finally, emission from iCOMs becomes more compact with increasing $E_{\mathrm{u}}$
    
    \item A rotational diagram analysis showed that for \ce{CH3NH2}, \ce{CH3CHO} and \ce{NH2CHO}, two gas components are needed depending on the region, for which we applied two different beam-filling factors, one for a region of emission at the scale of GMCs and the other at the scale of pSSCs.

    \item All $T_{\text{rot}}$ derived are below 100 K and the majority lie in the range $10-40$ K, indicating that iCOMs are sub-thermally excited. The column densities of \ce{CH3CHO} and \ce{C2H5OH} are relatively constant throughout the regions. The N-bearing species, on the other hand, show a decrease towards GMC\,7.

    \item The LVG analysis we performed on \ce{CH2NH} shows that both optical depths and non-LTE effects are at play for this species. The derived gas temperature and density show that \ce{CH2NH} emits in a gas of at least 20 K and at a density of $ 6\times10^4$ \pcmc\ for GMC\,7, and up to $6\times10^6$ \pcmc\ for GMC\,6. The emission size of \ce{CH2NH} is more compact than GMC scales ($\sim 0.15\arcsec-0.6\arcsec$ or $\sim 2.5-10$ pc).

    \item The cold extended component of \ce{CH3CHO} ($E_{\mathrm{u}}< 40$ K) and \ce{C2H5OH} are well correlated in all four regions, which could imply a chemical link between the two species. Gas phase production of \ce{CH3CHO} from \ce{C2H5OH} through the \say{ethanol tree} could be possible, although we cannot exclude grain formation for the two species. 

    \item The cold extended component of \ce{NH2CHO} ($E_{\mathrm{u}}<70$ K) and \ce{H2CO} have similar peak of emission within the CMZ of NGC\,253. If the two species are chemically linked, as proposed in the literature, then \ce{NH2CHO} could have formed in the gas phase via the reaction \ce{H2CO} + \ce{NH2}. 

    \item Both gas components of \ce{CH3NH2} are well correlated with \ce{CH2NH}, respectively, which could indicate that they are either tracing a similar physical process or that they formed through the same precursor on the ice grain mantle or in the gas phase. Our current results do not allow us to conclude between the two possibilities.
    
    \item If the iCOMs emit at GMC scales, the most favourable scenario to explain the widespread emission of the iCOMs is if they are tracing large-scale shocks within the GMCs. This resembles what is seen in the centre of our own Galaxy.

    \item Part of the iCOM emission may come from smaller scales, such as pSSC scales. Both thermal and non-thermal desorption may be thus at play, due to the heating from the forming stars. Not all iCOMs may trace the same gas component, some of them likely tracing heating processes (e.g. \ce{CH3NH2}, \ce{NH2CHO}) whilst the others may rather trace shock processes due to active star-formation (e.g. \ce{CH2NH}, \ce{CH3CHO}, \ce{C2H5OH}). We can however not firmly conclude on the origin of the iCOMs at pSSC scales without higher angular resolution or chemical modelling.

    \item At GMC scales (governed by a cold extended component of the iCOMs), GMC\,7 shows a deficiency in N-bearing iCOMs compared to the other regions. The cause may be a different evolutionary stage of GMC\, or to the fact that star formation is simply less active in that region. 

    \item pSSC\,5 and pSSC\,2 show different excitation conditions for  \ce{CH3CHO} and \ce{NH2CHO}. Whilst \ce{NH2CHO} does not have a warm component towards pSSC\,5, \ce{CH3CHO} does not have a warm component towards pSSC\,2. The difference could be partially explained if the two species are tracing different features, with \ce{CH3CHO} tracing the outflows driven by the pSSCs and \ce{NH2CHO} tracing the heating sources. However, we cannot explain the significant narrow line widths of \ce{NH2CHO} that are solely present towards pSSC\,2.
\end{itemize}

This first study of iCOMs towards the CMZ of NGC\,253 shows that several processes could be involved in their emission. The current angular resolution allows us to put some constraints on the region of emission of the iCOMs and their formation pathways, but chemical modelling and higher angular resolution observations will be crucial to address all the questions raised in this work. In addition, while in Galactic studies  high angular resolution helps to determine precisely where iCOMs are emitted, this remains a challenge in extragalactic studies.

\begin{acknowledgements}
We thank the anonymous referee who helped improve the clarity of the paper. This work is founded by the European Research Council (ERC) Advanced Grant MOPPEX 833460. V.M.R. and L.C. acknowledge support from the grant PID2022-136814NB-I00 by the Spanish Ministry of Science, Innovation and Universities/State Agency of Research MICIU/AEI/10.13039/501100011033 and by ERDF, UE. V.M.R. also aknowledges the grant RYC2020-029387-I funded by MICIU/AEI/10.13039/501100011033 and by "ESF, Investing in your future", and from the Consejo Superior de Investigaciones Cient{\'i}ficas (CSIC) and the Centro de Astrobiolog{\'i}a (CAB) through the project 20225AT015 (Proyectos intramurales especiales del CSIC); and from the grant CNS2023-144464 funded by MICIU/AEI/10.13039/501100011033 and by “European Union NextGenerationEU/PRTR”. This paper makes use of the following ALMA data: ADS/JAO.ALMA\#2017.1.00161.L and ADS/JAO.ALMA\#2018.1.00162.S. ALMA is a partnership of ESO (representing its member states), NSF (USA) and NINS (Japan), together with NRC (Canada), NSTC  and ASIAA (Taiwan),
and KASI (Republic of Korea), in cooperation with the Republic of Chile. The Joint ALMA Observatory is operated by ESO, AUI/NRAO and NAOJ. This research has made use of spectroscopic and collisional data from the EMAA database (https://emaa.osug.fr and https://dx.doi.org/10.17178/EMAA). 
\end{acknowledgements}

%
%-------------------------------------------------------------
%               Appendices have to be placed at the end, after
%                                        \end{thebibliography}
%-------------------------------------------------------------

\bibliographystyle{aa} 
\bibliography{aanda}

\onecolumn
\FloatBarrier
\begin{appendix} %First appendix

\section{Integrated Intensity Images}
The velocity integrated maps for \ce{CH3NH2}, \ce{CH2NH}, \ce{CH3CHO}, \ce{NH2CHO}, and \ce{C2H5OH}, are presented here. The velocities used for the integration are within the range 70-380 \kms.

\begin{figure}[ht]
    \centering
    \includegraphics[width=0.8\linewidth]{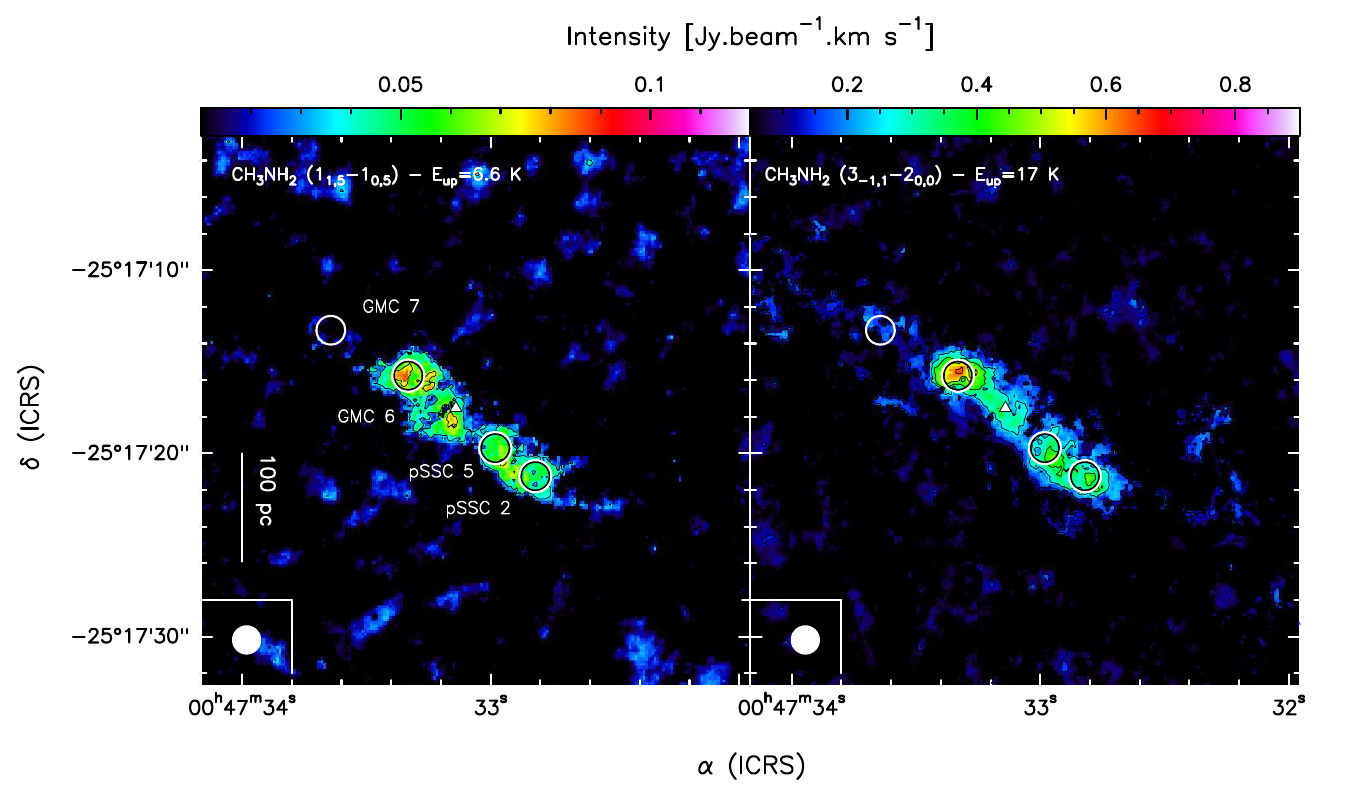}
    \caption{Velocity-integrated images of \ce{CH3NH2}. The kinematic centre \citep{muller-sanchez2010} is labelled by a white triangle. The regions where the spectra were extracted are shown by the black and white circles and are labelled in the left-most panel. The beam is depicted in the lower left corner of each plot. The scale bar of 100 pc corresponds to $\sim 6\arcsec$. \textit{Left:} Velocity-integrated map of \ce{CH3NH2} ($1_{1,5}-1_{0,5}$). Levels start at 3$\sigma$ (1$\sigma=$42 \mjybeamkms) with steps of 3$\sigma$. \textit{Right:} Velocity-integrated map of \ce{CH3NH2} ($3_{-1,1}-2_{0,0}$). Levels start at 3$\sigma$ (1$\sigma=$42 \mjybeamkms) with steps of 3$\sigma$. }
    \label{fig:ch3nh2-maps}
\end{figure}

\begin{figure}
    \centering
    \includegraphics[width=0.8\linewidth]{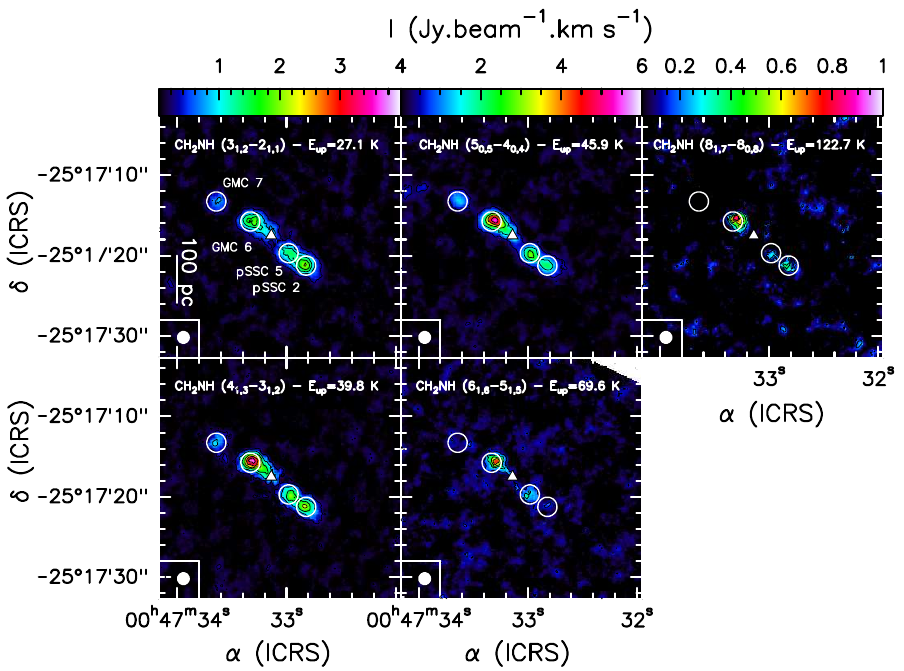}
    \caption{Same as Figure~\ref{fig:ch3nh2-maps} but for \ce{CH2NH}. \textit{Top Left:} Velocity-integrated map of \ce{CH2NH} ($3_{1,2}-2_{1,1}$). Levels start at 3$\sigma$ (1$\sigma=$90 \mjybeamkms) with steps of 5$\sigma$. \textit{Bottom Left:} Velocity-integrated map of \ce{CH2NH} ($4_{1,3}-3_{1,2}$). Levels start at 3$\sigma$ (1$\sigma=$78 \mjybeamkms) with steps of 8$\sigma$. \textit{Middle Top:} Velocity-integrated map of \ce{CH2NH} ($5_{0,5}-4_{0,4}$). Levels start at 3$\sigma$ (1$\sigma=$140 \mjybeamkms) with steps of 8$\sigma$. \textit{Bottom Middle:} Velocity-integrated map of \ce{CH2NH} ($6_{1,6}-5_{1,5}$). Levels start at 3$\sigma$ (1$\sigma=$210 \mjybeamkms) with steps of 5$\sigma$. \textit{Top Right:} Velocity-integrated map of \ce{CH2NH} ($8_{1,7}-8_{0,8}$). Levels start at 3$\sigma$ (1$\sigma=$61 \mjybeamkms) with steps of 3$\sigma$. }
    \label{fig:ch2nh-maps}
\end{figure}

\begin{figure}
    \centering
    \includegraphics[width=0.9\linewidth]{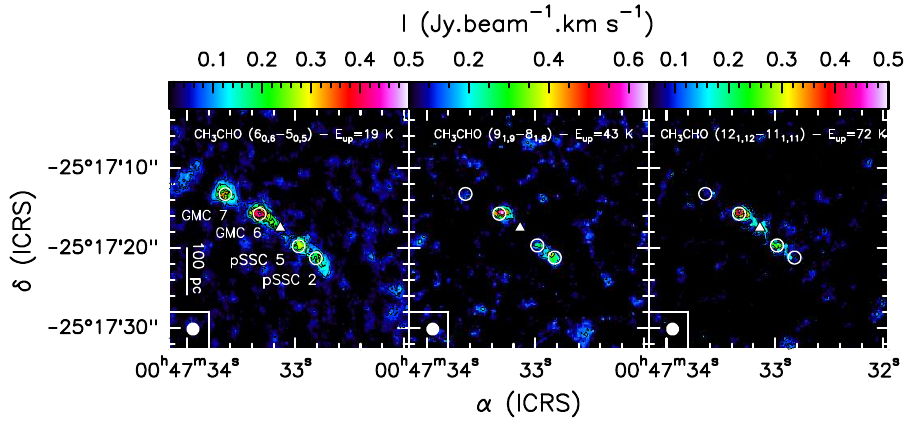}
    \caption{Same as Figure~\ref{fig:ch3nh2-maps} but for \ce{CH3CHO}. \textit{Left:} Velocity-integrated map of \ce{CH3CHO} ($1_{1,5}-1_{0,5}$ (E+A)). Levels start at 3$\sigma$ (1$\sigma=25$ \mjybeamkms) with steps of 3$\sigma$. \textit{Middle:} Velocity-integrated map of \ce{CH3CHO} ($9_{2,9}-8_{1,8}$ (E+A)). Levels start at 3$\sigma$ (1$\sigma=55$ \mjybeamkms) with steps of 3$\sigma$. \textit{Right:} Velocity-integrated map of \ce{CH3CHO} ($12_{1,12}-11_{1,11}$ (E+A)). Levels start at 3$\sigma$ (1$\sigma=36$ \mjybeamkms) with steps of 3$\sigma$.}
    \label{fig:ch3cho-maps}
\end{figure}

\begin{figure}
    \centering
    \includegraphics[width=1\linewidth]{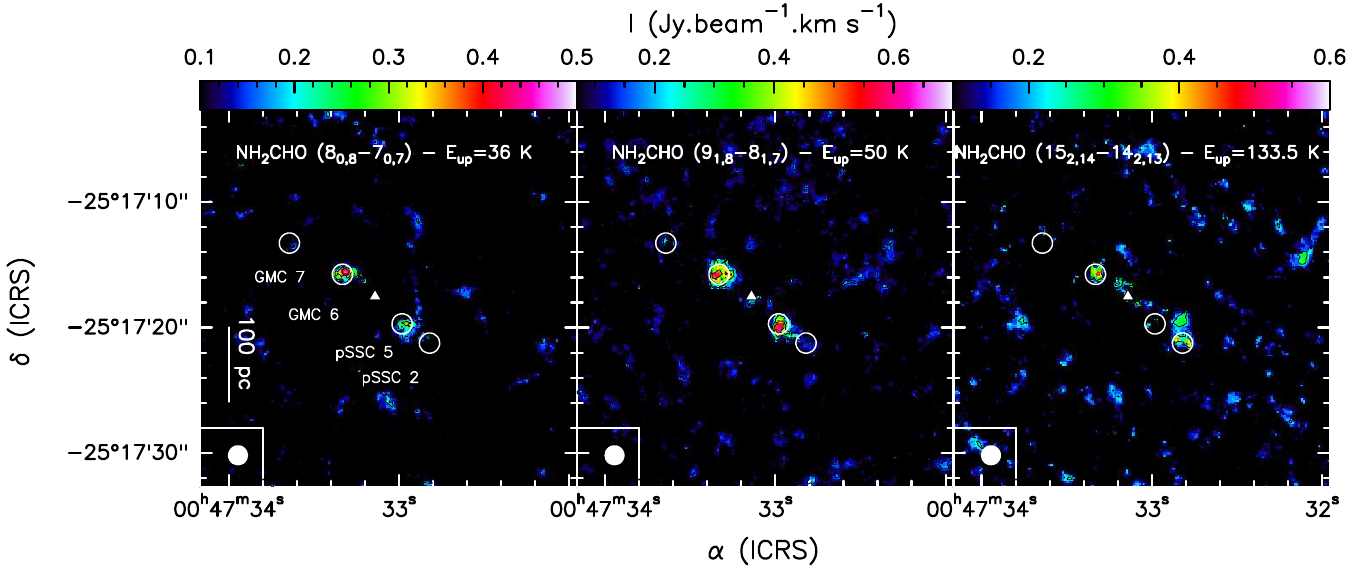}
    \caption{Same as Figure~\ref{fig:ch3nh2-maps} but for \ce{NH2CHO}. \textit{Left:} Velocity-integrated map of \ce{NH2CHO} ($8_{0,8}-7_{0,7}$). Levels start at 3$\sigma$ (1$\sigma=$58 \mjybeamkms) with steps of 2$\sigma$. \textit{Middle:} Velocity-integrated map of \ce{NH2CHO} ($9_{1,8}-8_{1,7}$). Levels start at 3$\sigma$ (1$\sigma=$52 \mjybeamkms) with steps of 3$\sigma$. \textit{Right:} Velocity-integrated map of \ce{NH2CHO} ($15_{2,14}-14_{2,13}$). Levels start at 3$\sigma$ (1$\sigma=$78 \mjybeamkms) with steps of 3$\sigma$. }
    \label{fig:nh2cho-maps}
\end{figure}

\begin{figure}
    \centering
    \includegraphics[width=0.8\linewidth]{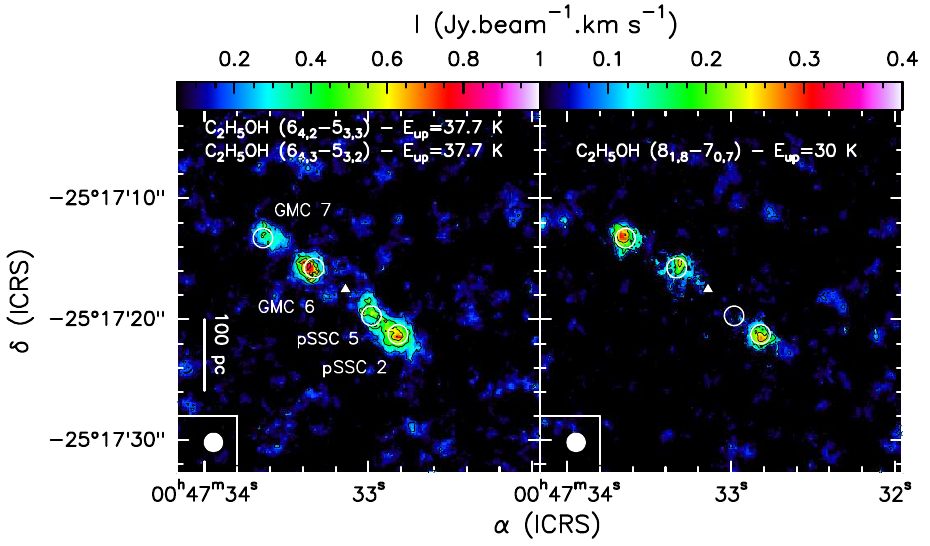}
    \caption{Same as Figure~\ref{fig:ch3nh2-maps} but for \ce{t-C2H5OH}. \textit{Left:} Velocity-integrated map of \ce{t-C2H5OH} ($6_{4,2}-5_{3,3}$ and $6_{4,3}-5_{3,2}$; blended together). Levels start at $3\sigma$ (1$\sigma=$64 \mjybeamkms) with steps of 3$\sigma$. \textit{Right:} Velocity-integrated map of \ce{t-C2H5OH} ($8_{1,8}-7_{0,7}$). Levels start at 3$\sigma$ (1$\sigma=$26 \mjybeamkms) with steps of 3$\sigma$.}
    \label{fig:c2h5oh-maps}
\end{figure}

\FloatBarrier

\section{Spectra}
\begin{figure}
    \centering
    \includegraphics[width=1\linewidth]{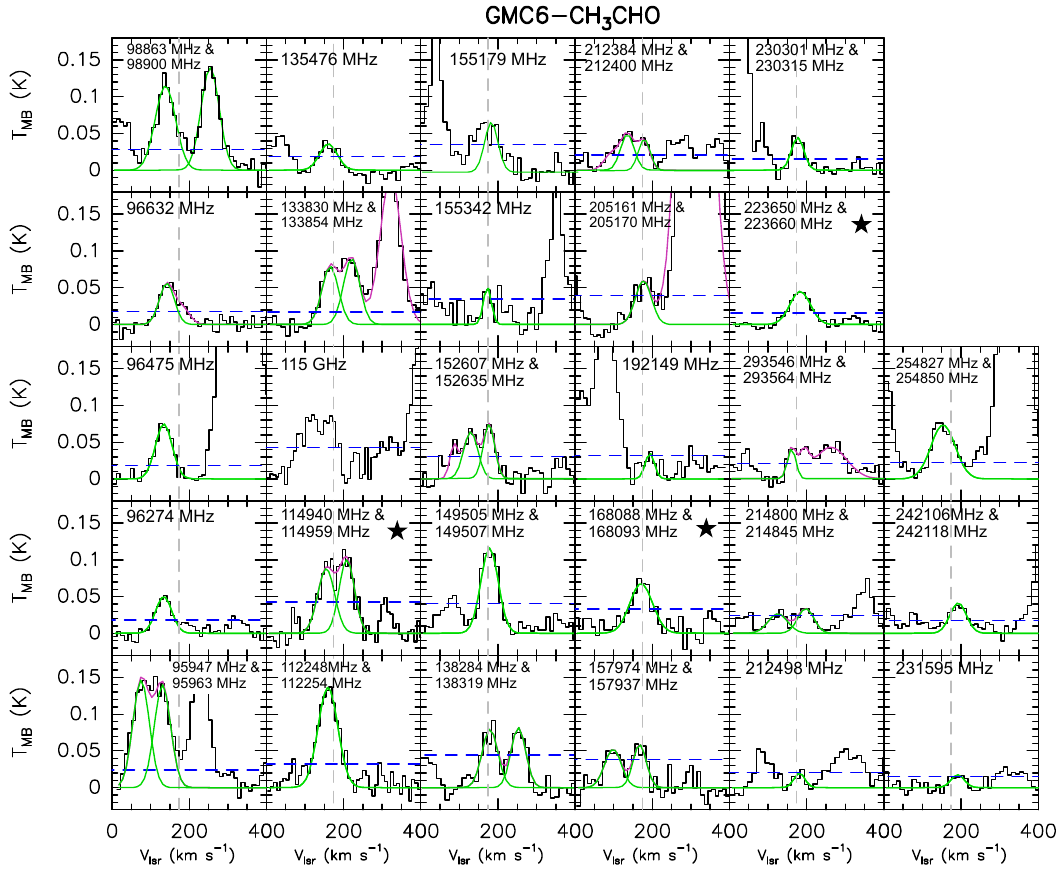}
    \caption{Spectra of \ce{CH3CHO} towards GMC 6. The frequency of the lines are indicated in each box. If multiple transitions of \ce{CH3CHO} are partially or totally blended together, the frequency of each transition is indicated in the same box. The green lines show the Gaussian fit performed for each transition of \ce{CH3CHO}. If a multiple Gaussian fit has been performed, the overall fit is shown by a magenta continuous line. The vertical dashed grey line represent the mean velocity at which the \ce{CH3CHO} lines are peaking towards GMC 6. The dashed blue line shows  the 3$\sigma$ level. The transitions used to perform the moment maps are identified with a black filled star. Several lines of \ce{CH3CHO} (with different $E_{\mathrm{u}}$ around 115 GHz are blended together so we did not perform a Gaussian fit and we did use them in the analysis. }
    \label{fig:spec_ch3cho_GMC6}
\end{figure}

\begin{figure}
    \centering
    \includegraphics[width=0.8\linewidth]{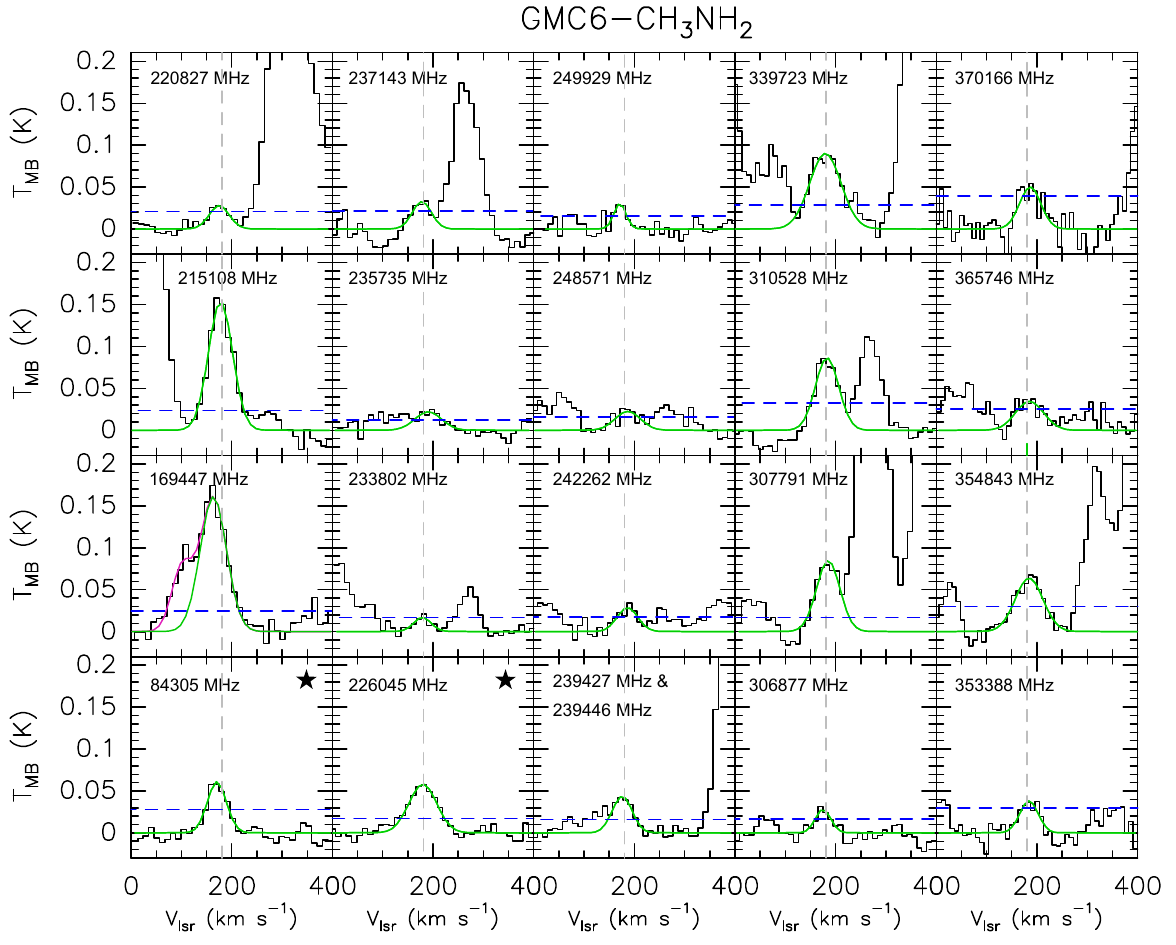}
    \caption{Same as Figure~\ref{fig:spec_ch3cho_GMC6} but for \ce{CH3NH2} towards  GMC\,6.}
    \label{fig:spec_ch3nh2_GMC6}
\end{figure}

\begin{figure}
    \centering
    \includegraphics[width=0.9\linewidth]{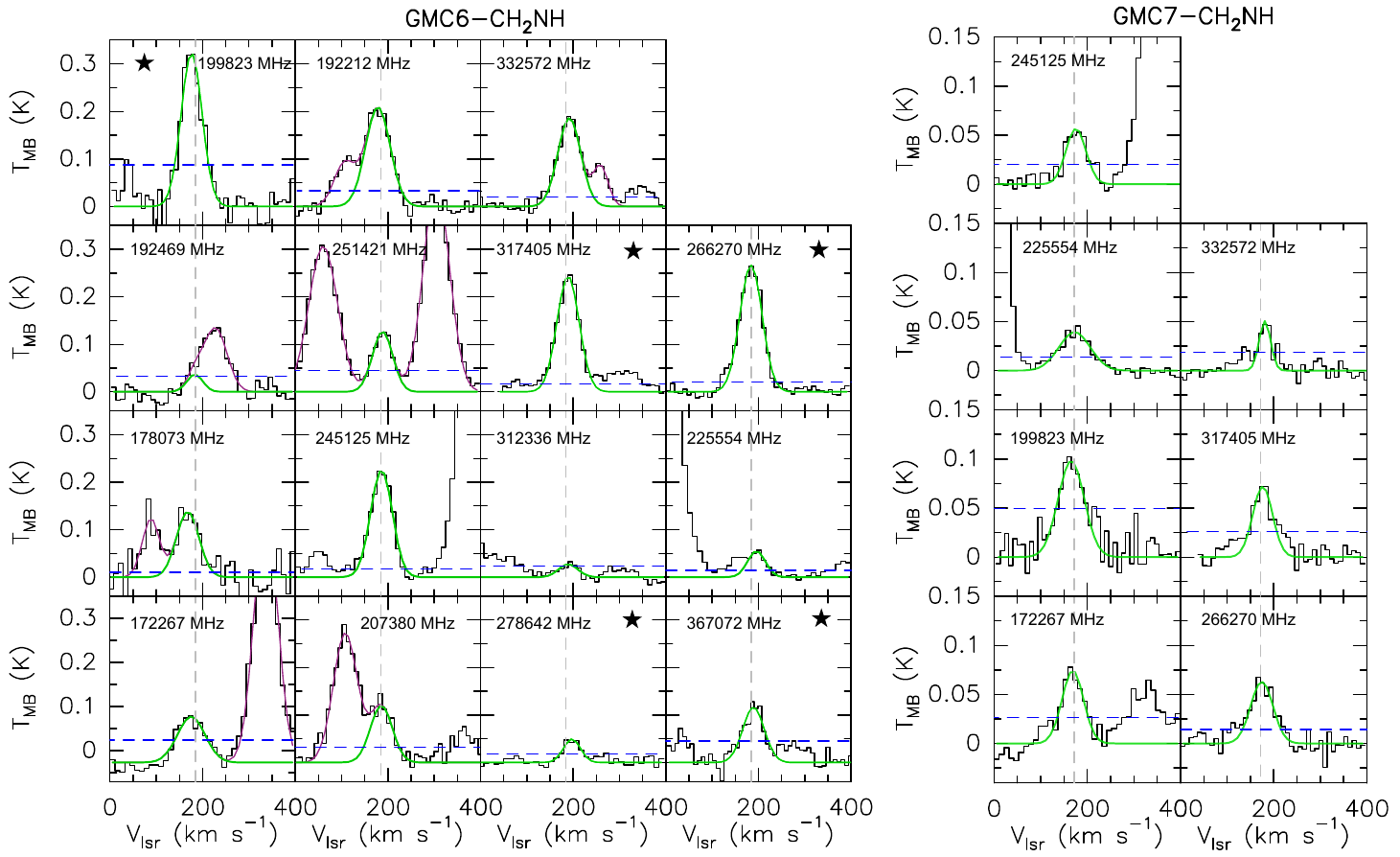}
    \caption{Same as Figure~\ref{fig:spec_ch3cho_GMC6} but for \ce{CH2NH} towards GMC\,6 and GMC\,7.}
    \label{fig:spec_ch2nh_GMC6_GMC7}
\end{figure}

\begin{figure}
    \centering
    \includegraphics[width=0.9\linewidth]{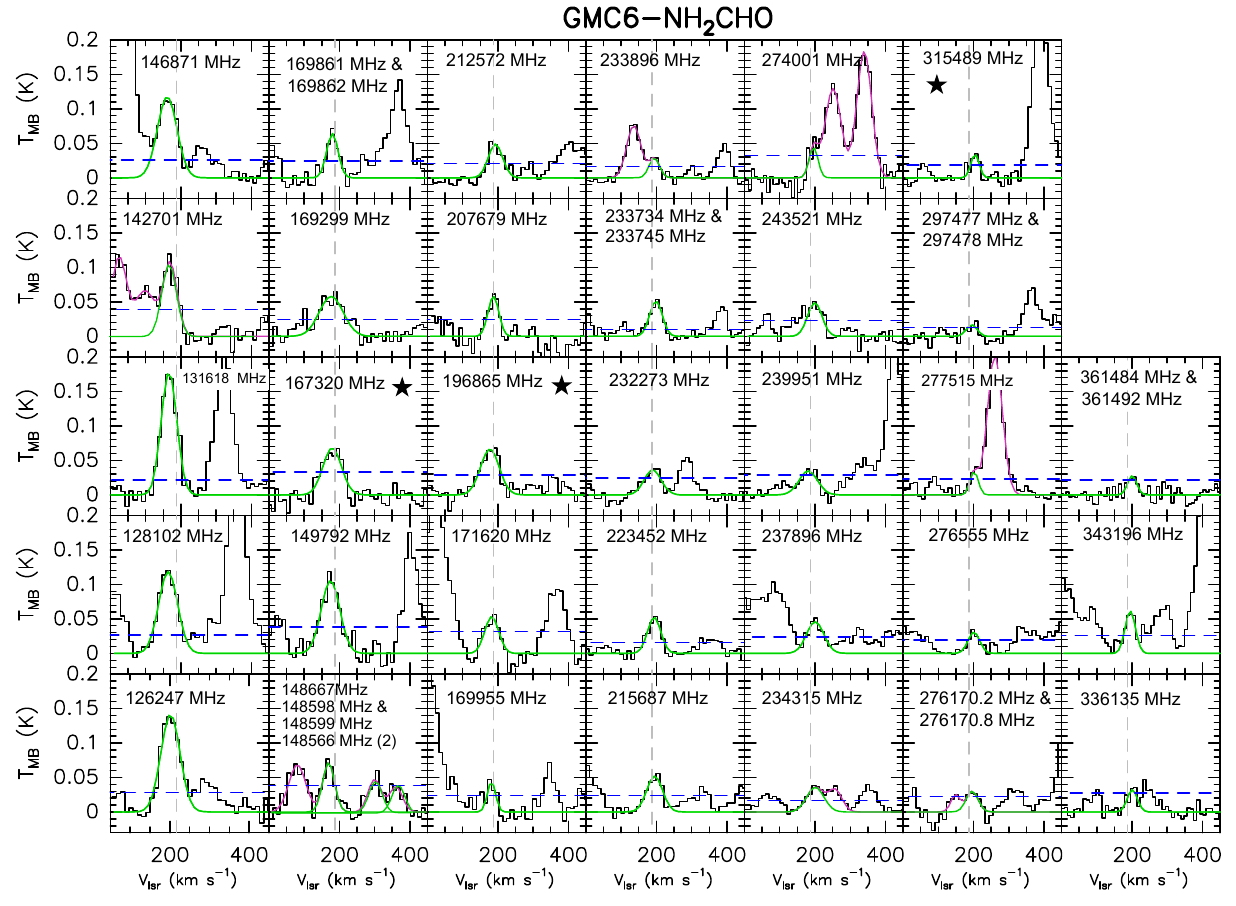}
    \caption{Same as Figure~\ref{fig:spec_ch3cho_GMC6} but for \ce{NH2CHO} towards  GMC\,6.}
    \label{fig:spec_nh2cho_GMC6}
\end{figure}

\begin{figure}
    \centering
    \includegraphics[width=0.8\linewidth]{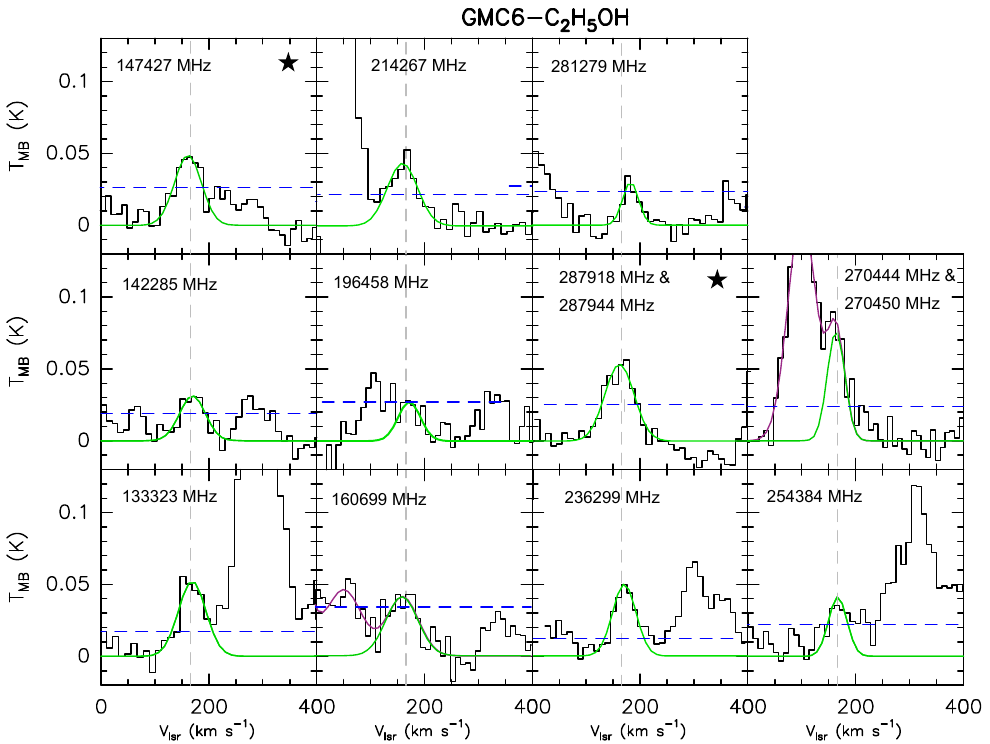}
    \caption{Same as Figure~\ref{fig:spec_ch3cho_GMC6} but for \ce{C2H5OH} towards  GMC\,6.}
    \label{fig:spec_c2h5oh_GMC6}
\end{figure}

\begin{figure}
    \centering
    \includegraphics[width=0.6\linewidth]{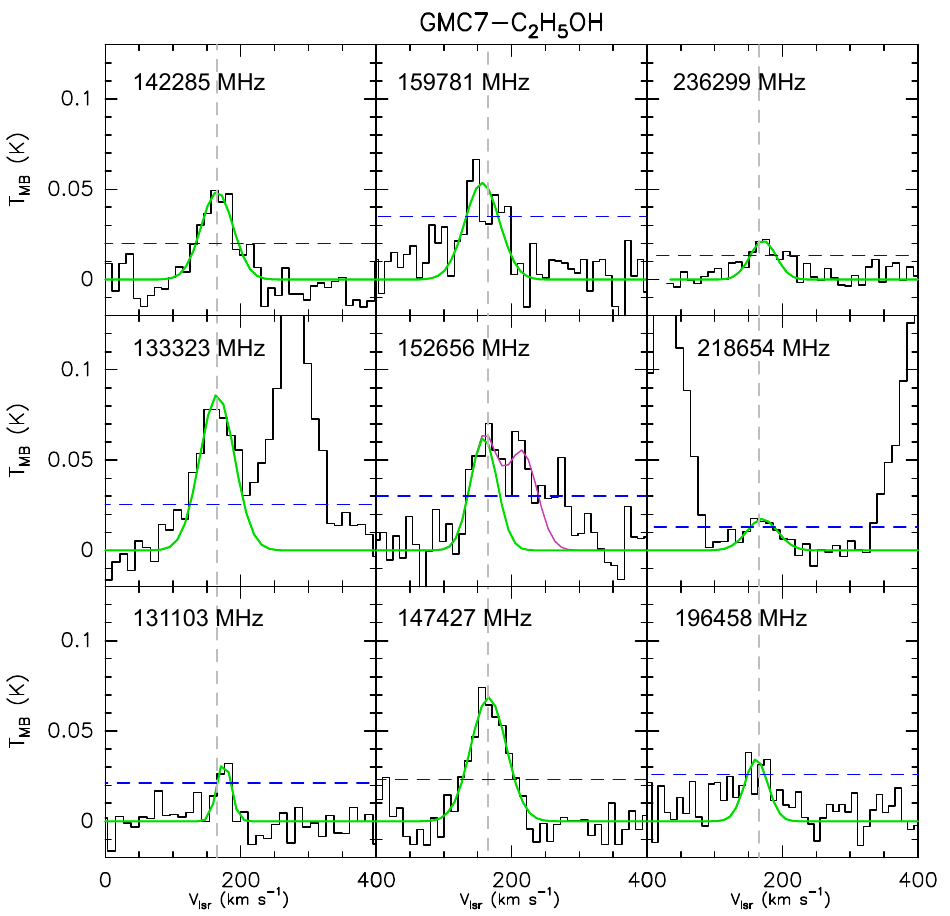}
    \caption{Same as Figure~\ref{fig:spec_ch3cho_GMC6} but for \ce{C2H5OH} towards GMC\,7.}
    \label{fig:spec_c2h5oh_GMC7}
\end{figure}

\begin{figure}
    \centering
    \includegraphics[width=0.8\linewidth]{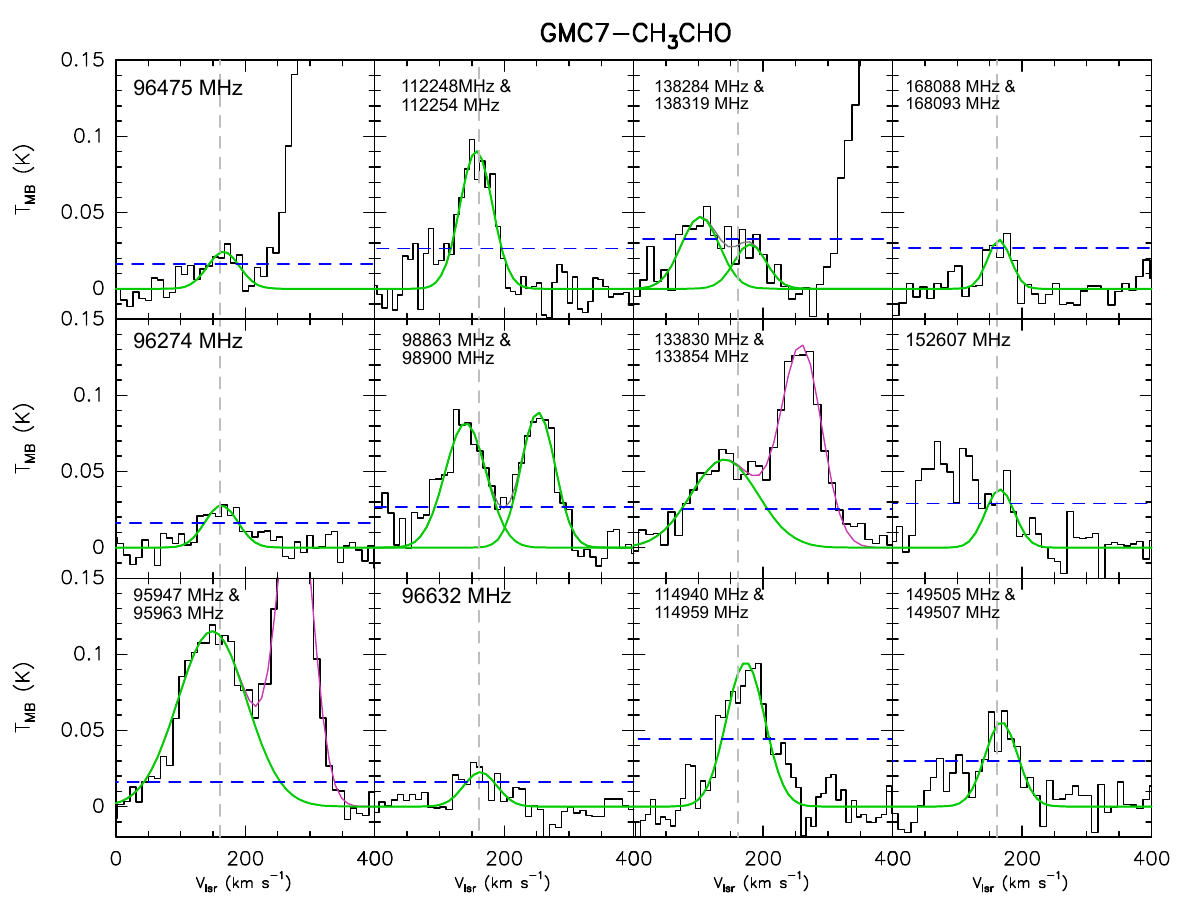}
    \caption{Same as Figure~\ref{fig:spec_ch3cho_GMC6} but towards GMC\,7.}
    \label{fig:spec_ch3cho_GMC7}
\end{figure}

\begin{figure}
    \centering
    \includegraphics[width=0.5\linewidth]{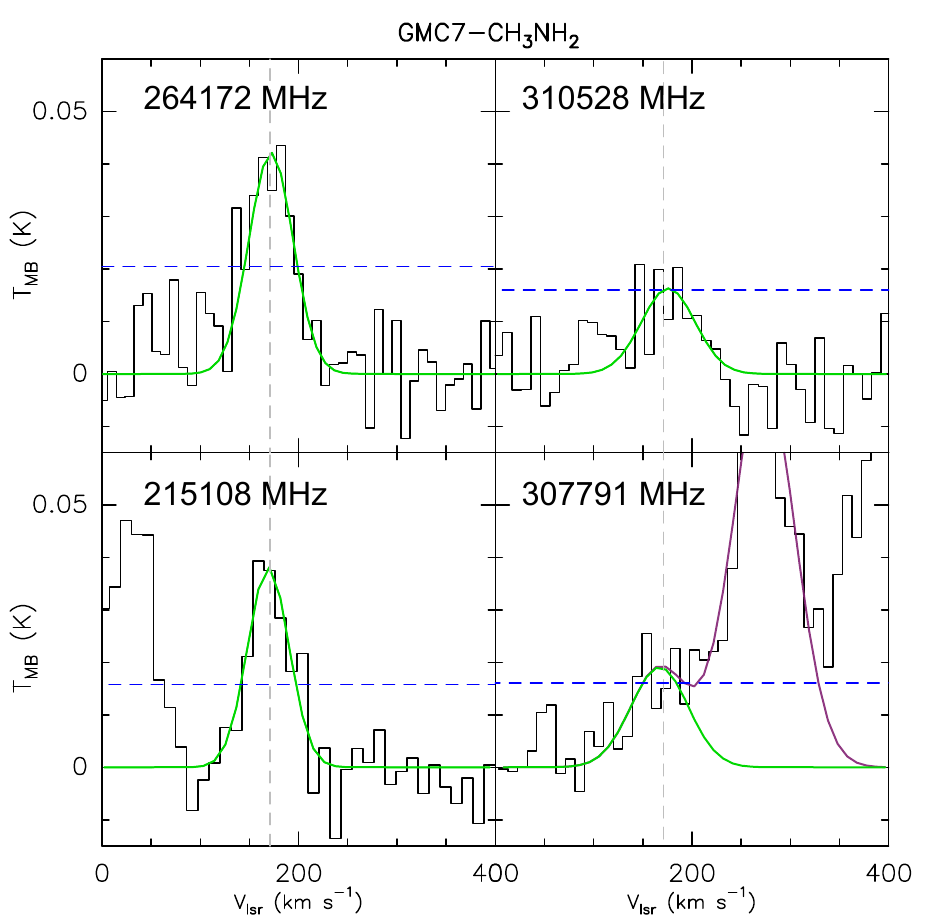}
    \caption{Same as Figure~\ref{fig:spec_ch3cho_GMC6} but for \ce{CH3NH2} towards GMC\,7.}
    \label{fig:spec_ch3nh2_GMC7}
\end{figure}

\begin{figure}
    \centering
    \includegraphics[width=0.7\linewidth]{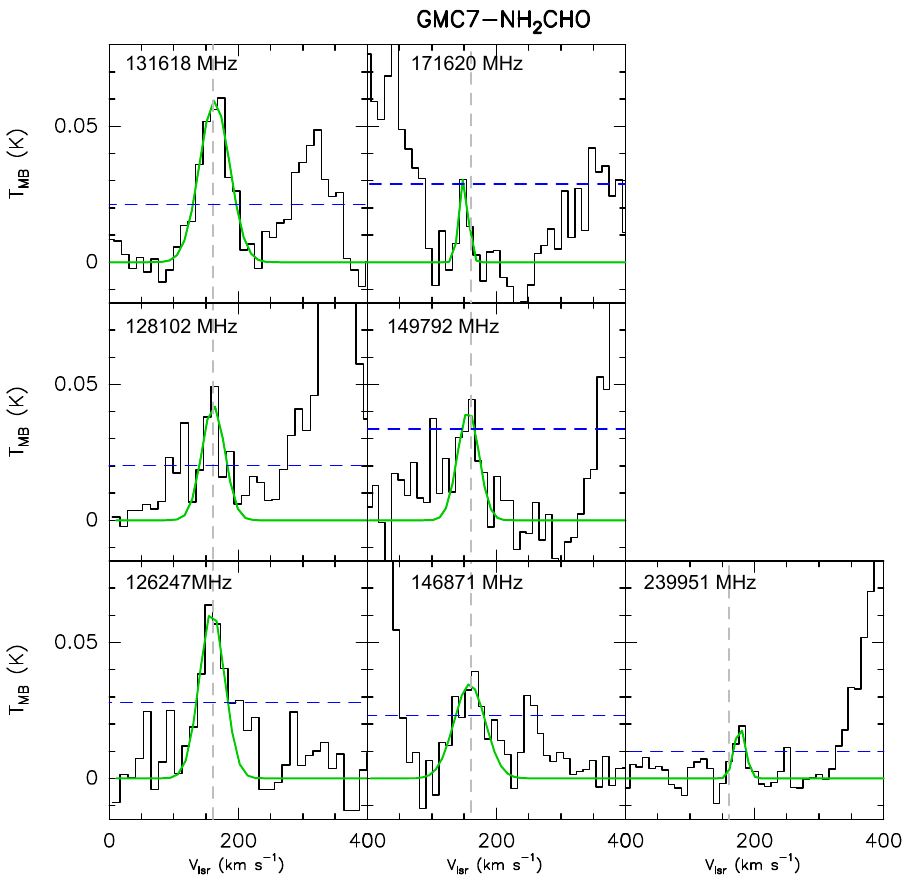}
    \caption{Same as Figure~\ref{fig:spec_ch3cho_GMC6} but for \ce{NH2CHO} towards GMC\,7.}
    \label{fig:spec_nh2cho_GMC7}
\end{figure}

\begin{figure}
    \centering
    \includegraphics[width=1\linewidth]{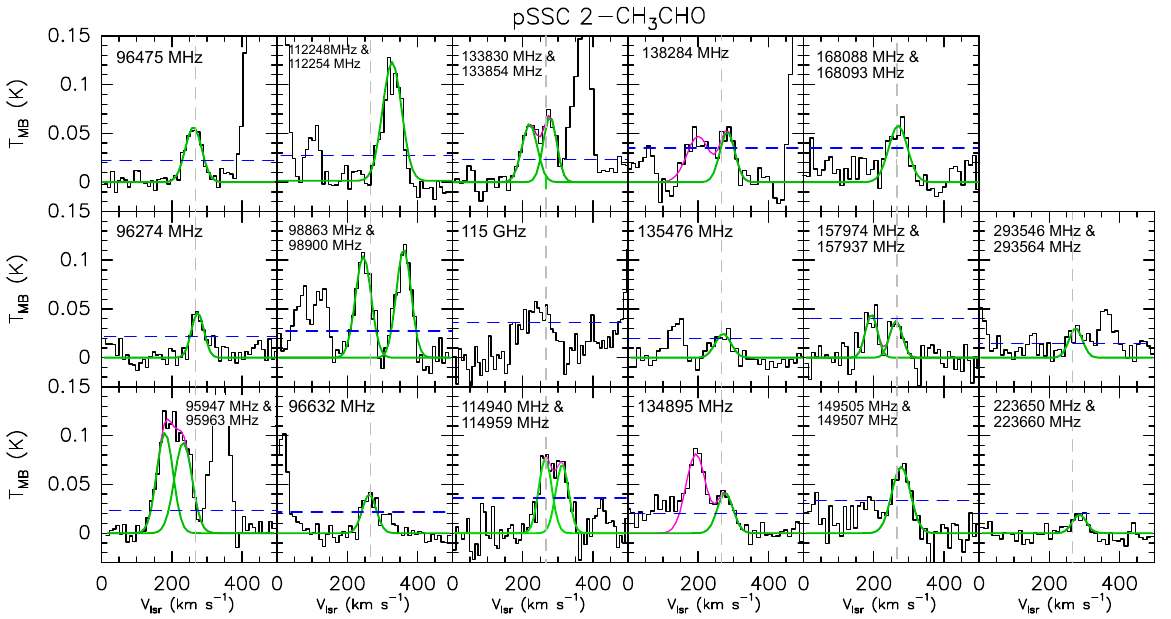}
    \caption{Same as Figure~\ref{fig:spec_ch3cho_GMC6} but towards pSSC\,2. Several lines of \ce{CH3CHO} (with different $E_{\mathrm{u}}$ around 115 GHz are blended together so we did not perform a Gaussian fit and we did use them in the analysis. }
    \label{fig:spec_ch3cho_SSC2}
\end{figure}

\begin{figure}
    \centering
    \includegraphics[width=0.7\linewidth]{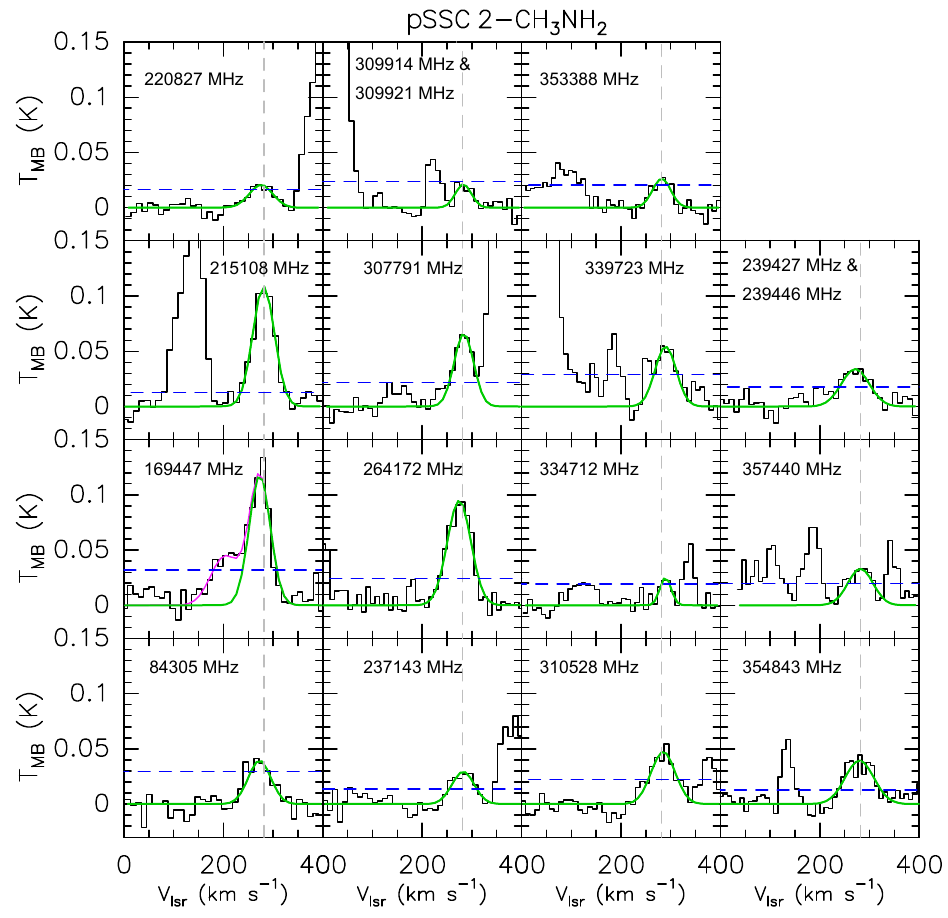}
    \caption{Same as Figure~\ref{fig:spec_ch3cho_GMC6} but for \ce{CH3NH2} towards pSSC\,2.}
    \label{fig:spec_ch3nh2_SSC2}
\end{figure}

\begin{figure}
    \centering
    \includegraphics[width=1\linewidth]{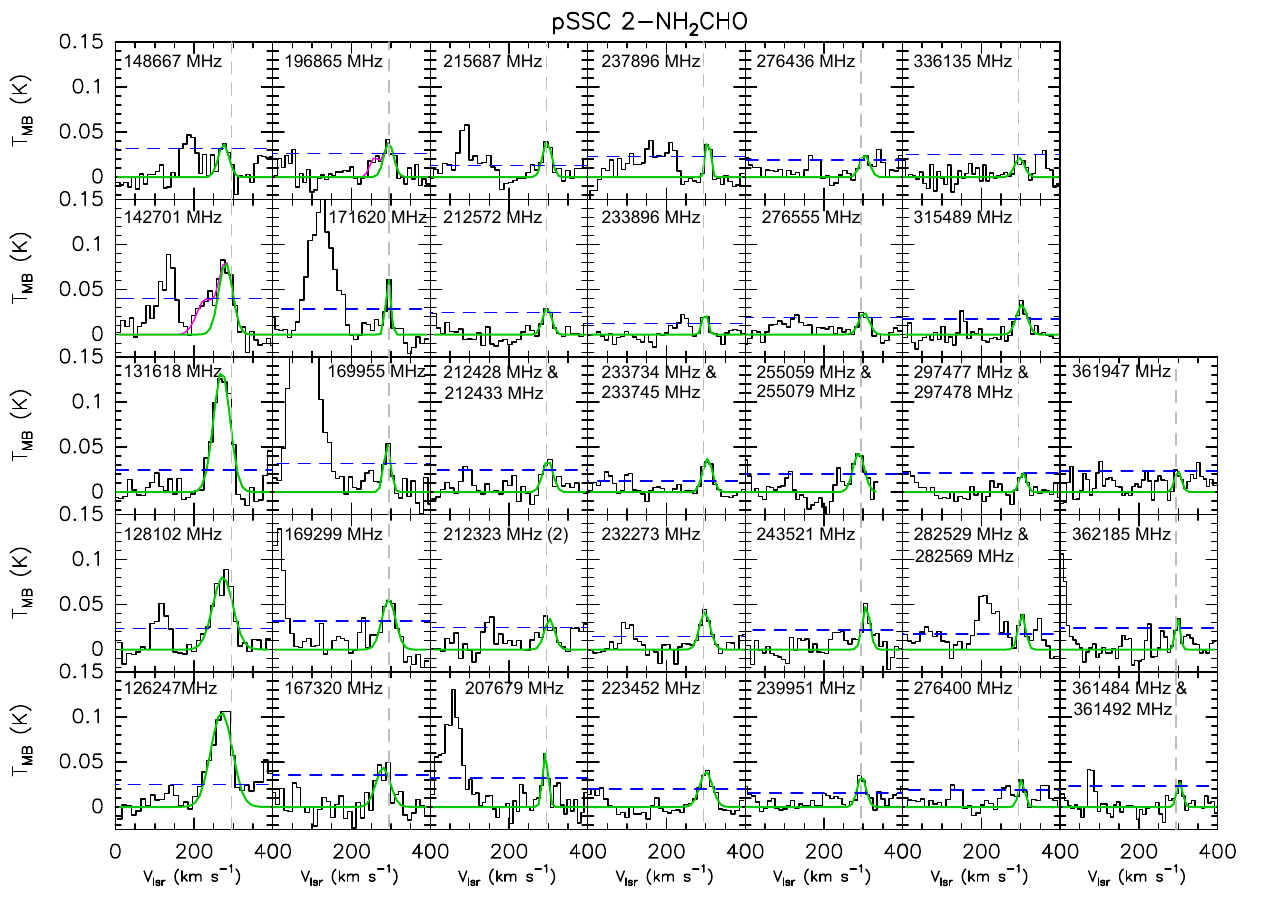}
    \caption{Same as Figure~\ref{fig:spec_ch3cho_GMC6} but for \ce{NH2CHO} towards pSSC\,2.}
    \label{fig:spec_nh2cho_SSC2}
\end{figure}

\begin{figure}
    \centering
    \includegraphics[width=0.8\linewidth]{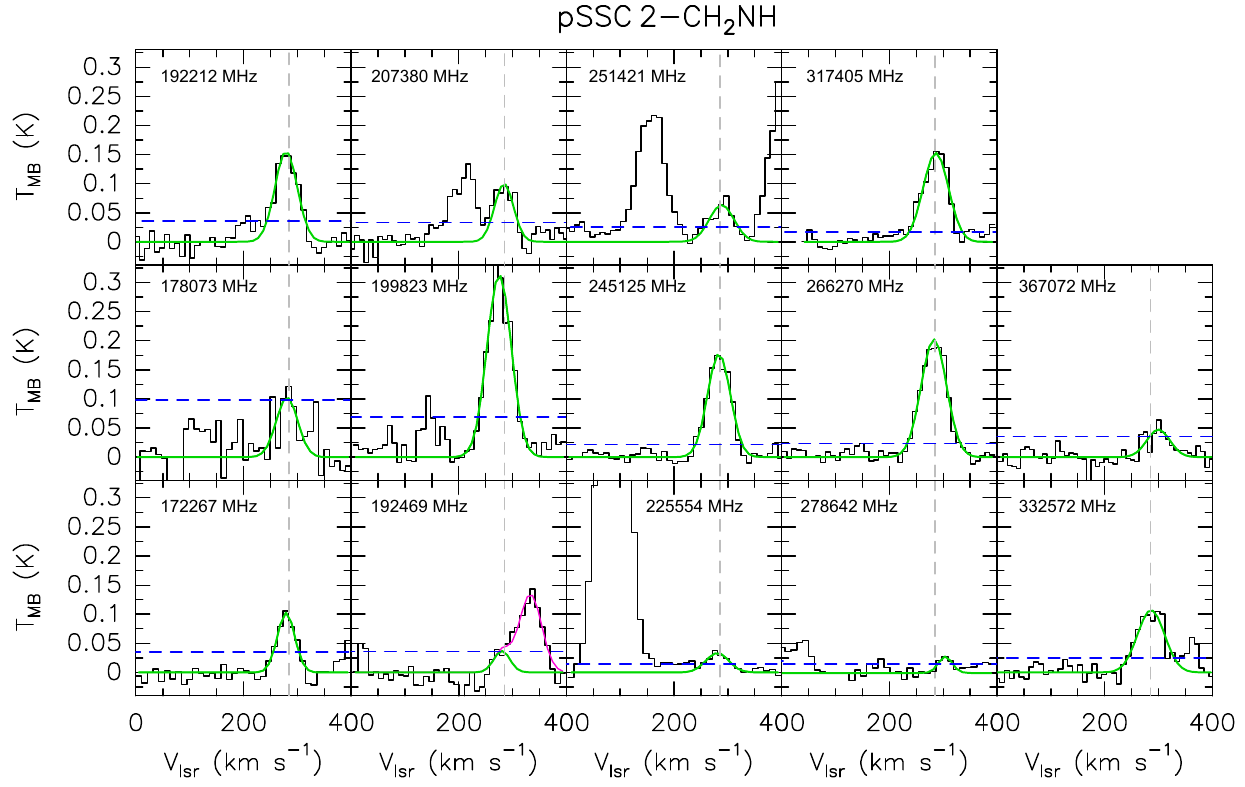}
    \caption{Same as Figure~\ref{fig:spec_ch3cho_GMC6} but for \ce{CH2NH} towards pSSC\,2.}
    \label{fig:spec_ch2nh_SSC2}
\end{figure}

\begin{figure}
    \centering
    \includegraphics[width=1\linewidth]{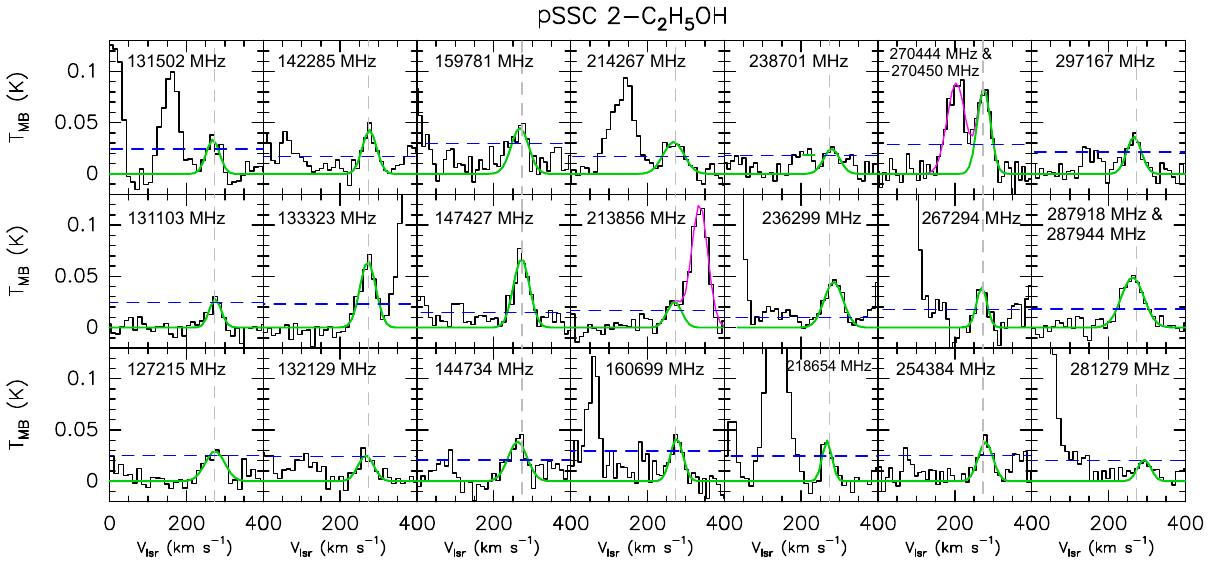}
    \caption{Same as Figure~\ref{fig:spec_ch3cho_GMC6} but for \ce{C2H5OH} towards pSSC\,2.}
    \label{fig:spec_c2h5oh_SSC2}
\end{figure}

\begin{figure}
    \centering
    \includegraphics[width=0.8\linewidth]{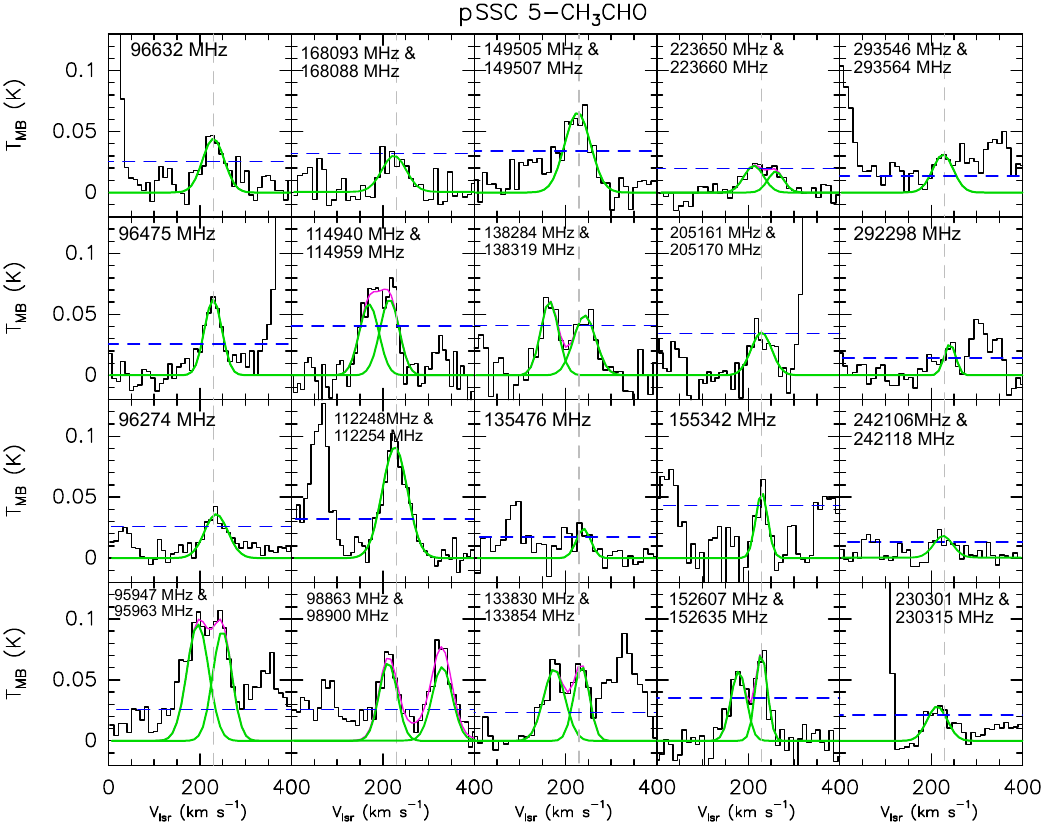}
    \caption{Same as Figure~\ref{fig:spec_ch3cho_GMC6} but towards pSSC\,5.}
    \label{fig:spec_ch3cho_SSC5}
\end{figure}

\begin{figure}
    \centering
    \includegraphics[width=1\linewidth]{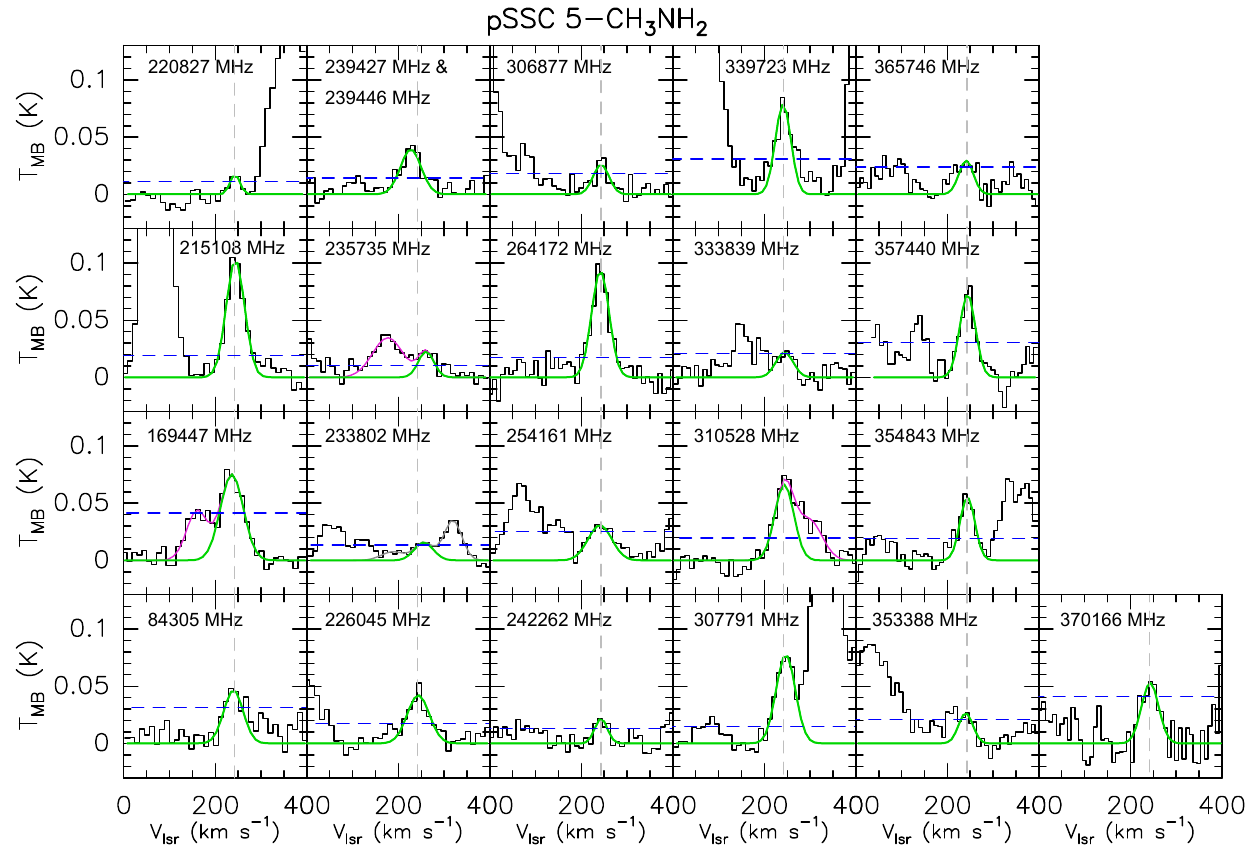}
    \caption{Same as Figure~\ref{fig:spec_ch3cho_GMC6} but for \ce{CH3NH2} towards pSSC\,5.}
    \label{fig:spec_ch3nh2_SSC5}
\end{figure}

\begin{figure}
    \centering
    \includegraphics[width=0.8\linewidth]{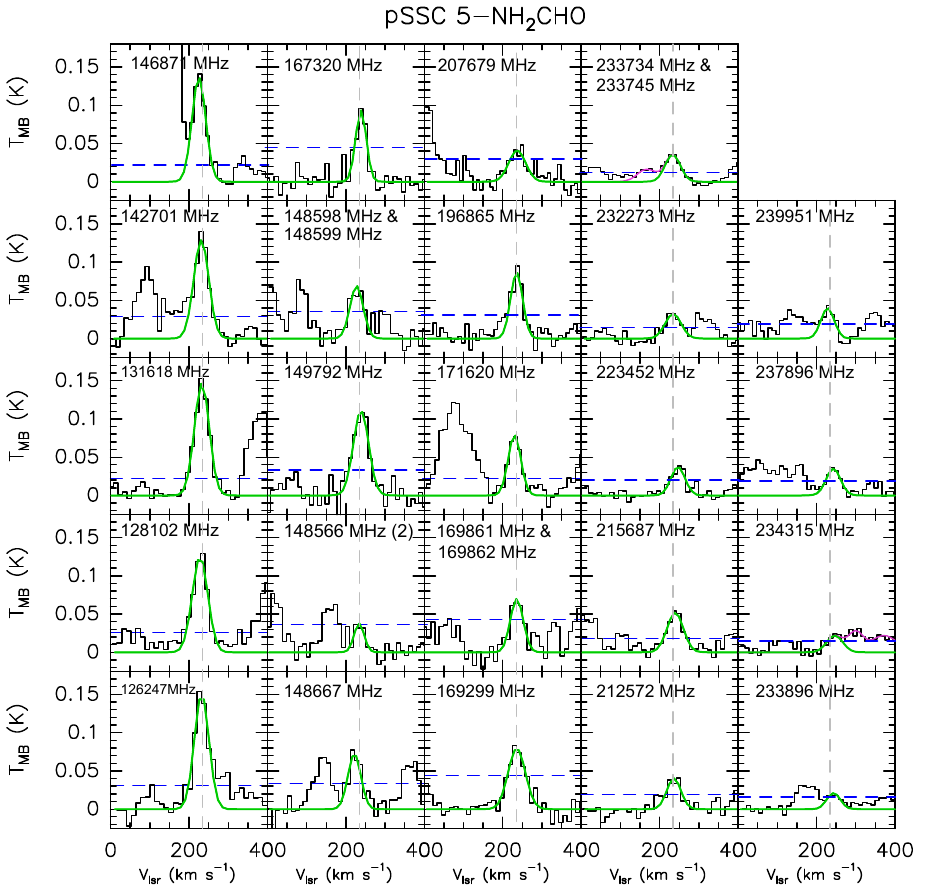}
    \caption{Same as Figure~\ref{fig:spec_ch3cho_GMC6} but for \ce{NH2CHO} towards pSSC\,5.}
    \label{fig:spec_nh2cho_SSC5}
\end{figure}

\begin{figure}
    \centering
    \includegraphics[width=0.7\linewidth]{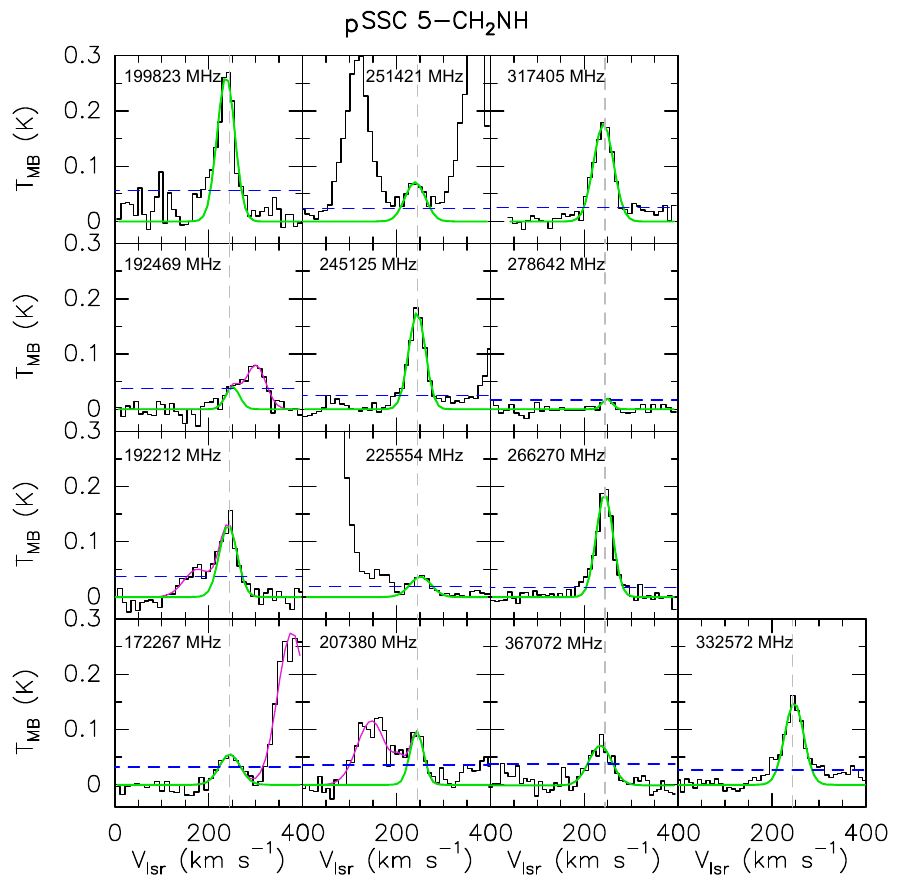}
    \caption{Same as Figure~\ref{fig:spec_ch3cho_GMC6} but for \ce{CH2NH} towards pSSC\,5.}
    \label{fig:spec_ch2nh_SSC5}
\end{figure}

\begin{figure}
    \centering
    \includegraphics[width=0.7\linewidth]{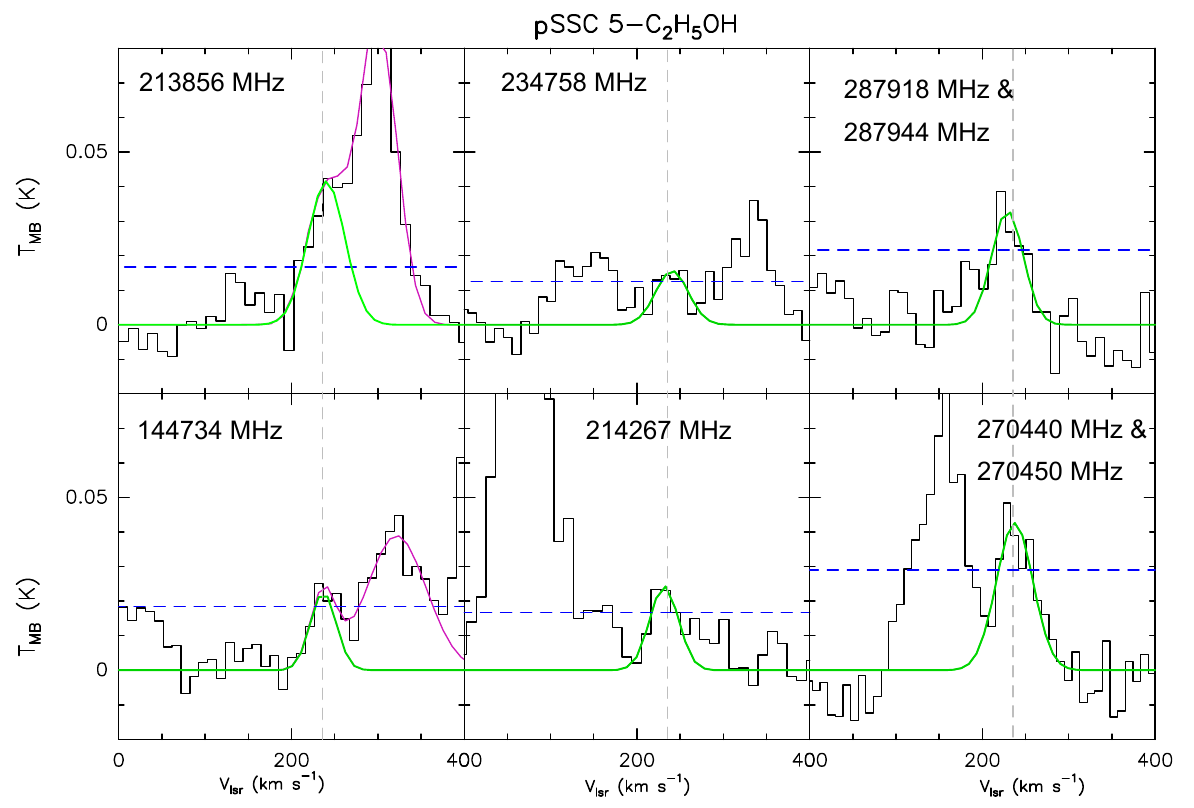}
    \caption{Same as Figure~\ref{fig:spec_ch3cho_GMC6} but for \ce{C2H5OH} towards pSSC\,5.}
    \label{fig:spec_c2H5OH_SSC5}
\end{figure}

Table~\ref{tab:lines} shows all the transitions used in the analyses, for each species and the best Gaussian fit. The calibration error of 15\% is not included in the uncertainties of the integrated intensities and the spectral resolution of 10 \kms is not included in the error of $V_{\mathrm{peak}}$ and $FWHM$. Those quantities are however taken into account in the analysis. Frequencies and spectroscopic parameters are taken from CDMS \citep[e.g.,][]{muller_2005, endres2016} for \ce{CH2NH} (TAG 029518; version 2$^*$; \citealt{dore2010, dore2012}), \ce{NH2CHO} (TAG 045516, version 2$^*$; e.g., \citealt{kryvda2009}), and \ce{t-C2H5OH} (TAG 046524, version 1$^*$; \citealt{pearson2008, muller2016}) and from JPL \citep{pickett_1998} for \ce{CH3CHO} (TAG 44003, version 3$^*$; \citealt{KLG1996} and references therein) and \ce{CH3NH2} (TAG 31008, version 1$^*$; \citealt{ilyushin2005}).
The mean FHWM and $V_\mathrm{peak}$ used in Sec.~\ref{subsec:rds} are presented in Table \ref{tab:av_GF}.
The line profiles and Gaussian fits are shown in Figures~\ref{fig:spec_ch3cho_GMC6} to \ref{fig:spec_c2H5OH_SSC5}. The full line identification is available in \cite{martin_alchemi_2021}.

\begin{longtable}{crcrcr|rccr}
\caption{Lines used in the analysis, their spectroscopic parameters and the Gaussian fit results and rms per region. }
\label{tab:lines}\\
    %\begin{tabular}{cccccc|cccc}
\hline \hline
Species &Frequency & Transition & $E_{\mathrm{u}}$ & $A_{ij}$& $g_{\mathrm{u}}$& $\int T_BdV$ & $V_\mathrm{peak}$ & $FWHM$ & rms \\ 
&(MHz) & & (K) & (s$^{-1}$)& & (K.\kms)& (\kms) & (\kms) & (mK)\\
\hline
\endfirsthead
    \caption{continued.}\\
    \hline \hline
  Species &Frequency & Transition & $E_{\mathrm{u}}$ & $A_{ij}$& $g_{\mathrm{u}}$& $\int T_BdV$ & $V_\mathrm{peak}$ & $FWHM$ & rms \\ 
    &(MHz) & & (K) & (s$^{-1}$)& & (K.\kms)& (\kms) & (\kms) & (mK)\\
    \hline
    \endhead
    \hline
    \endfoot
\multicolumn{10}{c}{GMC\,7}\\
\hline
\ce{CH2NH}& 225554 &$1_{1,1}-0_{0,0}$ & 10.8 & $9.63\times 10^{-5}$ & 9& $3.32 \pm 0.17$ & $172.6 \pm 1.9$ & $79.6 \pm 4.8$ &4.7\\
 & 172267 & $2_{1,1}-2_{0,2}$ & 17.5 & $6.12\times 10^{-5}$&15 &$4.31 \pm 0.72$ & $168.5 \pm 4.5$& $54.4 \pm 10.6$& 8.9  \\
& 199823 & $3_{1,2}-2_{1,1}$ & 27.1 & $6.35\times 10^{-5}$ & 21& $6.91 \pm 0.61$ & $165.4 \pm 2.6$ & $66.2 \pm 7.7$ & 16.4\\
& 245125 & $4_{1,4}-3_{1,3}$ & 37.3 & $1.28\times 10^{-4}$ & 27& $3.04 \pm 0.25$ & $175.5 \pm 1.9$ & $50.5 \pm 5.0$ & 6.7 \\
& 266270 &$4_{1,3}-3_{1,2}$ & 39.8 & $1.64\times 10^{-4}$ & 27 &$3.79 \pm 0.31$ & $174.4 \pm 2.0$ & $56.5 \pm 5.9$ &4.8\\
& 317405 &$5_{0,5}-4_{0,4}$ & 45.9 & $3.03\times 10^{-4}$ & 33& $3.83 \pm 0.23$ & $176.3 \pm 2.2$ & $ 50.0 \pm 10.0$ & 8.7\\
& 332572 &$5_{1,4}-4_{1,3}$ & 55.8 & $3.35\times 10^{-4}$ & 33 & $2.05 \pm 0.16$ & $170.0 \pm 10.0$ & $50.0 \pm 10.0$ &6.2\\
\end{longtable}


\begin{table}[ht]
    \centering
        \caption{Mean FWHM and $V_\mathrm{peak}$ derived from the Gaussian fits (see Sec.~\ref{subsec:rds}).}
    \label{tab:av_GF}
    \begin{tabular}{c|c|cc|cc|cc|c}
    \hline \hline
Region & \ce{CH2NH} & \multicolumn{2}{c|}{\ce{CH3NH2}} & \multicolumn{2}{c|}{\ce{NH2CHO}} & \multicolumn{2}{c|}{\ce{CH3CHO}} & \ce{C2H5OH}\\
      Component  & 1 & 1 &  2 & 1 & 2 &1 & 2 & 1 \\
         \hline 
         \multicolumn{9}{c}{FWHM (\kms)}\\
         \hline
         GMC\,7 & 55 & 55.5 & ... & 39 & ... & 62 & ...& 51 \\
         GMC\,6 & 54 & 53.5 & 44 & 59 & 45 &  50 & 39 &53 \\
         pSSC\,5 & 44 & 47 & 35 & 43 & ...&  45 & 54 &40\\
         pSSC\,2 & 49 & 46.5 & 53 & 40 & 25 & 44 & ...& 47 \\
         \hline\multicolumn{9}{c}{$V_\mathrm{peak}$ (\kms)}\\
         \hline
         GMC\,7 & 171 & 172 & ...& 161 & ...&161.5 & ...& 166 \\
         GMC\,6 & 186 &  179 & 184& 176 & 194 & 154 & 187 &168 \\
         pSSC\,5 & 243 &  242 & 243 & 235 & ...&  229 & 221 & 238\\
         pSSC\,2 & 287 & 284 & 281 &281 & 300 & 264 & ...&  273.5\\
         \hline
    \end{tabular}
\end{table}

\FloatBarrier

\section{Rotation diagrams}\label{appx:RDs}
All the rotation diagrams for GMC\,7, pSSC\,5 and pSSC\,2 are presented here. If one temperature component is fitted, we performed the rotation diagrams for the two possible sizes of emission, i.e. $1.6\arcsec$ for GMC scales and $0.12\arcsec$ for pSSC scales. We show here only the results of the fit for the GMC scales. The beam filling factor, (see Sec.~\ref{subsec:methodology}), is 0.5 for GMC scales and $\sim 0.01$ for pSSC scales. Results for both source sizes are reported in Table \ref{tab:results_RDs}.

\begin{figure}[ht]
    \centering
    \includegraphics[width=1\linewidth]{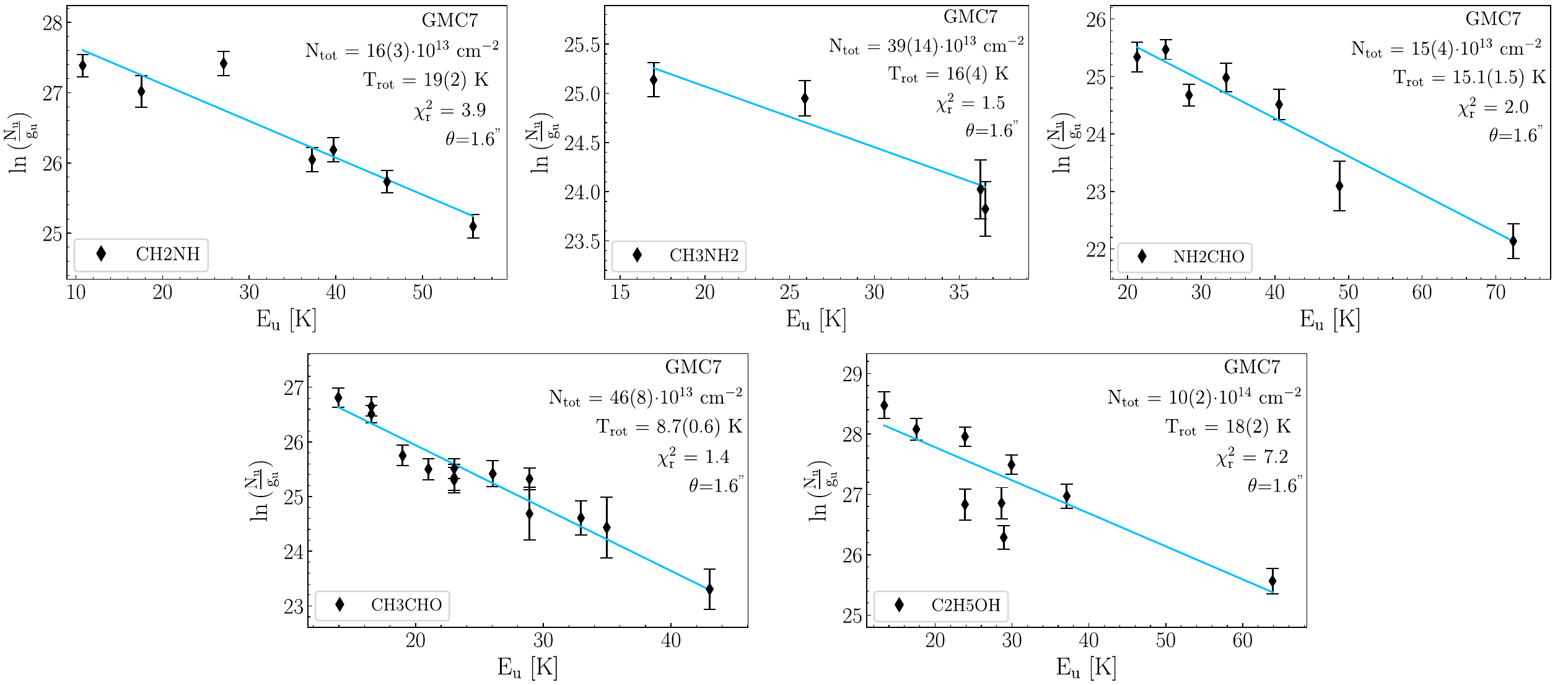}
    \caption{Rotation diagrams for each species towards GMC\,7. The parameters $E_{\mathrm{u}}$, $N_{\mathrm{u}}$, and $g_{\mathrm{u}}$ are the level energy (with respect to the ground state), column density and degeneracy of the upper level, respectively. The error bars on $\text{ln}(N_{\mathrm{u}}/g_{\mathrm{u}}$) include a calibration error of 15\% (see Sec.~\ref{sec:obs}). The blue solid line represent the best fits and the dashed grey lines are the extrapolations of the fit for the full range of $E_{\mathrm{u}}$ covered. The source size assumed is 1.6$\arcsec$ (see Sec.~\ref{subsec:methodology}).}
    \label{fig:rds_GMC7}
\end{figure}

\begin{figure}
    \centering
    \includegraphics[width=1\linewidth]{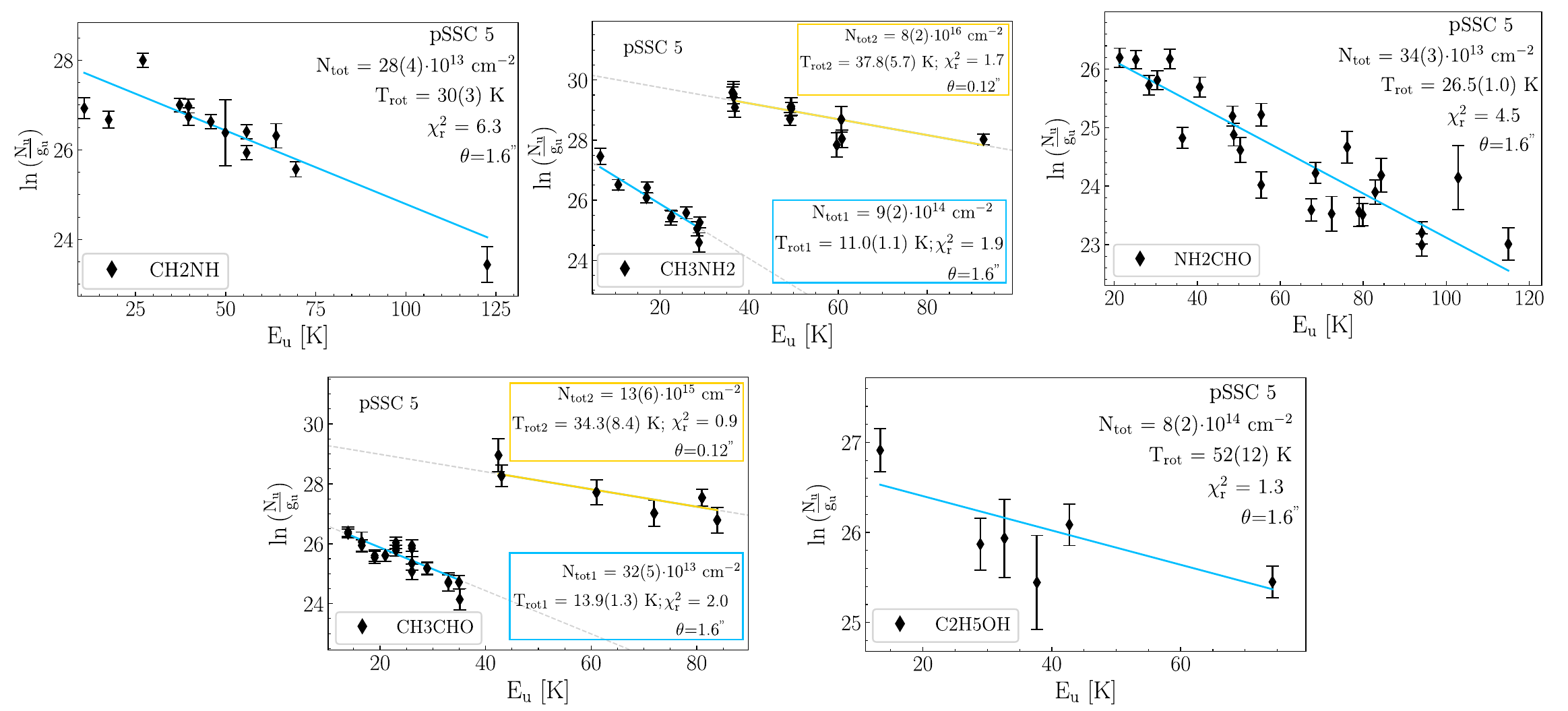}
    \caption{The parameters $E_{\mathrm{u}}$, $N_{\mathrm{u}}$, and $g_{\mathrm{u}}$ are the level energy (with respect to the ground state), column density and degeneracy of the upper level, respectively. The error bars on $\text{ln}(N_{\mathrm{u}}/g_{\mathrm{u}}$) include a calibration error of 15\% (see Sec.~\ref{sec:obs}). The blue (and orange if a second temperature component is fitted) solid line represent the best fits and the dashed grey lines are the extrapolations of the fit for the full range of $E_{\mathrm{u}}$ covered. The source size is indicated for each component (see Sec.~\ref{subsec:methodology}).}
    \label{fig:rds_SSC5}
\end{figure}

\begin{figure}
    \centering
    \includegraphics[width=1\linewidth]{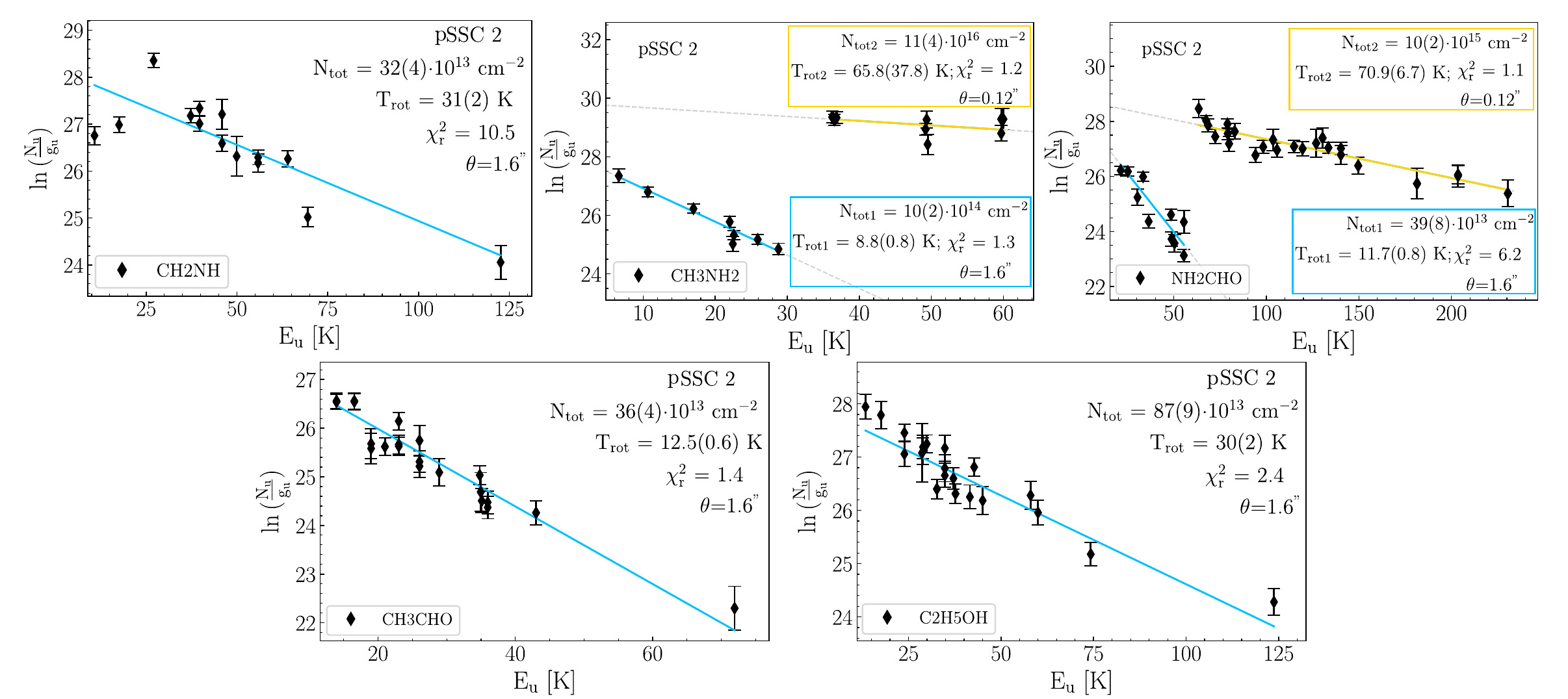}
    \caption{Same as Figure~\ref{fig:rds_SSC5} for pSSC\,2.}
    \label{fig:rds_SSC2}
\end{figure}

\begin{table}
    \centering
        \caption{Results from the rotation diagrams.}
    \label{tab:results_RDs}
    \begin{tabular}{ccccc}
    \hline \hline
        Species & Component & $T_{\text{rot}}$ & $N_{\text{tot}}$ & $\theta_S$ \\
        & & (K) & (\pcms) & ($\arcsec$)\\
         \hline
         \multicolumn{5}{c}{GMC\,7}\\
         \hline
         \multirow{2}{*}{\ce{CH2NH}} & \multirow{2}{*}{1} & \multirow{2}{*}{ $19.0 \pm 2.0$} &  $(1.6 \pm 0.3)\times 10^{14}$ & 1.6\\
         & &   &  $(1.5 \pm 0.2)\times 10^{16}$ & 0.12\\
         \multirow{2}{*}{\ce{CH3NH2}} &\multirow{2}{*}{ 1} &\multirow{2}{*}{ $16.0 \pm 4.0$} &  $(3.9 \pm 1.4)\times 10^{14}$ & 1.6 \\
         & & &  $(3.5 \pm 1.3)\times 10^{16}$ & 0.12 \\
         \multirow{2}{*}{\ce{NH2CHO}} &\multirow{2}{*}{ 1} &\multirow{2}{*}{ $15.1 \pm 1.5$} &  $(1.5 \pm 0.4)\times 10^{14}$ & 1.6\\
         & & &  $(1.3 \pm 0.3)\times 10^{16}$ & 0.12\\
         \multirow{2}{*}{\ce{CH3CHO}} &\multirow{2}{*}{ 1} &\multirow{2}{*}{ $8.7 \pm 0.6$ }&  $(4.6 \pm 0.8)\times 10^{14}$ & 1.6\\
         & & &  $(4.1 \pm 0.7)\times 10^{16}$ & 0.12\\
         \multirow{2}{*}{\ce{C2H5OH}} & \multirow{2}{*}{1} & \multirow{2}{*}{$18.0 \pm 2.0$} &  $(10.0 \pm 2.0)\times 10^{14}$ & 1.6\\
         & & &  $(8.5 \pm 1.4)\times 10^{16}$ & 0.12\\
         \hline
         \multicolumn{5}{c}{GMC\,6}\\
         \hline
         \multirow{2}{*}{\ce{CH2NH}} & \multirow{2}{*}{ 1 }&  \multirow{2}{*}{ $34.0 \pm 2.0$} &  $(5.3 \pm 0.5)\times 10^{14}$ & 1.6\\
           &  &  &  $(4.8 \pm 0.6)\times 10^{16}$ & 0.12\\
         \ce{CH3NH2} & 1 &  $9.2 \pm 0.8$ &  $(17.0 \pm 3.0)\times 10^{14}$ & 1.6\\
          & 2 &  $25.5 \pm 3.8$ &  $(10.0 \pm 3.0)\times 10^{16}$ & 0.12\\
         \ce{NH2CHO} & 1 & $18.3 \pm 1.4$ &  $(5.0 \pm 0.9)\times 10^{14}$ & 1.6\\
         & 2 & $55.3 \pm 4.0$ &  $(2.0 \pm 0.3)\times 10^{16}$ & 0.12\\
         \ce{CH3CHO} & 1 & $10.7 \pm 0.6$ &  $(5.1 \pm 0.7)\times 10^{14}$ & 1.6\\
         & 2 & $25.9 \pm 2.5$ &  $(2.6 \pm 0.7)\times 10^{16}$ & 0.12\\
         \multirow{2}{*}{\ce{C2H5OH}} &\multirow{2}{*}{ 1} &\multirow{2}{*}{$35.0 \pm 4.0$} &  $(11 \pm 2)\times 10^{14}$ & 1.6\\
         & & &  $(10.0 \pm 1.5)\times 10^{16}$ & 0.12\\
         \hline
         \multicolumn{5}{c}{pSSC\,5}\\
         \hline
         \multirow{2}{*}{\ce{CH2NH}} & \multirow{2}{*}{1} & \multirow{2}{*}{ $30.0 \pm 3.0$} &  $(2.8 \pm 0.4)\times 10^{14}$ & 1.6\\
         &  &  &  $(2.5 \pm 0.3)\times 10^{16}$ & 0.12\\
         \ce{CH3NH2} & 1 & $11.0 \pm 1.1$ &  $(9.0 \pm 2.0)\times 10^{14}$ & 1.6 \\
         & 2 & $37.8 \pm 5.7$ &  $(8.0 \pm 2.0)\times 10^{16}$ & 0.12 \\
         \multirow{2}{*}{\ce{NH2CHO} }& \multirow{2}{*}{1} &\multirow{2}{*}{ $26.5 \pm 1.0$} &  $(3.4 \pm 0.3)\times 10^{14}$ & 1.6\\
          &  &  &  $(3.0 \pm 0.3)\times 10^{16}$ & 0.12\\
         \ce{CH3CHO} & 1 & $13.9 \pm 1.3$ &  $(3.2 \pm 0.5)\times 10^{14}$ & 1.6\\
         & 2 & $34.3 \pm 8.4$ &  $(1.3 \pm 0.6)\times 10^{16}$ & 0.12\\
        \multirow{2}{*}{ \ce{C2H5OH} }& \multirow{2}{*}{1} &\multirow{2}{*}{ $52.0 \pm 12.0$} &  $(8.0 \pm 2.0)\times 10^{14}$ & 1.6\\
          &  &  &  $(7.0 \pm 2.0)\times 10^{16}$ & 0.12\\
         \hline
         \multicolumn{5}{c}{pSSC\,2}\\
         \hline
         \multirow{2}{*}{\ce{CH2NH}} & \multirow{2}{*}{1 }& \multirow{2}{*}{$31.0 \pm 2.0$} &  $(3.2 \pm 0.4)\times 10^{14}$ & 1.6 \\
          &  &  &  $(2.8 \pm 0.3)\times 10^{16}$ & 0.12 \\
         \ce{CH3NH2} & 1 & $8.8 \pm 0.8$ &  $(10.0 \pm 2.0)\times 10^{14}$ & 1.6 \\
         & 2 & $65.8 \pm 37.8$ &  $(11.0 \pm 4.0)\times 10^{16}$ & 0.12 \\
         \ce{NH2CHO} & 1 & $11.7 \pm 0.8$ &  $(3.9 \pm 0.8)\times 10^{14}$ & 1.6\\
         & 2 & $70.9 \pm 6.7$ &  $(1.0 \pm 0.2)\times 10^{16}$ & 0.12\\
         \multirow{2}{*}{\ce{CH3CHO}} & \multirow{2}{*}{1} &\multirow{2}{*}{ $12.5 \pm 0.6$} &  $(3.6 \pm 0.4)\times 10^{14}$ & 1.6\\
          &  &  &  $(3.3 \pm 0.4)\times 10^{16}$ & 0.12\\
         \multirow{2}{*}{\ce{C2H5OH}} & \multirow{2}{*}{1} & \multirow{2}{*}{$30.0 \pm 2.0$} &  $(8.7 \pm 0.9)\times 10^{14}$ & 1.6\\
          &  &&  $(7.8 \pm 0.8)\times 10^{16}$ & 0.12\\
         \hline
    \end{tabular}
    \tablefoot{For the species for which we fitted only one temperature component, we performed the rotational diagrams for the two possible sizes of emission, i.e. $1.6\arcsec$ (GMC scale) and $0.12\arcsec$ (SSC scale). If two gas components could be fitted, we considered that the first component emits from GMC scales whilst the second component emits from pSSC scales (see text). }
\end{table}

\FloatBarrier

\subsection{LTE population diagram for \texorpdfstring{\ce{CH2NH}}{CH2NH}}\label{appx:pop_diagrams}
We performed population diagrams \citep{goldsmith_1999} on \ce{CH2NH} to understand which of the optical depth or non-LTE effects affect the most the rotational diagrams we performed in Sec.~\ref{subsec:rds} for this species. We used the optical depths and the emission size derived from the LVG analysis (see Sec.~\ref{sec: nonlte}). The population diagrams are displayed below. As mentioned in Sec.~\ref{sec: nonlte}, the transitions at 225554 MHz and 172267 MHz were not included, as they were left out of the LVG analysis (hence, we do not have the optical depth values for these transitions).

\begin{figure}[ht]
    \centering
    \includegraphics[width=0.9\linewidth]{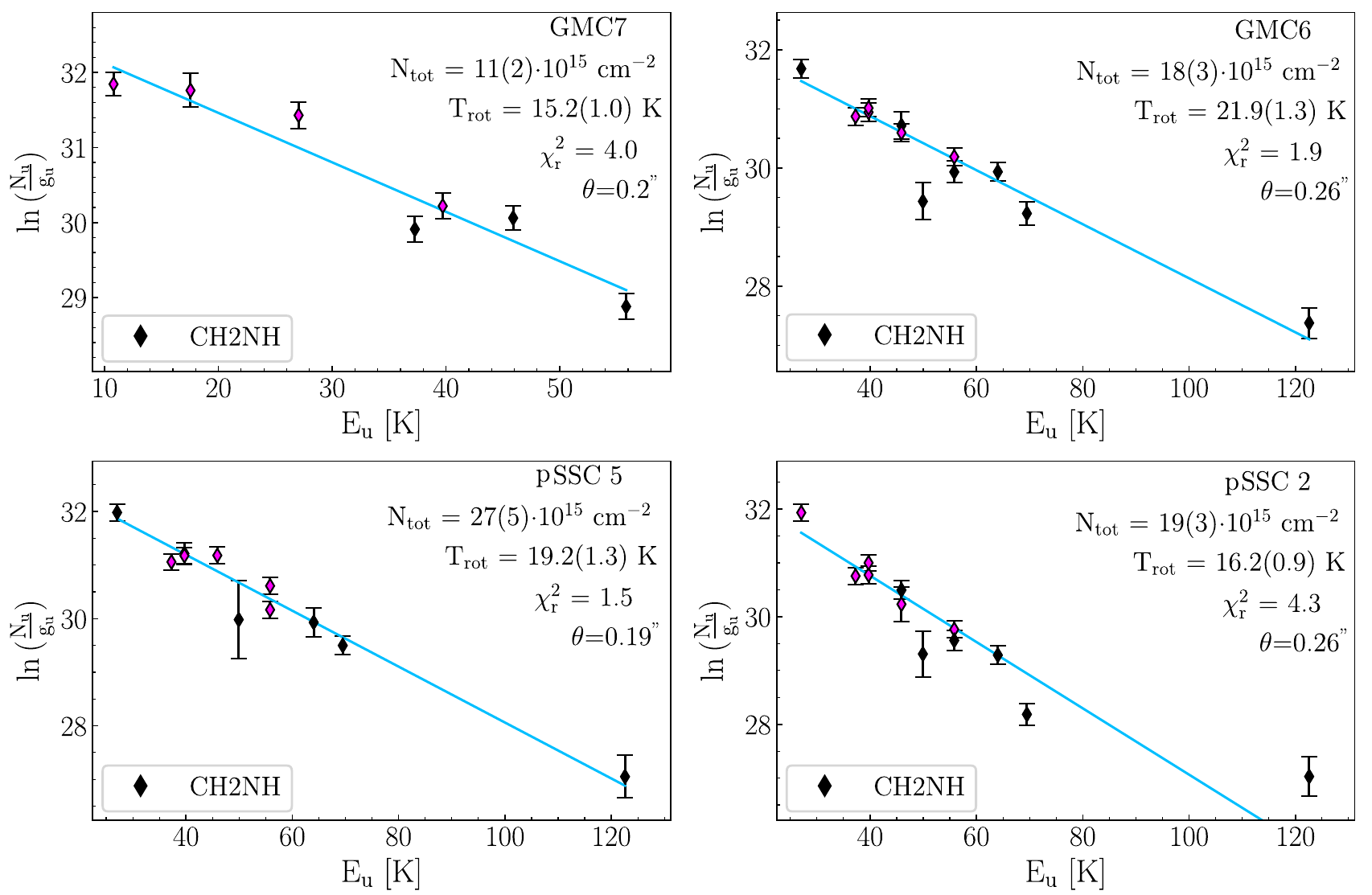}
    \caption{LTE population diagrams of \ce{CH2NH} for each region. The best-fit to the data point is shown by a filled blue line. Lines which were corrected for the optical depth are shown in magenta. Deviations from a straight lines are still present, and $T_{\text{rot}}$ are lower than the $T_{\text{gas}}$ derived from the LVG analysis. This indicates that non-LTE effects are present.}
    \label{fig:pop_diagrams}
\end{figure}

\section{Contour maps between the surveyed iCOMs and other species}
We show here the various contour superposition maps as a support for the discussion in Sections~\ref{sec:chemical_links} and \ref{sec:origin}.

\begin{figure}
    \centering
    \includegraphics[width=1\linewidth]{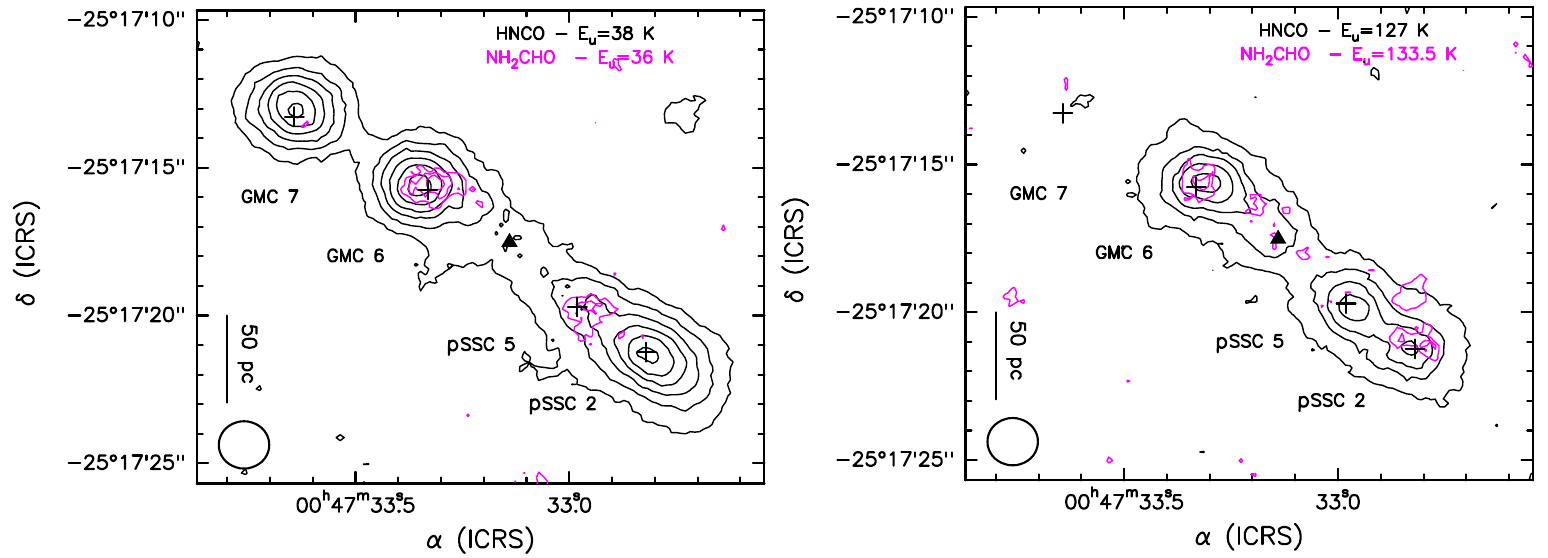}
    \caption{Overlap of contours corresponding to the emission of \ce{NH2CHO} (magenta) and HNCO (black). A spatial scale of 50 pc corresponds to $\sim 3\arcsec$. The kinematic centre ($\alpha_{\text{ICRS}}=00^h47^m33.14^s$ and $\delta_{\text{ICRS}}=-25^\circ 17'17.52''$; \citealt{muller-sanchez2010}) is labelled by a filled black triangle. The black crosses mark the position of the positions analysed in this work and corresponds to GMC\,7, GMC\,6, pSSC\,5 and pSSC\,2. The beam size if depicted in the lower left corner of each  box. \textit{Left} Contours overlap of \ce{NH2CHO} ($8_{0,8}-7_{0,7}$; $E_{\mathrm{u}}=36$ K) and HNCO ($8_{0,8}-7_{0,7}$; $E_{\mathrm{u}}=38$ K). Contours start at $3\sigma$ with steps of 2$\sigma$ and $5\sigma$, respectively. The $1\sigma$ level are 58 and 210 \mjybeamkms, respectively. \textit{Right:} Contours overlap of \ce{NH2CHO} ($15_{2,14}-14_{2,13}$; $E_{\mathrm{u}}=133.5$ K) and HNCO ($15_{0,15}-14_{0,14}$; $E_{\mathrm{u}}=127$ K). Contours start at $3\sigma$ with steps of 2$\sigma$ and $10\sigma$, respectively. The $1\sigma$ level are 78 and 204 \mjybeamkms, respectively. }
    \label{fig:nh2cho_hnco_contour_maps}
\end{figure}

\begin{figure}
    \centering
    \includegraphics[width=1\linewidth]{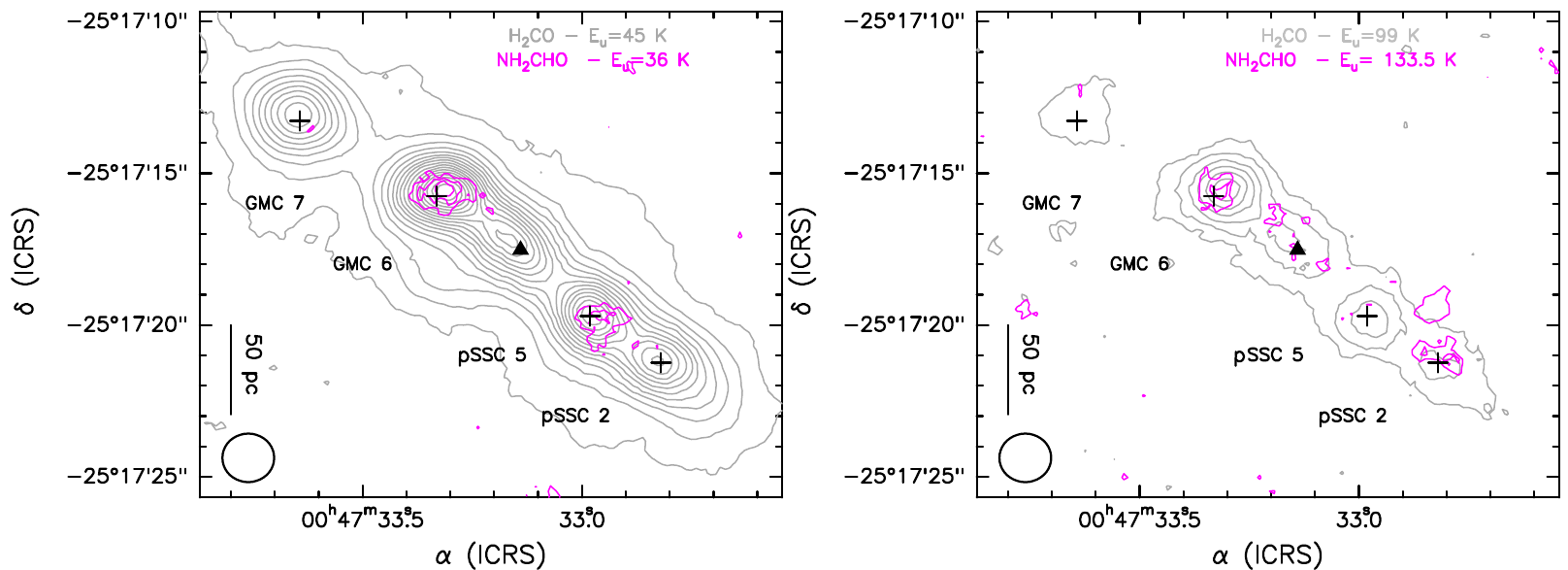}
    \caption{Same as Figure~\ref{fig:nh2cho_hnco_contour_maps} but with \ce{NH2CHO} (magenta) and \ce{H2CO} (grey). \textit{Left} Contours overlap of \ce{NH2CHO} ($8_{0,8}-7_{0,7}$; $E_{\mathrm{u}}=36$ K) and \ce{H2CO} ($4_{1,4}-3_{1,3}$; $E_{\mathrm{u}}=45$ K). Contours start at $3\sigma$ with steps of 2$\sigma$ and $10\sigma$, respectively. The $1\sigma$ level are 58 and 96.5 \mjybeamkms, respectively. \textit{Right:} Contours overlap of \ce{NH2CHO} ($15_{2,14}-14_{2,13}$; $E_{\mathrm{u}}=133.5$ K) and \ce{H2CO} ($5_{2,3}-4_{2,2}$; $E_{\mathrm{u}}=100$ K). Contours start at $3\sigma$ with steps of 2$\sigma$ and $8\sigma$, respectively. The $1\sigma$ level are 78 and 140 \mjybeamkms, respectively. }
    \label{fig:nh2cho_h2co_contour-maps}
\end{figure}

\begin{figure}
    \centering
    \includegraphics[width=1\linewidth]{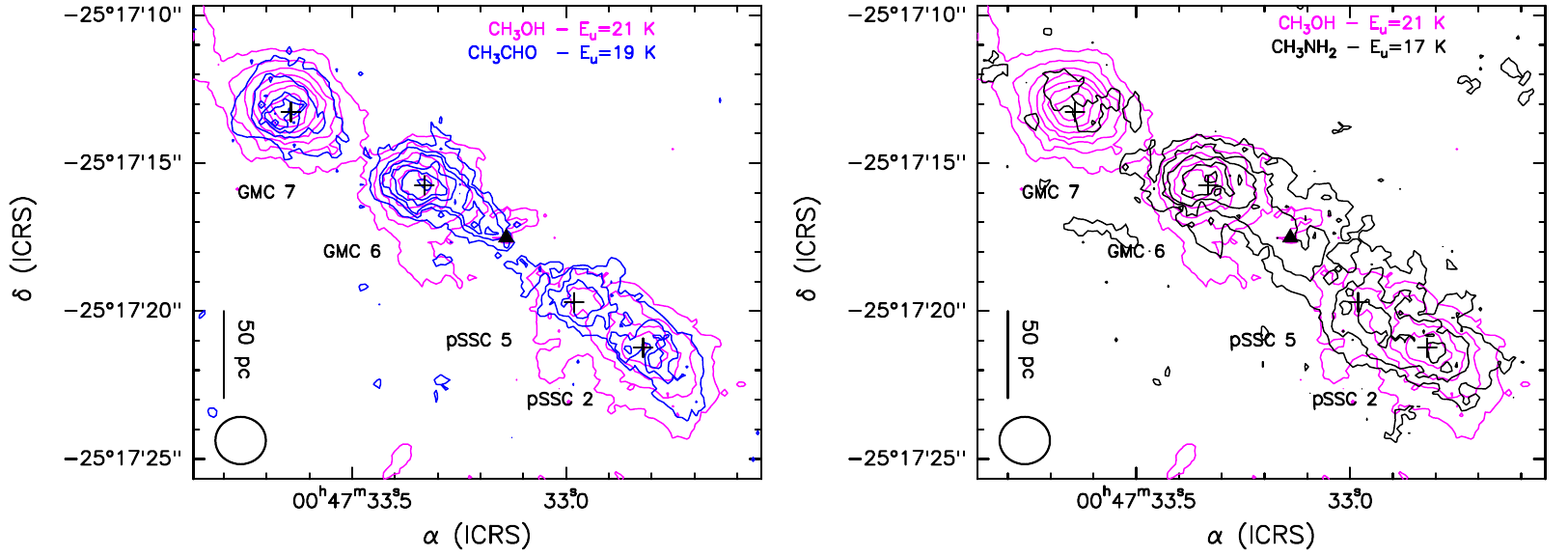}
    \caption{Same as Figure~\ref{fig:nh2cho_hnco_contour_maps} but with \ce{CH3OH} (magenta), \ce{CH3CHO} (blue) and \ce{CH3NH2} (black). \textit{Left} Contours overlap of \ce{CH3OH} ($2_{-1}-1_{-1}$ A; $E_{\mathrm{u}}=21.6$ K) and \ce{CH3CHO} ($1_{1,5}-1_{0,5}$ (E+A); $E_{\mathrm{u}}=19$ K). Contours start at $3\sigma$ with steps of 5$\sigma$ and $3\sigma$, respectively. The $1\sigma$ level are 10.7 and 25 \mjybeamkms, respectively. \textit{Right:} Contours overlap of \ce{CH3OH} ($2_{-1}-1_{-1}$ A; $E_{\mathrm{u}}=21.6$ K) and \ce{CH3NH2} ($3_{-1,1}-2_{0,0}$; $E_{\mathrm{u}}=17$ K). Contours start at $3\sigma$ with steps of 5$\sigma$ and $3\sigma$, respectively. The $1\sigma$ level are 10.7 and 42 \mjybeamkms, respectively.  }
    \label{fig:ch3oh_sup_mols}
\end{figure}

\end{appendix}
\end{document}